\documentclass[%
superscriptaddress,
reprint,
amsmath,amssymb,
aps,
prx,
longbibliography,
]{revtex4-2}

\usepackage[pdftex]{graphicx}
\graphicspath{{./}}

\usepackage{braket}

\usepackage{comment}
\usepackage{dcolumn}
\usepackage{bm}
\usepackage{hyperref}
\usepackage{amsmath}
\begin{document}
	
	\title{Remote Entanglement of Superconducting Qubits via Solid-State Spin Quantum Memories}
	
	
	\author{Hodaka Kurokawa}

	\address{Institute of Advanced Sciences, Yokohama National University, 79-5 Tokiwadai, Hodogaya, Yokohama 240-8501, Japan}
	\author{Moyuki Yamamoto}
	\address{Department of Physics, Graduate School of Engineering Science, Yokohama National University,  79-5 Tokiwadai, Hodogaya, Yokohama 240-8501, Japan }
	\author{Yuhei Sekiguchi}
	\address{Institute of Advanced Sciences, Yokohama National University, 79-5 Tokiwadai, Hodogaya, Yokohama 240-8501, Japan}
	\author{Hideo Kosaka}
	\email[E-mail: ]{kosaka-hideo-yp@ynu.ac.jp}
	\address{Institute of Advanced Sciences, Yokohama National University, 79-5 Tokiwadai, Hodogaya, Yokohama 240-8501, Japan}
	\address{Department of Physics, Graduate School of Engineering Science, Yokohama National University,  79-5 Tokiwadai, Hodogaya, Yokohama 240-8501, Japan }

	
	\begin{abstract}
		Quantum communication between remote superconducting systems is being studied intensively to increase the number of integrated superconducting qubits and to realize a distributed quantum computer. Since optical photons must be used for communication outside a dilution refrigerator, the direct conversion of microwave photons to optical photons has been widely investigated. However, the direct conversion approach suffers from added photon noise, heating due to a strong optical pump, and the requirement for large cooperativity. Instead, for quantum communication between superconducting qubits, we propose an entanglement distribution scheme using a solid-state spin quantum memory that works as an interface for both microwave and optical photons. The quantum memory enables quantum communication without significant heating inside the refrigerator, in contrast to schemes using high-power optical pumps. Moreover, introducing the quantum memory naturally makes it possible to herald entanglement and parallelization using multiple memories.
	
\end{abstract}
	
	\maketitle
	
	\section{Introduction}

	A superconducting qubit is one of the most promising building blocks for a quantum computer \cite{Kjaergaard2020}. 
	Recent developments in superconducting qubits enabled the integration of several tens of qubits onto a single chip and well-controlled quantum operation on them \cite{Arute2019}. However, for practical calculations, integration on the order of one million qubits is estimated to be required \cite{Ahsan2015}, which is an extremely challenging task considering the current technologies. Although quantum memories may alleviate this requirement by several orders of magnitude \cite{Gouzien2021},  tens of thousands of qubits are still needed. To mitigate the  severe demands concerning the integration of qubits, distributed quantum computing is suggested as an alternative \cite{Yimsiriwattana2004,VanMeter2016,Cuomo2020}. Similar to a distributed classical computer, a distributed quantum computer is realized by linking small quantum computers using classical and quantum communication channels.

	To realize quantum communication between superconducting qubits in remote dilution refrigerators, an optical photon must be used for communication outside the dilution refrigerators. Thus, the direct conversion of microwave photons to optical photons has been intensively investigated in the last decade \cite{Lambert2020,Lauk2020}. However, in the direct conversion scheme, high conversion efficiency is generally not compatible with the low optical pump power necessary to avoid heating up inside the refrigerator and to avoid generating noise. 
	Instead, schemes using entangled photons to generate entanglements between superconducting qubits have recently gained attention \cite{Zhong2020, Krastanov2021, Wu2021}. These schemes use entangled-photon pairs, which have already been widely adapted for entanglement distribution at telecom wavelengths \cite{Yamamoto2003,Li2004,Peng2005}. For quantum communication with optical photons, the application of a quantum memory is another choice to distribute entanglement between remote nodes \cite{Duan2001,Moehring2007,Hofmann2012,Bernien2013,Delteil2016}.  
	Thus, using an analogy with communication at telecom wavelengths, we propose a quantum memory as an interface for microwave and optical photons, which works as a quantum repeater to generate entanglements between superconducting qubits in different dilution refrigerators.

	In this article, we introduce a scheme to generate entanglement between superconducting qubits in different refrigerators via a quantum memory. To analyze the scheme's performance, we consider a specific situation : generating entanglement between remote superconducting qubits using a quantum interface \cite{Neuman2021} consisting of a color center in diamond, a microwave resonator, and an optomechanical cavity. Assuming the current state-of-the-art technologies, we estimate the entanglement generation rate of the memory-based scheme. Since it is not trivial whether the memory-based scheme has any advantages over other schemes, we compare its performance with those of the direct conversion scheme and the entangled-photon scheme. We confirm that as long as the conversion efficiency of the direct conversion scheme does not exceed a certain threshold, it is beneficial to introduce the memory for the entanglement generation. Besides, though the entanglement generation rates of the memory-based and entangled-photon schemes are estimated to be on the same order of magnitude, heat generation inside the refrigerator could be reduced more in the memory-based scheme than in the entangled-photon scheme. Since decreasing heating in the refrigerator is crucial for scaling quantum devices, using a memory could not only deliver entanglement but also help the thermal design inside a refrigerator, contributing to the establishment of a distributed quantum computer.
	\newpage
	
		\begin{figure*}
		\begin{center}
			\includegraphics[width=150 mm]{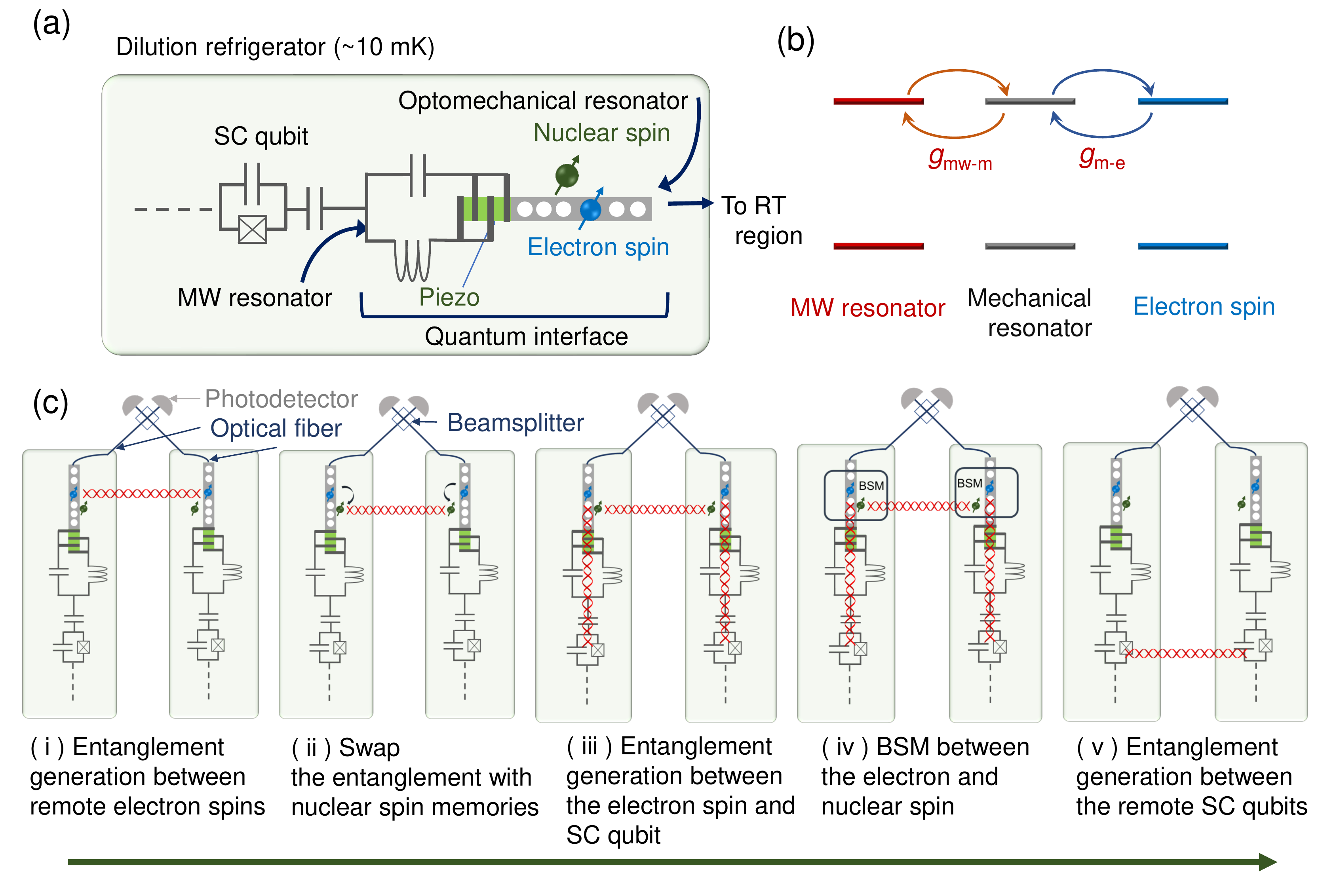}
		\end{center}
		\caption{(a) Schematic of a quantum interface inside a dilution refrigerator. The interface consists of a color center spin, a nuclear spin, a microwave resonator, and an optomechanical resonator. The microwave resonator is connected to a superconducting qubit via a waveguide. The optomechanical crystal is linked to an optical fiber, and an optical photon is transmitted outside the refrigerator.  (b) Schematic of the energy level of cascaded resonators  (qubits)  with electro-mechanical interaction, $g_\textrm{mw-m}$, and strain(mechanical)-spin coupling, $g_\textrm{m-e}$. (c) Entanglement generation protocol between remote superconducting qubits. (i) Entanglement is generated between remote electron spins. (ii) The entanglement is swapped to nearby nuclear spins. (iii) Entanglement is generated between the electron spin and the superconducting qubit in each node. (iv) Bell-state measurement (BSM) is performed between the electron and the nuclear spin. (v) Entanglement is generated between remote superconducting qubits.
		}
		\label{fig:schematic}
	\end{figure*}

	\section{entanglement generation protocol between remote superconducting qubits}

	This section overviews the entanglement generation protocol between superconducting systems in remote dilution refrigerators using a quantum interface based on the solid-state spin quantum memory. Figure \ref{fig:schematic} (a) shows a schematic of a quantum interface, consisting of a color center with nuclear spin memories in diamond and several resonators, that is connected to an optical fiber at the room temperature region and to a superconducting qubit via a waveguide. The device works as an interface for both microwave and optical photons. To obtain a single microwave photon coupling that is strong enough to transfer the quantum state efficiently, different kinds of resonators are integrated into the interface. Figure  \ref{fig:schematic} (b) shows a simplified model of the coupled resonators. Since a color center with the orbital degree of freedom strongly interacts with the strain (phonon) due to the spin-orbit interaction \cite{Lee2016,Meesala2018}, 
	a microwave photon is converted to a microwave phonon via the piezoelectric effect. Then, the phonon confined inside the one-dimensional optomechanical crystal cavity interacts with a color center through the strain (mechanical)-electron spin coupling. Here, we consider NV$^{-}$, NV$^0$, and SiV$^-$ as candidates that have high strain sensitivity in their ground or excited state.
	Similarly, coupling between a color center and an optical photon is enhanced by the confinement of the optical field inside the optomechanical cavity. 
	
	Figure \ref{fig:schematic} (c) shows a protocol for the remote entanglement generation between superconducting qubits using the quantum interface. We consider nodes inside dilution refrigerators separated by several meters. First, entanglement between remote electron spins is generated using a beamsplitter that erases optical path information. Then, the entanglement is swapped to nearby nuclear spin memories. Next, entanglement between the superconducting qubit and the electron spin is generated in each node. Finally, Bell-state measurement (BSM) is performed between the electron and nuclear spins in each node, generating entanglement between remote superconducting qubits. In the following, we will analyze and discuss each part of the protocol in terms of the entanglement generation rate. 
	
	\section{remote entanglement between solid-state spins}
	Entanglement generation between remote color centers in diamonds has already been demonstrated using two-photon and single-photon protocols \cite{Bernien2013,Humphreys2018,Pompili2021,Levonian2021} (see Appendix \ref{app-sec:single-or-two-photon-protocol} for details on single- and two- photon protocols). The two-photon protocol projects spins within different nodes (A and B) on one of the two Bell states,
	\begin{equation}
		\frac{1}{\sqrt{2}}(\ket{0}_\textrm{e}^\textrm{A}\ket{1}_\textrm{e}^\textrm{B}\pm\ket{1}_\textrm{e}^\textrm{A}\ket{0}_\textrm{e}^\textrm{B}),
	\end{equation}
	where the sign depends on whether the same detector ($+$) or different detector ($-$) is clicked in the two-photon emission rounds \cite{Bernien2013}. $\ket{0}$ ($\ket{1}$) denotes the spin state in the computational basis, the subscript e indicates the electron spin, and the superscripts A and B correspond to the two nodes. Since the two-photon protocol is tolerant to the error caused by photon loss and does not require optical path length stabilization, in the following we consider only the two-photon protocol.

	Then, we consider a practical limit of the entanglement generation rate between color centers separated by several meters. 
	The entanglement generation rate between remote color centers, $R_\textrm{e-e}$, is expressed as 
	\begin{equation}\label{eq:entangle-rate}
		R_\textrm{e-e}=\frac{1}{2}r_\textrm{e-e}(\eta_\textrm{e}^\textrm{opt})^2,
	\end{equation}
	where $\eta_\textrm{e}^\textrm{opt}$ is the detection efficiency of the optical photon emitted from a color center considering all losses during the transmission and $r_\textrm{e-e}$ is the entanglement generation attempt rate of the electron spins. The factor 2 in the denominator of eq.(\ref{eq:entangle-rate}) indicates that only two of the four Bell states can be heralded using a beamsplitter. Note that $\eta_\textrm{e}^\textrm{opt}\sim1$ is already feasible considering the current state-of-the-art technologies (see Appendix \ref{app-sec:trans-eff}).

	For color centers separated by 10 m, the propagation time of an optical photon in an optical fiber is 25 ns from a color center to a photodetector placed at the midpoint of the refrigerators (5 m). Then, the ideal entanglement attempt rate will be 40 MHz. However, in such a case, the actual entanglement attempt rate is limited by the gate operation time for the emission of a coherent photon. In a two-photon protocol that used the NV$^{-}$ center, the typical operation time to generate a coherent photon was $\sim10$ $\mu$s \cite{Bernien2013,Humphreys2018}, corresponding to $r_\textrm{e-e}=100$ kHz. Thus, assuming $\eta_\textrm{e}^\textrm{opt}=$ 0.9-1, $R_\textrm{e-e}$ would be 80-100 kHz. Actually, considering that the initialization of a color center and nuclear spin takes several tens of $\mu$s \cite{Humphreys2018}, the whole entanglement generation rate between the electron spins would be limited to a few tens of kilohertz.

	\section{Entanglement generation between the superconducting qubit and the spin}\label{sec:entanglement-sc-spin}
	
	\begin{figure*}
		\begin{center}
			\includegraphics[width=150mm]{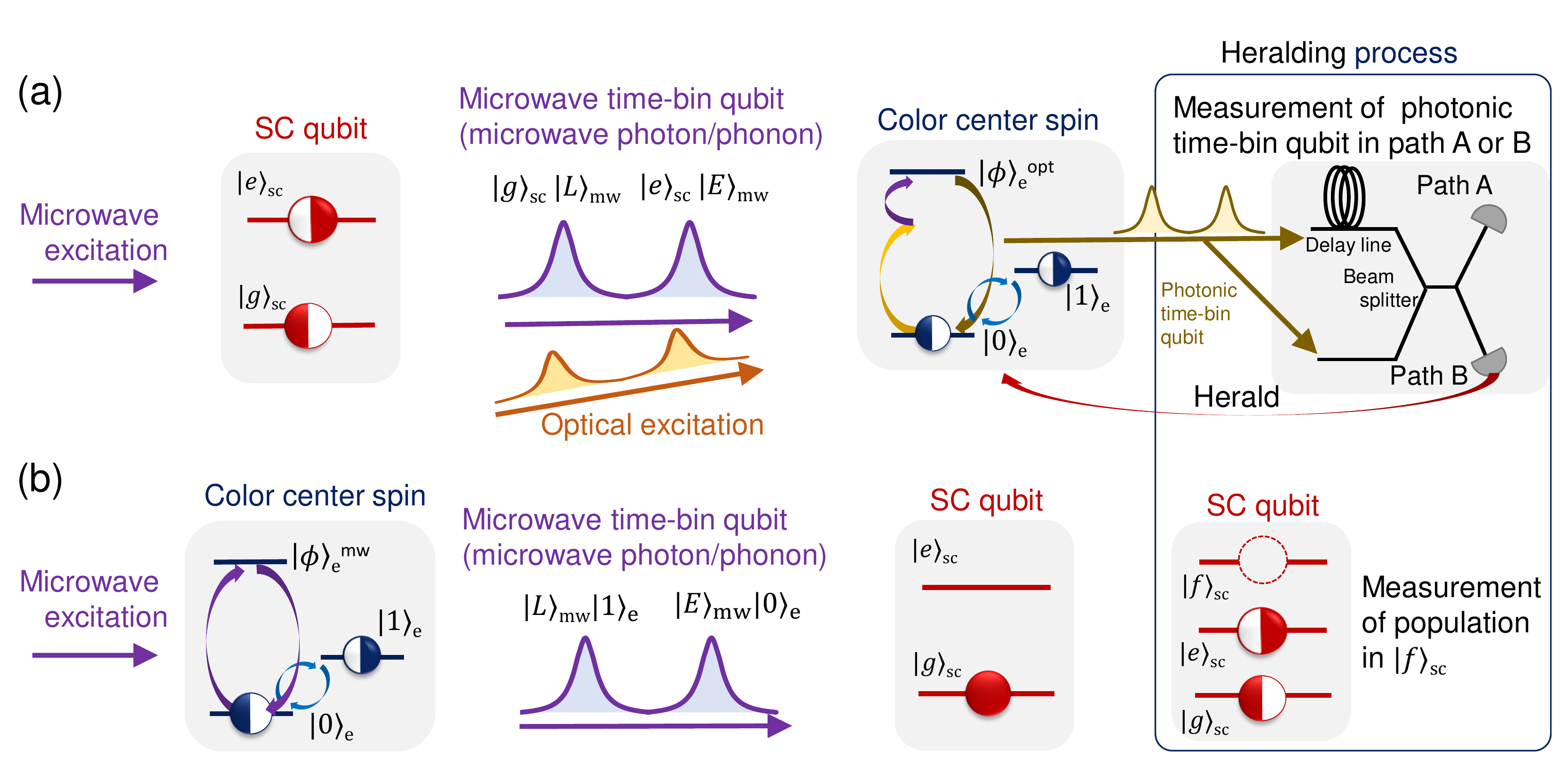}
		\end{center}
		\caption{Protocols to generate entanglement between a superconducting qubit and spin. (a) The superconducting qubit, which is in the superposition state, $(1/\sqrt{2})(\ket{g}_\textrm{sc}+\ket{e}_\textrm{sc})$,  generates a microwave time-bin qubit. Here, for simplicity, the superconducting qubit is depicted as a two-level system, though another energy level and a coupled resonator are actually used to generate the microwave time-bin qubit \cite{Kurpiers2019}. The electron spin is prepared in the state  $(1/\sqrt{2})(\ket{0}_\textrm{e}+\ket{1}_\textrm{e})$. The electron spin, which is in $\ket{0}_\textrm{e}$, receives the microwave early time-bin qubit with a simultaneous optical excitation. The successful excitation to the optically excited state, $\ket{\phi}_\textrm{e}^\textrm{opt}$, results in the emission of an optical early time-bin qubit. The repetition of the procedure with bit-flip between $\ket{0}_\textrm{e}$ and $\ket{1}_\textrm{e}$ generates the optical time-bin qubit, which is measured using a delay line and a beamsplitter. The detection of the optical photon heralds the successful absorption of the microwave time-bin qubit. 	Applying a correction to the computational basis of the spin depending on the measurement outcome teleports the state of the microwave time-bin qubit to the spin, generating entanglement between the superconducting qubit and the spin.  (b) The electron spin is prepared in the state  $(1/\sqrt{2})(\ket{0}_\textrm{e}+\ket{1}_\textrm{e})$. A microwave driving pulse is applied to excite $\ket{0}_\textrm{e}$ to $\ket{\phi}_\textrm{e}^\textrm{mw}$, which couples to a mechanical mode of the optomechanical crystal. $\ket{\phi}_\textrm{e}^\textrm{mw}$ emits the microwave early time-bin qubit via the optomechanical crystal and microwave cavity. Repeating the emission with bit-flip between $\ket{0}_\textrm{e}$ and $\ket{1}_\textrm{e}$ generates a microwave time-bin qubit.  The superconducting qubit receives the microwave time-bin qubit, generating entanglement between the spin and the superconducting qubit. The successful absorption of the microwave time-bin qubit can be heralded by confirming the absence of population in the second excited state, $\ket{f}_\textrm{sc}$, of the superconducting qubit \cite{Kurpiers2019}.}
		\label{fig:spin-sc-scheme1and2_1}
	\end{figure*}

	In this section, we propose two protocols to generate entanglement between the superconducting qubit and the electron spin in diamond. We then consider the entanglement generation rate based on these protocols. In the following, we consider that $\ket{0}_\textrm{e}$ is the spin ground state in the computational basis that is excited by the microwave time-bin qubit.

	Figure \ref{fig:spin-sc-scheme1and2_1} (a) shows the first protocol, which uses an optical photon to herald the successful entanglement. The protocol can be applied to a color center that has high strain susceptibility in its ground or excited state. In Appendix \ref{app-sec:absorb_emit_microwave_using_spin}, we discuss the feasibility of the protocol considering NV$^{-}$ and NV$^0$ as an example.

	 First, the superconducting qubit, which is prepared in the state, $(1/\sqrt{2})(\ket{g}_\textrm{sc}+\ket{e}_\textrm{sc})$, emits a coherent microwave time-bin qubit \cite{Kurpiers2018,Kurpiers2019}, resulting in the state,
	\begin{equation}
		\frac{1}{\sqrt{2}}(\ket{g}_\textrm{sc}\ket{L}_\textrm{mw}+\ket{e}_\textrm{sc}\ket{E}_\textrm{mw}).
	\end{equation}
 	Here, $\ket{g}_\textrm{sc}$ and $\ket{e}_\textrm{sc}$ denote the superconducting qubit basis states, and $\ket{L}_\textrm{mw}$ ( $\ket{E}_\textrm{mw}$) is the microwave late (early) time-bin qubit. The electron spin is prepared in the state $(1/\sqrt{2})(\ket{0}_\textrm{e}+\ket{1}_\textrm{e})$. Then, $\ket{0}_\textrm{e}$ is excited by $\ket{E}_\textrm{mw}$, which is converted to a microwave phonon via the piezoelectric effect.  The spin is subjected to an optical pump at the same time as the absorption of $\ket{E}_\textrm{mw}$ by the spin (see Appendix \ref{app-sec:absorb_emit_microwave_using_spin} for details). The detuning of the optical pump is equal to the frequency of $\ket{E}_\textrm{mw}$. As a result of the successful absorption of $\ket{E}_\textrm{mw}$, the spin is excited to an optically excited state, $\ket{\phi}_\textrm{e}^\textrm{opt}$. Then, the spin emits a photonic early time-bin qubit, $\ket{E}_\textrm{opt}$. After the absorption of $\ket{E}_\textrm{mw}$, $\ket{0}_\textrm{e}$ and $\ket{1}_\textrm{e}$ are flipped. The procedure of photon emission is repeated for $\ket{L}_\textrm{mw}$. The spin emits a photonic late time-bin qubit, $\ket{L}_\textrm{opt}$, conditioned on the successful absorption of $\ket{L}_\textrm{mw}$. After that, $\ket{0}_\textrm{e}$ and $\ket{1}_\textrm{e}$ are flipped again. Consequently, the spin entangles with $\ket{E}_\textrm{opt}$ and $\ket{L}_\textrm{opt}$ as 
		\begin{equation}
		\frac{1}{\sqrt{2}}(\ket{g}_\textrm{sc}\ket{L}_\textrm{opt}\ket{1}_\textrm{e}+\ket{e}_\textrm{sc}\ket{E}_\textrm{opt}\ket{0}_\textrm{e}).
	\end{equation}
	Using a delay line and a beamsplitter, photonic time-bin qubits are measured on the basis $(1/\sqrt{2})(\ket{0}_\textrm{opt}^\textrm{A}\ket{1}_\textrm{opt}^\textrm{B}\pm\ket{1}_\textrm{opt}^\textrm{A}\ket{0}_\textrm{opt}^\textrm{B})$. Detection of a single photon at the photodetectors heralds the successful interaction between the microwave time-bin qubit and the spin. The spin state is then corrected depending on the measurement outcome. Consequently, the state of the microwave time-bin qubit is teleported into the spin state with the heralding of the optical photon as 
	\begin{equation}
		\frac{1}{\sqrt{2}}(\ket{g}_\textrm{sc}\ket{1}_\textrm{e}+\ket{e}_\textrm{sc}\ket{0}_\textrm{e}).
	\end{equation}

	It is noteworthy that different protocols have been suggested to transfer the state of a superconducting qubit to the electron spin (NV$^-$) in diamond \cite{Douce2015,Li2018}. The scheme proposed in Ref.\cite{Douce2015} uses a superconducting flux qubit coupled to a single NV$^-$.  To maximize the interaction between the superconducting qubit and NV$^-$, NV$^-$ is placed in close proximity to the flux qubit ($\sim15$ nm). However, such a configuration may hinder integration of an optical cavity, since the electrode that is too close to the NV$^-$ center contribute to the loss of the optical photon, degrading the quality factor of the optical cavity. Thus, the protocol is not compatible with our scheme, which uses a photonic (optomechanical) cavity to enhance spin-photon interaction. In another approach, a microwave photon emitted from a superconducting qubit is converted to an optical photon using an electro-optic effect and driving laser \cite{Li2018,Park2006,Larsson2009,Schietinger2008,Li2011}. Using the converted optical photon and a driving laser, the ground state of NV$^-$ is controlled through the interaction between a $\Lambda$-type three-level system ($\ket{m_s=\pm1}$ and $\ket{A_2}$ of NV$^-$). Since their scheme already integrates whispering-gallery-mode photonic cavity into the device structure, it could be another choice to transfer the state of a superconducting qubit to the electron spin.

	Figure \ref{fig:spin-sc-scheme1and2_1} (b) shows the second protocol for the entanglement generation, which uses a microwave time-bin qubit emitted from the spin. In this scheme, a color center having high strain susceptibility in the ground state is required (e.g., NV$^0$ and SiV$^-$). In the Appendix \ref{app-sec:absorb_emit_microwave_using_spin}, we consider the detail of the protocol, using NV$^0$ as an example.

	The initial state of the spin is prepared in $(1/\sqrt{2})(\ket{0}_\textrm{e}+\ket{1}_\textrm{e})$. Then, $\ket{0}_\textrm{e}$ is excited to $\ket{\phi}_\textrm{e}^\textrm{mw}$ by a microwave driving pulse supplied from an external signal generator. According to the model shown in Fig. \ref{fig:schematic} (b), the excited spin emits a microwave phonon to the optomechanical cavity (See Appendix \ref{app-sec:numerical_simulation} for the detailed model). Repeating the emission with bit-flip between $\ket{0}_\textrm{e}$ and $\ket{1}_\textrm{e}$ generates the state (see Appendix \ref{app-sec:absorb_emit_microwave_using_spin} for details),  
	\begin{equation}
		\frac{1}{\sqrt{2}}(\ket{L}_\textrm{mw}\ket{1}_\textrm{e}+\ket{E}_\textrm{mw}\ket{0}_\textrm{e}).
	\end{equation}
	The microwave time-bin qubit is transferred to the superconducting qubit via a waveguide coupled to the microwave cavity. The superconducting qubit, which is prepared in $\ket{g}_\textrm{sc}$, absorbs the microwave time-bin qubit, resulting in the entangled state,
	\begin{equation}				    
		\frac{1}{\sqrt{2}}(\ket{g}_\textrm{sc}\ket{0}_\textrm{e}+\ket{e}_\textrm{sc}\ket{1}_\textrm{e}).
	\end{equation}
	By measuring population in the second excited state, $\ket{f}_\textrm{sc}$ of the superconducting qubit, we can confirm whether the superconducting qubit successfully absorbs the microwave time-bin qubit or not, heralding the generation of entanglement \cite{Kurpiers2019}.
	
	A difference between these two protocols is whether the superconducting qubit or the spin is the emitter (receiver) of the microwave time-bin qubit. Efficient ($>90\%$) microwave time-bin qubit generation and absorption by the superconducting qubit have been demonstrated \cite{Kurpiers2019,Ilves2020}, suggesting that the superconducting qubit can be used in both ways with similar efficiency. Thus, the efficiency of microwave emission or absorption by the color center spins determines which protocols to choose. However, this efficiency depends on various parameters: the choice of color center, coupling between the spin and the phonon, coupling between the microwave phonon and the microwave photon, and the quality factor of the optomechanical crystal and the microwave resonator. Thus, in the following, we will treat the microwave photon generation or absorption efficiency of the spin as a parameter. Regarding the efficiency based on feasible device parameters, we calculate and discuss it in the Appendix \ref{app-sec:numerical_simulation}.

	Another difference is that the first protocol uses the two-photon absorption process at the spin while the second protocol uses single-photon absorption at the superconducting qubit. In the first protocol, the frequencies of the superconducting qubit and the spin need not be in perfect resonance, since tuning the frequency of an optical driving laser can compensate the frequency difference between them. Besides, it is easier to detect a single optical photon than a single microwave photon. However, the use of an optical driving laser may cause heating inside the refrigerator without a high-quality photonic crystal cavity to enable the optical drive by a single optical photon. On the other hand, in the second protocol, an optical driving laser is not necessary, thus mitigating the possible heating due to the drive laser and simplifying the measurement system compared to the first protocol. We should note that measurement of the spin on $\ket{+}_\textrm{e}=(1/\sqrt{2})(\ket{0}_\textrm{e}+\ket{1}_\textrm{e})$ basis directly converts a microwave time-bin qubit to an optical time-bin qubit in the first protocol.  Here, however, we focus only on the protocol using the spin as a quantum memory.

	\begin{figure}
		\begin{center}
			\includegraphics[width=80mm]{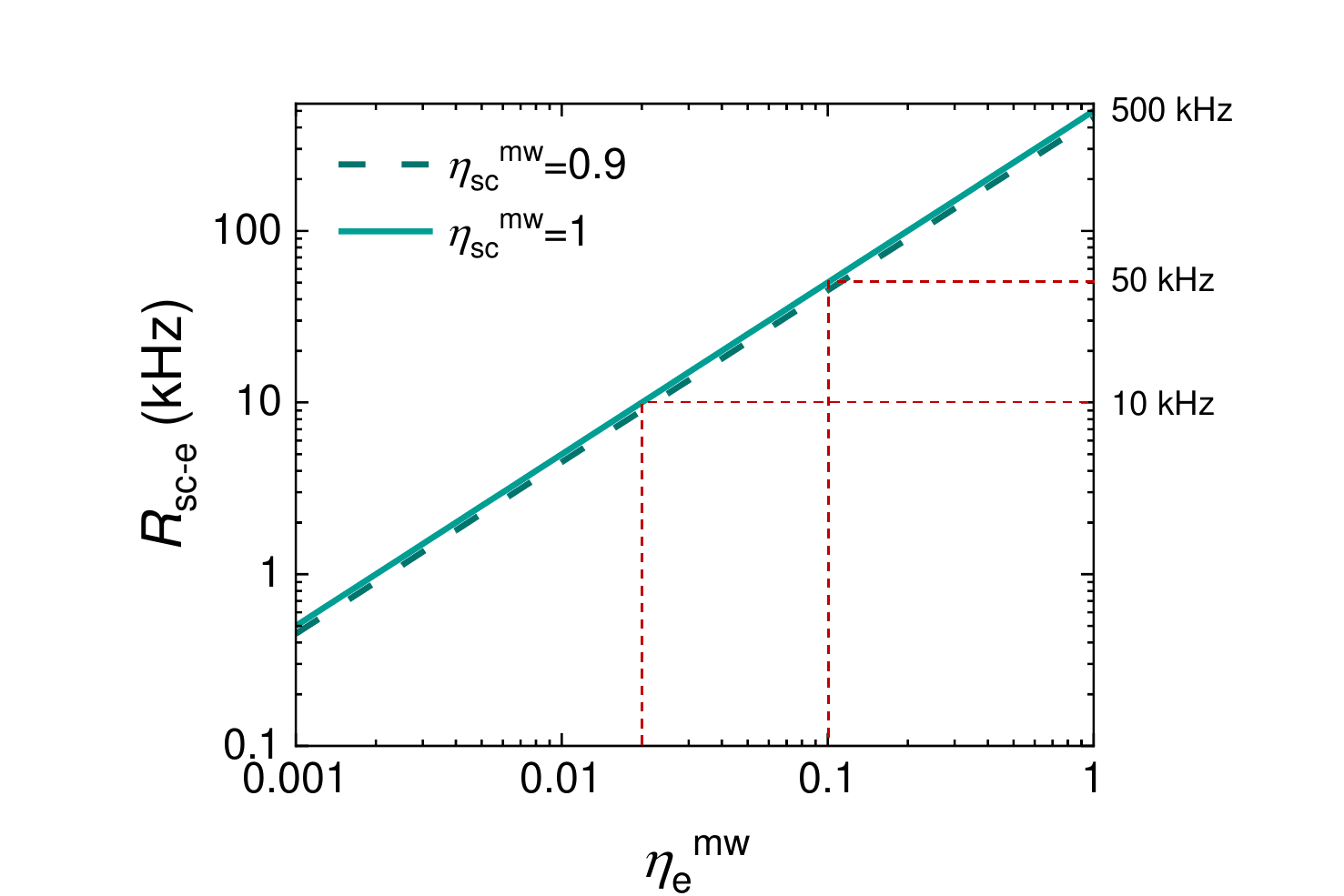}
		\end{center}
		\caption{The entanglement generation rate between the superconducting qubit and electron spin as a function of the absorption (emission) efficiency of a microwave time-bin qubit by the spin. The rate was calculated for different absorption (emission) efficiencies of a microwave time-bin qubit by the superconducting qubit, $\eta_\textrm{sc}^\textrm{mw}$= 0.9, 1.0. The red dashed lines in the graph are shown to clarify the entanglement generation rates of 10 kHz, and 50 kHz, which are considered necessary so as not to seriously degrade the whole entanglement generation rate between the superconducting qubits.}
		\label{fig:entanglement-spin-sc_1}
	\end{figure}
	
	On the basis of both schemes, we estimate an actual entanglement generation rate between the superconducting qubit and the spin. For a superconducting qubit coupled to a cavity, the generation or absorption of a microwave time-bin qubit takes several hundreds of nanoseconds to a few microseconds \cite{Kurpiers2019,Ilves2020}. Here, we assume that it takes 1 $\mu$s to generate or absorb a microwave time-bin qubit for the superconducting qubit. Similarly, for the electron spin, $\sim$1 $\mu$s is thought to be needed for the operation based on the simulation of the master equation (see Appendix \ref{app-sec:numerical_simulation}). Then, ignoring the propagation time in the waveguide between the superconducting qubit and the electron spin, it is possible to attempt entanglement generation at a rate of $r_\textrm{sc-e}\sim$500 kHz, where $r_\textrm{sc-e}$ is the entanglement trial rate. Considering the success probability of microwave photon emission or absorption, the successful entanglement generation rate between the superconducting qubit and spin, $R_\textrm{sc-e}$, can be expressed as
	\begin{equation}
		R_\textrm{sc-e}=r_\textrm{sc-e}\eta_\textrm{sc}^\textrm{mw}\eta_\textrm{e}^\textrm{mw},
	\end{equation}
	where $\eta_\textrm{sc}^\textrm{mw}$ ($\eta_\textrm{e}^\textrm{mw}$) is the probability of successful microwave emission or absorption operation by the superconductor (spin). Here, we can confirm the generation of entanglement with the heralding techniques mentioned above. Figure \ref{fig:entanglement-spin-sc_1} shows $R_\textrm{sc-e}$ as a function of $\eta_\textrm{e}^\textrm{mw}$. Although we calculate $R_\textrm{sc-e}$ for $\eta_\textrm{e}^\textrm{mw}$ = 0.9, 1.0, it does not seriously affect $R_\textrm{sc-e}$. 
	Assuming the entanglement generation rate between remote color centers to be 10 kHz based on the discussion in the previous section, $R_\textrm{sc-e}>10$ kHz do not seriously degrade the whole entanglement generation speed between the superconducting qubits. Thus, $\eta_\textrm{e}^\textrm{mw}=0.1$, corresponding to $R_\textrm{sc-e}\sim50$ kHz, could be sufficient to generate entanglement between the superconducting qubit and the spin. Even $\eta_\textrm{e}^\textrm{mw}\sim0.02$, corresponding to $R_\textrm{sc-e}\sim10$ kHz, could work if the decrease in the whole entanglement generation rate to 5 kHz is acceptable. Although the interaction between a microwave photon and single spin still needs proof-of-concept experiments, our numerical simulation indicates that $\eta_\textrm{e}^\textrm{mw}\sim0.1$ is feasible considering the current technologies and strain sensitivity of a specific color center (see Appendix \ref{app-sec:numerical_simulation}).

	\section{Entanglement generation between remote superconducting qubits}
	\begin{figure*}
		\begin{center}
			\includegraphics[width=150mm]{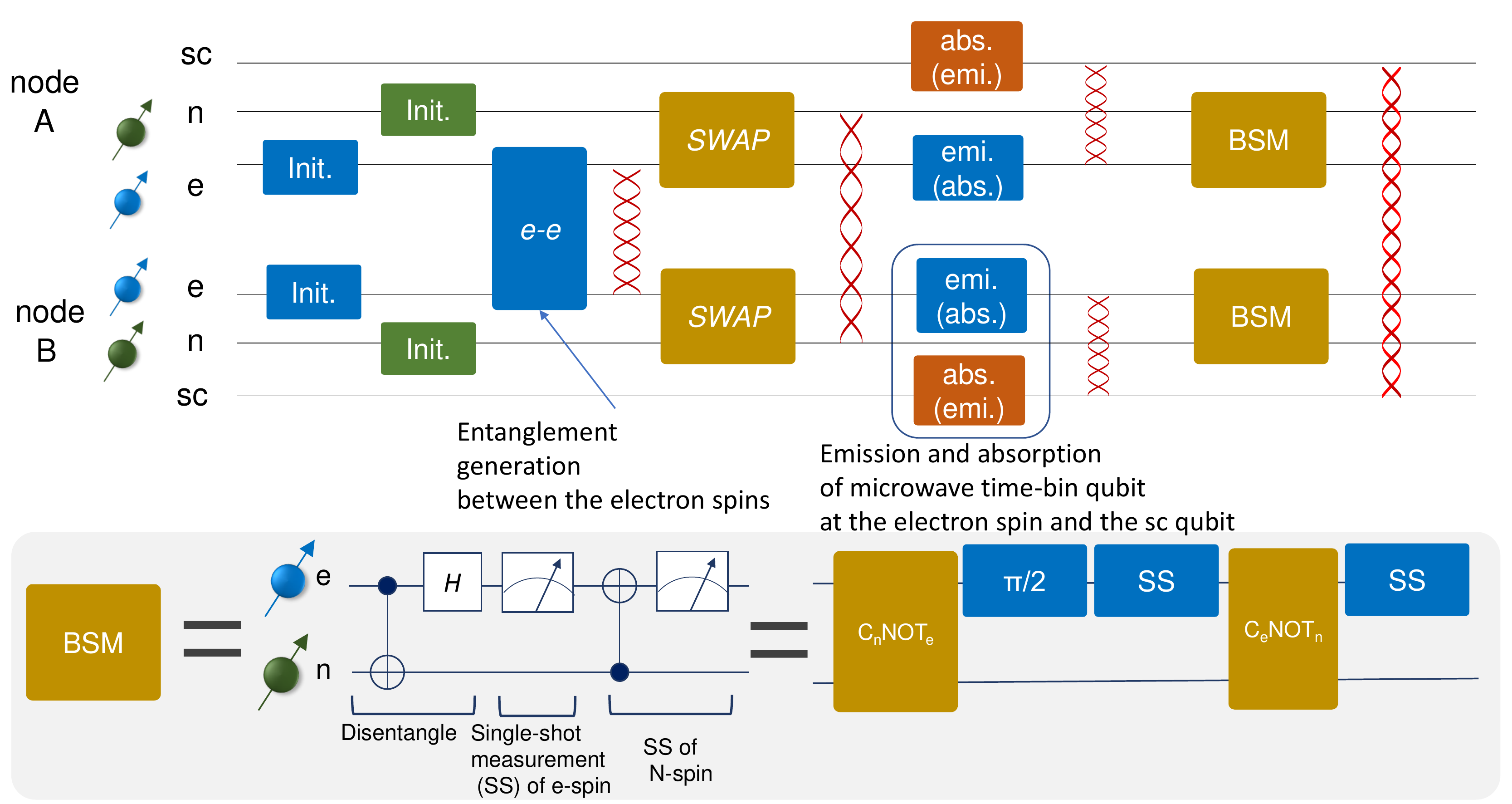}
		\end{center}
		\caption{A gate sequence for the entanglement generation between remote superconducting qubits. Initialization of the electron and nuclear spins (Init.) is followed by entanglement generation between the electron spins in the different nodes (e-e). Entanglement between the electron spins is then swapped to the nuclear spins (SWAP). Entanglement between the superconducting qubit and the electron spin is generated via the absorption (abs.) and emission (emi.) of the microwave time-bin qubit. Finally, BSM is performed for the electron and the nuclear spins to generate entanglement between the superconducting qubits. A BSM sequence is shown in the gray box below the whole gate sequence to estimate the BSM gate time. The state of the electron spin is measured by a single-shot measurement (SS). The times for each operation are summarized in Table \ref{tab:parameters}. }
		\label{fig:whole-gate-sequence}
	\end{figure*}
	After the successful entanglement generation between the superconducting qubit and the spin in nodes A and B, BSM is performed between the electron and nuclear spins in each node (Fig. \ref{fig:whole-gate-sequence}).  Depending on the BSM result, the quantum state of the superconducting qubits in nodes A and B is projected to the states, $(1/\sqrt{2})(\ket{g}_\textrm{sc}^\textrm{A}\ket{g}_\textrm{sc}^\textrm{B}\pm\ket{e}_\textrm{sc}^\textrm{A}\ket{e}_\textrm{sc}^\textrm{B}$) or  $(1/\sqrt{2})(\ket{g}_\textrm{sc}^\textrm{A}\ket{e}_\textrm{sc}^\textrm{B}\pm\ket{e}_\textrm{sc}^\textrm{A}\ket{g}_\textrm{sc}^\textrm{B}$). The overall gate sequence of the entanglement generation is shown in Fig. \ref{fig:whole-gate-sequence}. Also, in Table \ref{tab:parameters}, we summarize the parameters needed to control and read out the quantum states based on the above discussions and references shown in the caption. Here, we consider three situations: (A) moderate hyperfine coupling and weak microwave drive power, (B) strong hyperfine coupling and moderate microwave drive power, and (C) strong hyperfine coupling and strong microwave drive power. Based on these parameters, the total gate operation time for BSM is 20-30 $\mu$s, as can be seen from Fig. \ref{fig:whole-gate-sequence} and Table \ref{tab:parameters}. To efficiently generate entanglement, the time for the BSM must be shorter than the coherence time of the superconducting qubits. State-of-the-art superconducting qubits have sufficient performance, since their coherence time reaches 100 $\mu$s \cite{Kjaergaard2020} or even $\sim1$ ms \cite{Somoroff2021}, which is longer than the BSM time. However, a much quicker BSM is still needed to generate the entanglement with high fidelity.

	Considering the success probability of entanglement generation between the remote spins and that between the superconducting qubit and the spin, the average entanglement generation rate between the superconducting qubits, $R_\textrm{sc-sc}^\textrm{mem}$, depends on $R_\textrm{e-e}$ and $R_\textrm{sc-e}$ as 
	\begin{equation}\label{eq:sc-sc-rate-with-memory-considering-photon-loss}
		R_\textrm{sc-sc}^\textrm{mem}=\frac{1}{\tau^\textrm{init}+1/R_\textrm{e-e}+\tau_\textrm{n}^\textrm{SWAP}+\tau_\textrm{sc-e}+\tau^\textrm{BSM}},
	\end{equation}
	where $\tau^\textrm{init}$ is the time to initialize the electron and the nuclear spins, $\tau_\textrm{n}^\textrm{SWAP}$ is the time to swap the state of the electron spin with that of the nuclear spin, $\tau_\textrm{sc-e}$ is the average time to generate entanglement between the electron spin and the superconducting qubit, and $\tau^\textrm{BSM}$ is the time to perform BSM for the electron and nuclear spins.
	$\tau_\textrm{sc-e}$ can be calculated by
	\begin{equation}
		\begin{split}
			\tau_\textrm{sc-e}&=(R_\textrm{sc-e}^2+\sum_{i=0}^{\infty}(i+2)R_\textrm{sc-e}^2(1-R_\textrm{sc-e})^{i+1}\\
			&\quad\times[(1-n)^{i+1}+2\sum_{j=0}^{i}(1-R_\textrm{sc-e})^{j}])/r_\textrm{sc-e}.
		\end{split}
	\end{equation}
	In Table \ref{tab:parameters}, assuming the highest efficiency,  $\eta_\textrm{e}^\textrm{opt}=1$ and  $\eta_\textrm{sc}^\textrm{mw}=\eta_\textrm{e}^\textrm{mw}=1$,  we estimate $R_\textrm{sc-sc}^\textrm{mem}$ to be 10-20 kHz at maximum.

	\begin{table}[]
		\centering
		\caption{Parameters related to the entanglement generation protocol in three different situations. (A) Moderate hyperfine coupling to nuclear spin and weak microwave driving power. (B) Strong hyperfine coupling to nuclear spin and moderate microwave driving power. (C) Strong hyperfine coupling to nuclear spin and strong microwave driving power.
		The parameters are hyperfine coupling of electron and nuclear spin, $A$, the Rabi frequency of the electron spin, $\Omega_\textrm{e}^\textrm{Rabi}$, the Rabi frequency of the nuclear spin, $\Omega_\textrm{n}^\textrm{Rabi}$,  time to initialize the electron spin, $\tau_\textrm{e}^\textrm{init}$, time to initialize the nuclear spin, $\tau_\textrm{n}^\textrm{init}$, time to generate entanglement between remote electron spins, $\tau_\textrm{e-e}$, time to swap the electron spin state to that of the nuclear spin, $\tau_\textrm{n}^\textrm{SWAP}$, time to emit a microwave time-bin qubit from the superconducting qubit (spin), $\tau_\textrm{mw}^\textrm{emi}$, time to receive a microwave time-bin qubit at the spin (superconducting qubit), $\tau_\textrm{mw}^\textrm{abs}$, time to perform a Hadamard gate, $\tau_\textrm{e}^{\pi/2}$, time to perform a $\textrm{C}_\textrm{e}\textrm{NOT}_\textrm{n}$ gate, $\tau^{\textrm{C}_\textrm{e}\textrm{NOT}_\textrm{n}}$, time to perform a $\textrm{C}_\textrm{n}\textrm{NOT}_\textrm{e}$ gate, $\tau^{\textrm{C}_\textrm{n}\textrm{NOT}_\textrm{e}}$, time to perform a single-shot measurement of the electron spin, $\tau_\textrm{e}^\textrm{SS}$, and the maximum entanglement generation rate, $R_\textrm{sc-sc}^\textrm{mem}$, assuming $\eta_\textrm{e}^\textrm{opt}=1$, $\eta_\textrm{sc}^\textrm{mw}=\eta_\textrm{e}^\textrm{mw}=1$. The gate times for the electron and nuclear spins ($\tau_\textrm{e}^{\pi/2}$, $\tau^{\textrm{C}_\textrm{e}\textrm{NOT}_\textrm{n}}$, and $\tau^{\textrm{C}_\textrm{n}\textrm{NOT}_\textrm{e}}$) are calculated based on $\Omega_\textrm{e}^\textrm{Rabi}$ and $\Omega_\textrm{n}^\textrm{Rabi}$ except in the case of $A\sim1$ MHz. We consider to use a $\pi$ pulse to perform $\textrm{C}_\textrm{e}\textrm{NOT}_\textrm{n}$ and $\textrm{C}_\textrm{n}\textrm{NOT}_\textrm{e}$ like Ref.\cite{Shim2013}.
		When $A\sim1$ MHz, $\tau^{\textrm{C}_\textrm{e}\textrm{NOT}_\textrm{n}}$ is determined according to Ref.\cite{Hegde2020} and $\tau^{\textrm{C}_\textrm{n}\textrm{NOT}_\textrm{e}}$ is calculated from $\Omega_\textrm{e}^\textrm{Rabi}$ assuming the use of a geometric gate (2$\pi$ pulse) \cite{Zu2014,Nagata2018}. 		
		$\tau_\textrm{e}^\textrm{init}$ is determined referring to experimentally used values \cite{Bernien2013,Humphreys2018}. (For measurement-based initialization,  $\tau_\textrm{e}^\textrm{init}$=10-30 $\mu$s \cite{Humphreys2018, Bhaskar2020,Levonian2021}.)  $\tau_\textrm{n}^\textrm{init}=\tau^{\textrm{C}_\textrm{e}\textrm{NOT}_\textrm{n}}$+$\tau^{\textrm{C}_\textrm{n}\textrm{NOT}_\textrm{e}}	+\tau_\textrm{e}^\textrm{init}$, $\tau_\textrm{n}^\textrm{SWAP}=\tau^{\textrm{C}_\textrm{e}\textrm{NOT}_\textrm{n}}$+$\tau^{\textrm{C}_\textrm{n}\textrm{NOT}_\textrm{e}}$.  We refer to Ref. \cite{Bernien2013,Humphreys2018} for $\tau_\textrm{e-e}$ and to Ref. \cite{Humphreys2018,Nguyen2019a,Levonian2021} for $\tau_\textrm{e}^\textrm{SS}$.  
		}
		\label{tab:parameters}
		\begin{tabular}{lrrr}
			
			{}&\multicolumn{1}{c}{(A)}&\multicolumn{1}{c}{(B)}&\multicolumn{1}{c}{(C)}\\
			\hline
			$A$ & \multicolumn{1}{l}{$\sim$1 MHz} & \multicolumn{2}{c}{$\sim$100 MHz \cite{Shim2013}} \\
			$\Omega_\textrm{e}^\textrm{Rabi}$ & 0.5 MHz & 10 MHz & 100 MHz \\
			$\Omega_\textrm{n}^\textrm{Rabi}$ &  & 0.4 MHz & 4 MHz \\ \hline
			$\tau_\textrm{e}^\textrm{init}$ \cite{Bernien2013,Humphreys2018} & 5 $\mu$s & 5 $\mu$s & 5 $\mu$s \\
			$\tau_\textrm{n}^\textrm{init}$ & 17 $\mu$s & 6.3 $\mu$s & 5.13 $\mu$s \\
			$\tau_\textrm{e-e}$ \cite{Bernien2013,Humphreys2018} & 10 $\mu$s & 10 $\mu$s & 10 $\mu$s \\
			$\tau_\textrm{n}^\textrm{SWAP}$ & 12 $\mu$s & 1.3 $\mu$s & 0.13 $\mu$s \\
			$\tau_\textrm{mw}^\textrm{emi}$ & 1 $\mu$s & 1 $\mu$s & 1 $\mu$s \\
			$\tau_\textrm{mw}^\textrm{abs}$ & 1 $\mu$s & 1 $\mu$s & 1 $\mu$s \\
			$\tau_\textrm{e}^{\pi/2}$ & 0.5 $\mu$s & 0.025 $\mu$s & 0.0025 $\mu$s \\
			$\tau^{\textrm{C}_\textrm{e}\textrm{NOT}_\textrm{n}}$ & 10 $\mu$s \cite{Hegde2020} & 1.25 $\mu$s & 0.125 $\mu$s \\
			$\tau^{\textrm{C}_\textrm{n}\textrm{NOT}_\textrm{e}}$ & 2 $\mu$s & 0.05 $\mu$s & 0.005 $\mu$s \\
			$\tau_\textrm{e}^\textrm{SS}$\cite{Humphreys2018,Nguyen2019a,Levonian2021} & 10 $\mu$s & 10 $\mu$s & 10 $\mu$s \\
			$R_\textrm{sc-sc}^\textrm{mem}$ & 11 kHz & 18 kHz & 19 kHz \\ \hline
		\end{tabular}
	\end{table}

	\begin{figure}
		\begin{center}
			\includegraphics[width=80mm]{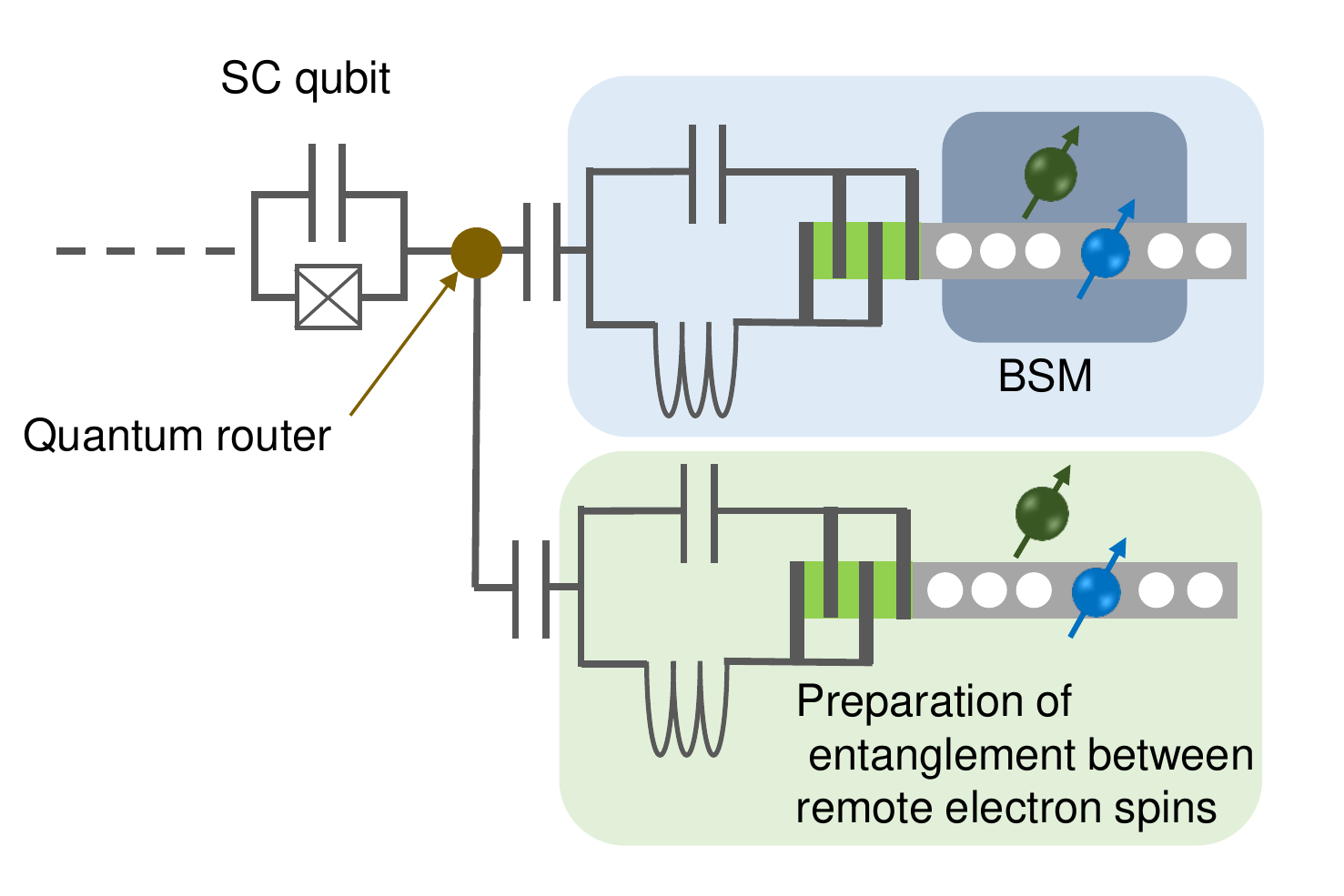}
		\end{center}
		\caption{Schematic of the quantum interface with two parallel memories. A quantum router \cite{Wang2021} is used to route a microwave photon to each interface.}
		\label{fig:multiple_memory_1}
	\end{figure}
	
	If we can use multiple quantum spin memories \cite{Neuman2021} as shown in Fig.\ref{fig:multiple_memory_1}, we can prepare the remote entanglement between electron spins during the BSM of another spin memory. Based on the assumption shown in Table \ref{tab:parameters}, a system with an additional parallel memory is sufficient to maintain the entanglement between the remote spins and to supply it to the superconducting qubits. The overhead time to generate electron spin-spin entanglement is again eliminated, resulting in the entanglement attempt rate of $\sim$40 kHz at maximum, which is limited by the speed of the BSM of the electron and nuclear spins. 
	

	\section{discussion}
	
	We estimate the entanglement generation rate between remote superconducting qubits using the quantum spin memory. Here, we compare the performance of the memory-based scheme to those of other protocols without memory. 
	
		\begin{figure}
		\begin{center}
			\includegraphics[width=80mm]{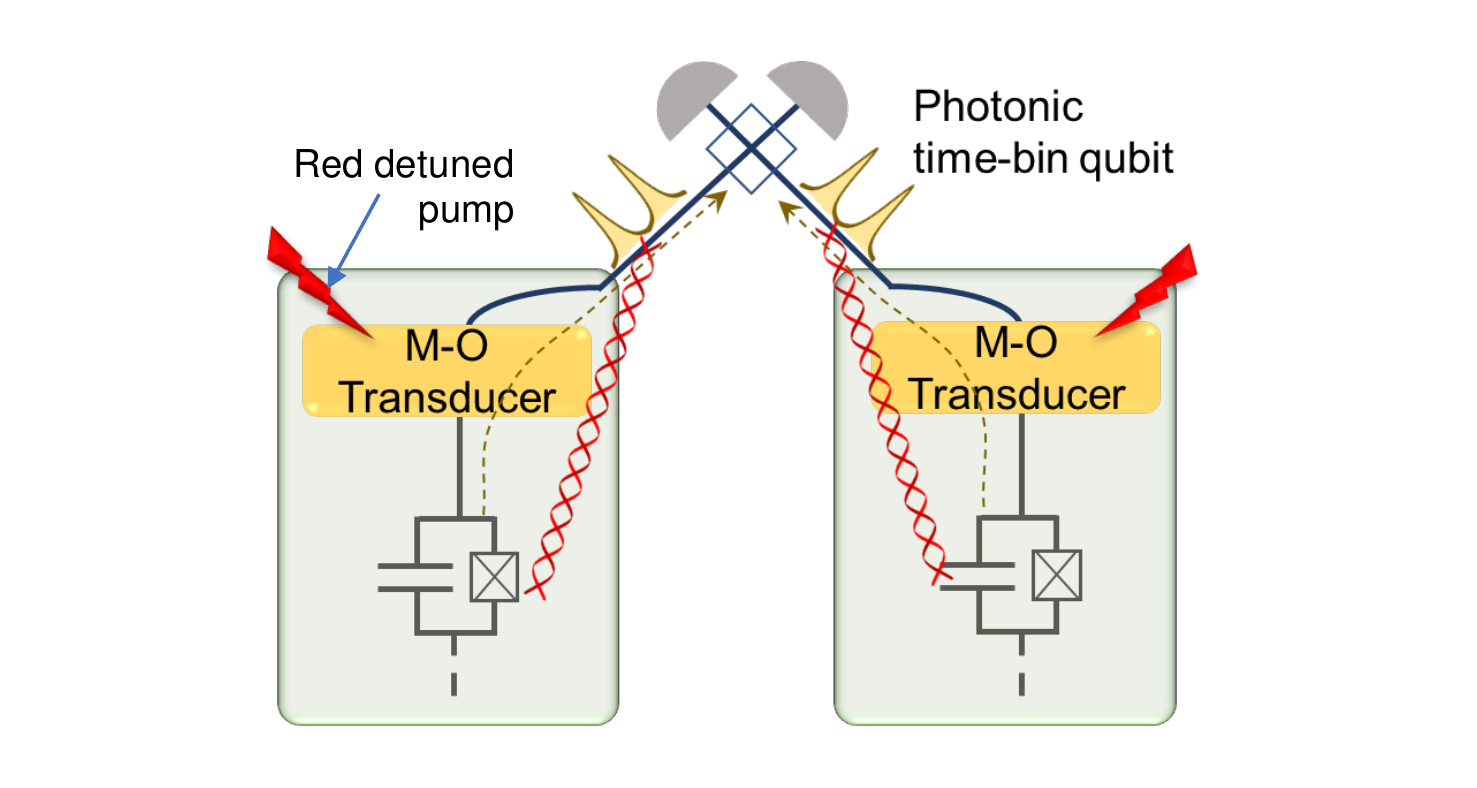}
		\end{center}
		\caption{Schematic to generate entanglement between remote superconducting qubits using a microwave-to-optical transducer. The two-photon protocol with a red detuned pump (the direct conversion scheme) is shown. The microwave time-bin qubit from the superconducting qubits is converted to a photonic time-bin qubit at the transducer. A beamsplitter and photodetectors are used to generate and herald the entanglement between the superconducting qubits.}
		\label{fig:scheme_DC_1}
	\end{figure}

	\subsection{Comparison to direct conversion scheme}
\begin{figure}
		\begin{center}
			\includegraphics[width=80mm]{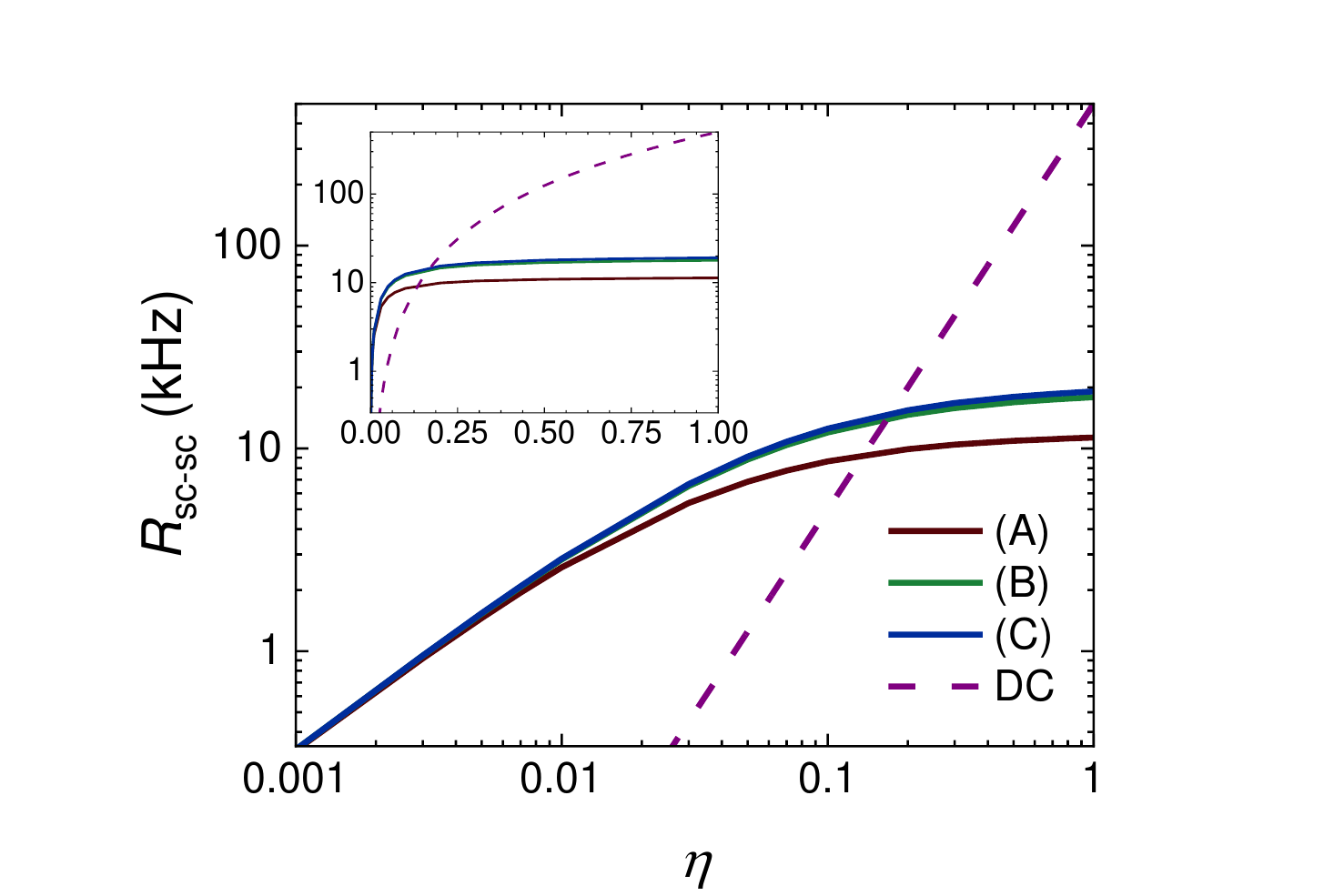}
		\end{center}
		\caption{Entanglement generation rate between remote superconducting qubits calculated using eq. (\ref{eq:sc-sc-rate-with-memory-considering-photon-loss}) and (\ref{eq:entangle-rate-DC}). The inset shows the same plot with a linear horizontal axis scale. The solid lines are those of the memory-based scheme. They are each plotted as a function of the absorption (emission)  efficiency of the microwave time-bin qubit by the spin, $\eta_\textrm{e}^\textrm{mw}$. A, B, and C in the legend correspond to the three cases summarized in Table \ref{tab:parameters}: (A) weak hyperfine coupling and weak microwave drive, (B) strong hyperfine coupling and moderate microwave drive, (C) strong hyperfine coupling and strong microwave drive. The dashed line corresponds to the direct conversion (DC) scheme. It is plotted as a function of microwave-to-optical transduction efficiency, $\eta_\textrm{DC}$. 
		}
		\label{fig:discussion-compare-DC_1}
	\end{figure}

	Direct conversion of a microwave to an optical photon is a straightforward approach transferring a quantum state of a superconducting qubit or microwave photon itself outside a dilution refrigerator. Direct conversion makes it possible to implement an entanglement generation protocol between remote superconducting qubits using  the two-photon protocol.  Here, though the single-photon protocol can be used, we consider only the two-photon protocol, in which stabilization of the optical path is not necessary. Figure \ref{fig:scheme_DC_1} shows a schematic of the protocol.  First, the superconducting qubits coupled to the cavity emit the coherent time-bin qubit, 
	\begin{equation}
		\frac{1}{\sqrt{2}}(\ket{g}_\textrm{sc}\ket{L}_\textrm{mw}+\ket{e}_\textrm{sc}\ket{E}_\textrm{mw}).
	\end{equation}
	The microwave time-bin qubit is then converted to the photonic time-bin qubit at the transducer. The photonic time-bin qubits from nodes A and B interfere at a beamsplitter and are measured by a photodetector. Consequently, the superconducting qubits in different nodes are projected to the entangled state as
	\begin{equation}
		\frac{1}{\sqrt{2}}(\ket{e}_\textrm{sc}^\textrm{A}\ket{g}_\textrm{sc}^\textrm{B}\pm\ket{g}_\textrm{sc}^\textrm{A}\ket{e}_\textrm{sc}^\textrm{B} ),
	\end{equation}
	where the sign depends on which detector is clicked. The same as in eq. (\ref{eq:entangle-rate}), the  entanglement generation rate in this direct conversion case, $R_\textrm{sc-sc}^\textrm{DC}$, can be expressed as
	\begin{equation}\label{eq:entangle-rate-DC}
		R_\textrm{sc-sc}^\textrm{DC}=\frac{1}{2}r_\textrm{sc-sc}\eta_\textrm{DC}^2(\eta_\textrm{loss}^\textrm{DC})^2,
	\end{equation}
	where $r_\textrm{sc-sc}$ is the entanglement attempt rate corresponding to the speed with which the microwave time-bin qubit is generated at a superconducting qubit,  $\eta_\textrm{DC}$ is the microwave-to-optical conversion efficiency, and  $\eta_\textrm{loss}^\textrm{DC}$ includes all losses during the propagation of the photon. For the calculation of $R_\textrm{sc-sc}^\textrm{DC}$, we assumed that the superconducting qubit coupled to the microwave cavity can generate a microwave time-bin qubit at a rate of $r_\textrm{sc-sc}=$1 MHz \cite{Kurpiers2019,Ilves2020}. Here, for simplicity, we do not consider the detail of the transducer. 
	Using eq.(\ref{eq:entangle-rate-DC}) and ignoring any transmission loss in the waveguide or fiber ($\eta_\textrm{loss}^\textrm{DC}$=1), we calculated an upper limit of $R_\textrm{sc-sc}^\textrm{DC}$. Figure \ref{fig:discussion-compare-DC_1} shows $R_\textrm{sc-sc}^\textrm{DC}$ as a function of $\eta_\textrm{DC}$. To compare the protocol with memory, $R_\textrm{sc-sc}^\textrm{mem}$ is also shown in the same graph. We calculated $R_\textrm{sc-sc}^\textrm{mem}$ using eq.(\ref{eq:sc-sc-rate-with-memory-considering-photon-loss}) and assuming $\eta_\textrm{e}^\textrm{opt}=\eta_\textrm{sc}^\textrm{mw}=0.9$ or 1 for the three cases shown in Table. \ref{tab:parameters}. Since putting $\eta_\textrm{e}^\textrm{opt}=\eta_\textrm{sc}^\textrm{mw}=0.9$ decreased $R_\textrm{sc-sc}^\textrm{mem}$ by less than 2 $\%$ when $\eta_\textrm{e}^\textrm{mw}>0.1$, we show only the result of $\eta_\textrm{e}^\textrm{opt}=\eta_\textrm{sc}^\textrm{mw}=1$ in Fig. \ref{fig:discussion-compare-DC_1}. It should be noted that the horizontal axis for $R_\textrm{sc-sc}^\textrm{mem}$ corresponds to the conversion efficiency of the microwave photon to the electron spin or vice versa ($\eta_\textrm{e}^\textrm{mw}$).  
	Figure \ref{fig:discussion-compare-DC_1} shows that when microwave-to-optical transduction efficiency ($\eta_\textrm{DC}$) is below 0.1 and microwave-to-spin conversion efficiency  ($\eta_\textrm{e}^\textrm{mw}$) is $\sim0.1$, the memory-based scheme exceeds the performance of the DC scheme. Besides, in practice, it is generally difficult to reach the regime, $\eta_\textrm{DC}>0.1$, while maintaining low optical pump power, low thermal noise, and sufficient external microwave (optical) coupling  \cite{Zhong2020}. 	Although using a color center in diamond as a single-microwave phonon emitter (receiver) still needs a proof-of-concept experiment, the numerical simulation (see Appendix \ref{app-sec:numerical_simulation}) suggests that $\eta_\textrm{e}^\textrm{mw}\sim$0.1 can be realized in the feasible design of the cavity and color center, promising the usefulness of this memory-based protocol.
	
		\subsection{Comparison to entangled-photon scheme without memory}

		\begin{figure}
		\begin{center}
			\includegraphics[width=80mm]{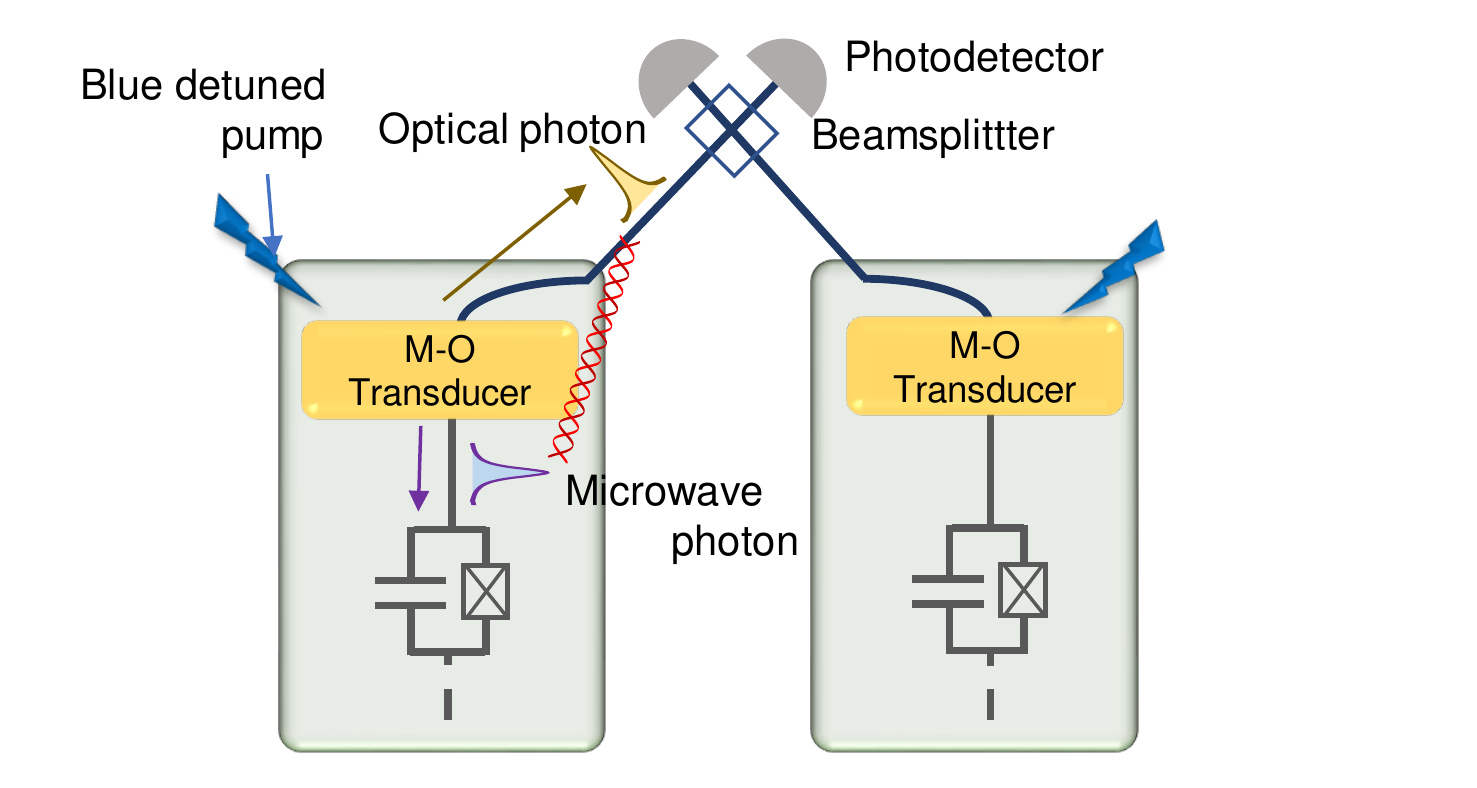}
		\end{center}
		\caption{Schematic of the generation of entanglement between remote superconducting qubits using microwave-to-optical transducers. The protocol with a blue detuned pump (the entangled-photon scheme) \cite{Krastanov2021} is shown. An entangled-photon pair is generated at the transducer by the spontaneous parametric downconversion. As with Fig.\ref{fig:scheme_DC_1}, a beamsplitter and photodetectors are used to generate and herald the entanglement between the microwave photons.}
		\label{fig:scheme_EP_1}
	\end{figure}

	To avoid the difficulties involved in the realization of a high-efficiency microwave-to-optical transducer, entangled-photon schemes have recently attracted attention for their ability to generate entanglement between a microwave photon inside a dilution refrigerator and an optical photon outside of it \cite{Zhong2020, Krastanov2021, Wu2021}. In these schemes, the parametric downconversion of the blue-detuned optical pump light generates entangled photons at the microwave-optical transducers (e.g., electro-optic transducers, optomechanical transducers). Figure \ref{fig:scheme_EP_1} shows a schematic of the protocol. When the transducers at nodes A and B are simultaneously subjected to the blue-detuned light, an entangled microwave-optical photon pair is generated in either node. The measurement of an optical photon using a beamsplitter and photodetectors heralds the generation of the entanglement between microwave photons in each node, 
	\begin{equation}
		\frac{1}{\sqrt{2}}(\ket{0}_\textrm{mw}^\textrm{A}\ket{1}_\textrm{mw}^\textrm{B}\pm\ket{1}_\textrm{mw}^\textrm{A}\ket{0}_\textrm{mw}^\textrm{B}),
	\end{equation}
	where the sign depends on the side on which the detector is clicked.

	Since the number of microwave photons in each node entangles, the absorption of a microwave photon at the superconducting qubits, which are initially at $\ket{g}_\textrm{sc}$, generates the entanglement as
	\begin{equation}
		\frac{1}{\sqrt{2}}(\ket{g}_\textrm{sc}^\textrm{A}\ket{e}_\textrm{sc}^\textrm{B}\pm\ket{e}_\textrm{sc}^\textrm{A}\ket{g}_\textrm{sc}^\textrm{B} ).
	\end{equation}
	
	In Ref. \cite{Krastanov2021}, the entanglement generation rate was estimated to be 1-10 kHz with in-refrigerator heating of 10 $\mu$W and 10-100 kHz with in-refrigerator heating of 100 $\mu$W based on state-of-the-art device parameters. This heating comes from the strong optical pump generating entangled photons at a high rate. Since heating on the order of 10 $\mu$W cannot be negligible considering the cooling power of the 10 mK region in a dilution refrigerator, the operation of higher-temperature plates with greater cooling power is proposed \cite{Zhong2020,Krastanov2021}.

	In the memory-based protocol, in-refrigerator heating mainly comes from the dissipation of the microwave drive pulse and laser light to control and read out the spin.  To achieve an electron Rabi frequency of 10 MHz, a flux density of $\sim$ 3 G is needed, considering the gyromagnetic ratio of the electron, $\gamma/2\pi\sim3$ MHz/G. From Ampere's law, $\sim$1 mA current is needed to induce the required magnetic field at a color center whose distance from the electrode is $\sim$1 $\mu$m. Then, the input power at the mixing chamber plate should be 150 $\mu$W with 50 $\Omega$ termination outside the dilution refrigerator. The attenuation of the commercially available superconducting coaxial waveguide is $\sim$0.01 dB/m at a few GHz below 4 K \cite{Kurpiers2017}. Thus, considering a superconducting waveguide with 1 m length below the mixing chamber ($\sim$10 mK) and microwave pulses with a duty cycle of 0.1 (the ratio of microwave (RF) gate time to the whole entanglement generation sequence time), the dissipated power is estimated to be 30 nW, which is more than two orders of magnitude smaller than the typical cooling power at the mixing chamber, $\sim$10 $\mu$W (e.g., LD-series, Bluefors). Even if we use 15 mW microwave pulses, a 10-fold increase in the gate speed owing to the high power decreases the duty cycle to 0.01, resulting in dissipation of about 400 nW. Besides, the loss at the cable is suggested to be lowered by two orders of magnitude using a waveguide with a well-prepared surface \cite{Kurpiers2017}, which may further reduce the heat generation. We should note that, we assumed here that the contact resistance at each joint or connector can be negligible. It should also be noted that, for color centers whose ground state is the eigenstate of both spin and orbit (e.g., NV$^0$, group-IV vacancy color centers), much more microwave power is required than the above estimation, since flipping both spin and orbital quantum numbers using only a magnetic field is forbidden in the first-order\cite{Ruf2021}.

	The typical optical power needed to control and read out a color center is below 100 nW. Moreover, a readout of the spin state in a single-photon regime (0.001-0.1 average photon in a pulse) was demonstrated to be feasible by using a photonic crystal structure \cite{Nguyen2019a,Bhaskar2020,Levonian2021}. Thus, the optical input power could also be well below the typical cooling power of the mixing chamber plate.

	Although it is difficult to predict heat generation exactly at low temperatures, based on the above rough estimation, the total heat generation below the mixing chamber is two to three orders of magnitude smaller than that is generated in the entangled-photon scheme without memory. In addition, the entanglement generation rate of the memory-based scheme (several tens of kilohertz) is comparable to that of the entangled-photon scheme without memory. Namely, introducing the quantum memory interfacing microwave and optical photon can decrease heat generation, maintaining an entanglement generation rate in the tens of kilohertz at the cost of fabrication and control overhead.

	\subsection{Perspectives}

	Though the memory-based entanglement generation scheme could be a promising option for operation in the 10 mK region, there are still several challenges to overcome. We must integrate all of the necessary functions (e.g., gate speed, memory time, strain sensitivity), which were demonstrated individually or at different color centers. Moreover, as mentioned above, the interaction between a single electron spin and a microwave photon needs a proof-of-principle experiment.

	In contrast to the application to long-distance communication, gate speed to control the electron spin limits the whole entanglement generation rate. Thus, a quantum memory enabling fast initialization, control, and readout with a nuclear spin memory that has moderate memory time ($\sim$ 10 ms) is desirable. It may be necessary to explore a color center that has suitable characteristics. Besides, considering the operation with a superconducting circuit, control of a color center under a zero-magnetic field \cite{Sekiguchi2016,Sekiguchi2017,Nagata2018} should be considered even though some superconducting devices may work in magnetic fields ranging from $\sim $mT to $\sim$ T \cite{Samkharadze2016,Krause2021}.  Since the proposed scheme can be applied to other quantum memories having both microwave and optical access, such as trapped atoms, ions, and quantum dots, these quantum memories might be alternatives to color centers.

	\section{conclusion}
	
	We proposed a protocol for entanglement generation between remote superconducting qubits using solid-state spin quantum memories, which can be used as interfaces for both microwave and optical photons. Based on the current state-of-the-art technologies, we estimated the maximum entanglement generation rate to be 10-40 kHz. The proposed memory-based scheme's entanglement generation rate can exceed that of the direct conversion (DC) scheme as long as the microwave-to-optical conversion efficiency of DC is below 0.1.  Moreover, the memory-based scheme can reduce the amount of heat generated below the mixing chamber compared to that generated by the entangled-photon scheme, which uses a strong optical pump to generate entangled-photon pairs. For quantum communication between remote superconducting qubits, the reduction of heat is a central issue considering that operation occurs inside a dilution refrigerator. This suggests the usefulness of introducing the quantum memory.

	\section*{Acknowledgements}
	This work was supported by the Japan Society for the Promotion of Science (JSPS) Grants-in-Aid for Scientific Research (20H05661, 20K20441); by a Japan Science and Technology Agency (JST) CREST grant (JPMJCR1773); and by a JST Moonshot R$\&$D grant (JPMJMS2062). We also acknowledge the Ministry of Internal Affairs and Communications (MIC) for funding, research and development for the construction of a global quantum cryptography network (JPMI00316).
	
		\section*{Authorship contribution}
	H. Kurokawa conceived the main protocol and performed the calculation. M. Y. and Y. S. contributed the design of partial protocol. H. Kosaka supervised the project.
	

\begin{thebibliography}{61}%
\makeatletter
\providecommand \@ifxundefined [1]{%
 \@ifx{#1\undefined}
}%
\providecommand \@ifnum [1]{%
 \ifnum #1\expandafter \@firstoftwo
 \else \expandafter \@secondoftwo
 \fi
}%
\providecommand \@ifx [1]{%
 \ifx #1\expandafter \@firstoftwo
 \else \expandafter \@secondoftwo
 \fi
}%
\providecommand \natexlab [1]{#1}%
\providecommand \enquote  [1]{``#1''}%
\providecommand \bibnamefont  [1]{#1}%
\providecommand \bibfnamefont [1]{#1}%
\providecommand \citenamefont [1]{#1}%
\providecommand \href@noop [0]{\@secondoftwo}%
\providecommand \href [0]{\begingroup \@sanitize@url \@href}%
\providecommand \@href[1]{\@@startlink{#1}\@@href}%
\providecommand \@@href[1]{\endgroup#1\@@endlink}%
\providecommand \@sanitize@url [0]{\catcode `\\12\catcode `\$12\catcode
  `\&12\catcode `\#12\catcode `\^12\catcode `\_12\catcode `\%12\relax}%
\providecommand \@@startlink[1]{}%
\providecommand \@@endlink[0]{}%
\providecommand \url  [0]{\begingroup\@sanitize@url \@url }%
\providecommand \@url [1]{\endgroup\@href {#1}{\urlprefix }}%
\providecommand \urlprefix  [0]{URL }%
\providecommand \Eprint [0]{\href }%
\providecommand \doibase [0]{https://doi.org/}%
\providecommand \selectlanguage [0]{\@gobble}%
\providecommand \bibinfo  [0]{\@secondoftwo}%
\providecommand \bibfield  [0]{\@secondoftwo}%
\providecommand \translation [1]{[#1]}%
\providecommand \BibitemOpen [0]{}%
\providecommand \bibitemStop [0]{}%
\providecommand \bibitemNoStop [0]{.\EOS\space}%
\providecommand \EOS [0]{\spacefactor3000\relax}%
\providecommand \BibitemShut  [1]{\csname bibitem#1\endcsname}%
\let\auto@bib@innerbib\@empty
\bibitem [{\citenamefont {Kjaergaard}\ \emph {et~al.}(2020)\citenamefont
  {Kjaergaard}, \citenamefont {Schwartz}, \citenamefont {Braum{\"{u}}ller},
  \citenamefont {Krantz}, \citenamefont {Wang}, \citenamefont {Gustavsson},\
  and\ \citenamefont {Oliver}}]{Kjaergaard2020}%
  \BibitemOpen
  \bibfield  {author} {\bibinfo {author} {\bibfnamefont {M.}~\bibnamefont
  {Kjaergaard}}, \bibinfo {author} {\bibfnamefont {M.~E.}\ \bibnamefont
  {Schwartz}}, \bibinfo {author} {\bibfnamefont {J.}~\bibnamefont
  {Braum{\"{u}}ller}}, \bibinfo {author} {\bibfnamefont {P.}~\bibnamefont
  {Krantz}}, \bibinfo {author} {\bibfnamefont {J.~I.-J.}\ \bibnamefont {Wang}},
  \bibinfo {author} {\bibfnamefont {S.}~\bibnamefont {Gustavsson}},\ and\
  \bibinfo {author} {\bibfnamefont {W.~D.}\ \bibnamefont {Oliver}},\ }\bibfield
   {title} {\bibinfo {title} {{Superconducting Qubits: Current State of
  Play}},\ }\href {https://doi.org/10.1146/annurev-conmatphys-031119-050605}
  {\bibfield  {journal} {\bibinfo  {journal} {Annual Review of Condensed Matter
  Physics}\ }\textbf {\bibinfo {volume} {11}},\ \bibinfo {pages} {369}
  (\bibinfo {year} {2020})},\ \Eprint {https://arxiv.org/abs/1905.13641}
  {arXiv:1905.13641} \BibitemShut {NoStop}%
\bibitem [{\citenamefont {Arute}\ \emph {et~al.}(2019)\citenamefont {Arute},
  \citenamefont {Arya}, \citenamefont {Babbush}, \citenamefont {Bacon},
  \citenamefont {Bardin}, \citenamefont {Barends}, \citenamefont {Biswas},
  \citenamefont {Boixo}, \citenamefont {Brandao}, \citenamefont {Buell},
  \citenamefont {Burkett}, \citenamefont {Chen}, \citenamefont {Chen},
  \citenamefont {Chiaro}, \citenamefont {Collins}, \citenamefont {Courtney},
  \citenamefont {Dunsworth}, \citenamefont {Farhi}, \citenamefont {Foxen},
  \citenamefont {Fowler}, \citenamefont {Gidney}, \citenamefont {Giustina},
  \citenamefont {Graff}, \citenamefont {Guerin}, \citenamefont {Habegger},
  \citenamefont {Harrigan}, \citenamefont {Hartmann}, \citenamefont {Ho},
  \citenamefont {Hoffmann}, \citenamefont {Huang}, \citenamefont {Humble},
  \citenamefont {Isakov}, \citenamefont {Jeffrey}, \citenamefont {Jiang},
  \citenamefont {Kafri}, \citenamefont {Kechedzhi}, \citenamefont {Kelly},
  \citenamefont {Klimov}, \citenamefont {Knysh}, \citenamefont {Korotkov},
  \citenamefont {Kostritsa}, \citenamefont {Landhuis}, \citenamefont
  {Lindmark}, \citenamefont {Lucero}, \citenamefont {Lyakh}, \citenamefont
  {Mandr{\`{a}}}, \citenamefont {McClean}, \citenamefont {McEwen},
  \citenamefont {Megrant}, \citenamefont {Mi}, \citenamefont {Michielsen},
  \citenamefont {Mohseni}, \citenamefont {Mutus}, \citenamefont {Naaman},
  \citenamefont {Neeley}, \citenamefont {Neill}, \citenamefont {Niu},
  \citenamefont {Ostby}, \citenamefont {Petukhov}, \citenamefont {Platt},
  \citenamefont {Quintana}, \citenamefont {Rieffel}, \citenamefont {Roushan},
  \citenamefont {Rubin}, \citenamefont {Sank}, \citenamefont {Satzinger},
  \citenamefont {Smelyanskiy}, \citenamefont {Sung}, \citenamefont
  {Trevithick}, \citenamefont {Vainsencher}, \citenamefont {Villalonga},
  \citenamefont {White}, \citenamefont {Yao}, \citenamefont {Yeh},
  \citenamefont {Zalcman}, \citenamefont {Neven},\ and\ \citenamefont
  {Martinis}}]{Arute2019}%
  \BibitemOpen
  \bibfield  {author} {\bibinfo {author} {\bibfnamefont {F.}~\bibnamefont
  {Arute}}, \bibinfo {author} {\bibfnamefont {K.}~\bibnamefont {Arya}},
  \bibinfo {author} {\bibfnamefont {R.}~\bibnamefont {Babbush}}, \bibinfo
  {author} {\bibfnamefont {D.}~\bibnamefont {Bacon}}, \bibinfo {author}
  {\bibfnamefont {J.~C.}\ \bibnamefont {Bardin}}, \bibinfo {author}
  {\bibfnamefont {R.}~\bibnamefont {Barends}}, \bibinfo {author} {\bibfnamefont
  {R.}~\bibnamefont {Biswas}}, \bibinfo {author} {\bibfnamefont
  {S.}~\bibnamefont {Boixo}}, \bibinfo {author} {\bibfnamefont {F.~G. S.~L.}\
  \bibnamefont {Brandao}}, \bibinfo {author} {\bibfnamefont {D.~A.}\
  \bibnamefont {Buell}}, \bibinfo {author} {\bibfnamefont {B.}~\bibnamefont
  {Burkett}}, \bibinfo {author} {\bibfnamefont {Y.}~\bibnamefont {Chen}},
  \bibinfo {author} {\bibfnamefont {Z.}~\bibnamefont {Chen}}, \bibinfo {author}
  {\bibfnamefont {B.}~\bibnamefont {Chiaro}}, \bibinfo {author} {\bibfnamefont
  {R.}~\bibnamefont {Collins}}, \bibinfo {author} {\bibfnamefont
  {W.}~\bibnamefont {Courtney}}, \bibinfo {author} {\bibfnamefont
  {A.}~\bibnamefont {Dunsworth}}, \bibinfo {author} {\bibfnamefont
  {E.}~\bibnamefont {Farhi}}, \bibinfo {author} {\bibfnamefont
  {B.}~\bibnamefont {Foxen}}, \bibinfo {author} {\bibfnamefont
  {A.}~\bibnamefont {Fowler}}, \bibinfo {author} {\bibfnamefont
  {C.}~\bibnamefont {Gidney}}, \bibinfo {author} {\bibfnamefont
  {M.}~\bibnamefont {Giustina}}, \bibinfo {author} {\bibfnamefont
  {R.}~\bibnamefont {Graff}}, \bibinfo {author} {\bibfnamefont
  {K.}~\bibnamefont {Guerin}}, \bibinfo {author} {\bibfnamefont
  {S.}~\bibnamefont {Habegger}}, \bibinfo {author} {\bibfnamefont {M.~P.}\
  \bibnamefont {Harrigan}}, \bibinfo {author} {\bibfnamefont {M.~J.}\
  \bibnamefont {Hartmann}}, \bibinfo {author} {\bibfnamefont {A.}~\bibnamefont
  {Ho}}, \bibinfo {author} {\bibfnamefont {M.}~\bibnamefont {Hoffmann}},
  \bibinfo {author} {\bibfnamefont {T.}~\bibnamefont {Huang}}, \bibinfo
  {author} {\bibfnamefont {T.~S.}\ \bibnamefont {Humble}}, \bibinfo {author}
  {\bibfnamefont {S.~V.}\ \bibnamefont {Isakov}}, \bibinfo {author}
  {\bibfnamefont {E.}~\bibnamefont {Jeffrey}}, \bibinfo {author} {\bibfnamefont
  {Z.}~\bibnamefont {Jiang}}, \bibinfo {author} {\bibfnamefont
  {D.}~\bibnamefont {Kafri}}, \bibinfo {author} {\bibfnamefont
  {K.}~\bibnamefont {Kechedzhi}}, \bibinfo {author} {\bibfnamefont
  {J.}~\bibnamefont {Kelly}}, \bibinfo {author} {\bibfnamefont {P.~V.}\
  \bibnamefont {Klimov}}, \bibinfo {author} {\bibfnamefont {S.}~\bibnamefont
  {Knysh}}, \bibinfo {author} {\bibfnamefont {A.}~\bibnamefont {Korotkov}},
  \bibinfo {author} {\bibfnamefont {F.}~\bibnamefont {Kostritsa}}, \bibinfo
  {author} {\bibfnamefont {D.}~\bibnamefont {Landhuis}}, \bibinfo {author}
  {\bibfnamefont {M.}~\bibnamefont {Lindmark}}, \bibinfo {author}
  {\bibfnamefont {E.}~\bibnamefont {Lucero}}, \bibinfo {author} {\bibfnamefont
  {D.}~\bibnamefont {Lyakh}}, \bibinfo {author} {\bibfnamefont
  {S.}~\bibnamefont {Mandr{\`{a}}}}, \bibinfo {author} {\bibfnamefont {J.~R.}\
  \bibnamefont {McClean}}, \bibinfo {author} {\bibfnamefont {M.}~\bibnamefont
  {McEwen}}, \bibinfo {author} {\bibfnamefont {A.}~\bibnamefont {Megrant}},
  \bibinfo {author} {\bibfnamefont {X.}~\bibnamefont {Mi}}, \bibinfo {author}
  {\bibfnamefont {K.}~\bibnamefont {Michielsen}}, \bibinfo {author}
  {\bibfnamefont {M.}~\bibnamefont {Mohseni}}, \bibinfo {author} {\bibfnamefont
  {J.}~\bibnamefont {Mutus}}, \bibinfo {author} {\bibfnamefont
  {O.}~\bibnamefont {Naaman}}, \bibinfo {author} {\bibfnamefont
  {M.}~\bibnamefont {Neeley}}, \bibinfo {author} {\bibfnamefont
  {C.}~\bibnamefont {Neill}}, \bibinfo {author} {\bibfnamefont {M.~Y.}\
  \bibnamefont {Niu}}, \bibinfo {author} {\bibfnamefont {E.}~\bibnamefont
  {Ostby}}, \bibinfo {author} {\bibfnamefont {A.}~\bibnamefont {Petukhov}},
  \bibinfo {author} {\bibfnamefont {J.~C.}\ \bibnamefont {Platt}}, \bibinfo
  {author} {\bibfnamefont {C.}~\bibnamefont {Quintana}}, \bibinfo {author}
  {\bibfnamefont {E.~G.}\ \bibnamefont {Rieffel}}, \bibinfo {author}
  {\bibfnamefont {P.}~\bibnamefont {Roushan}}, \bibinfo {author} {\bibfnamefont
  {N.~C.}\ \bibnamefont {Rubin}}, \bibinfo {author} {\bibfnamefont
  {D.}~\bibnamefont {Sank}}, \bibinfo {author} {\bibfnamefont {K.~J.}\
  \bibnamefont {Satzinger}}, \bibinfo {author} {\bibfnamefont {V.}~\bibnamefont
  {Smelyanskiy}}, \bibinfo {author} {\bibfnamefont {K.~J.}\ \bibnamefont
  {Sung}}, \bibinfo {author} {\bibfnamefont {M.~D.}\ \bibnamefont
  {Trevithick}}, \bibinfo {author} {\bibfnamefont {A.}~\bibnamefont
  {Vainsencher}}, \bibinfo {author} {\bibfnamefont {B.}~\bibnamefont
  {Villalonga}}, \bibinfo {author} {\bibfnamefont {T.}~\bibnamefont {White}},
  \bibinfo {author} {\bibfnamefont {Z.~J.}\ \bibnamefont {Yao}}, \bibinfo
  {author} {\bibfnamefont {P.}~\bibnamefont {Yeh}}, \bibinfo {author}
  {\bibfnamefont {A.}~\bibnamefont {Zalcman}}, \bibinfo {author} {\bibfnamefont
  {H.}~\bibnamefont {Neven}},\ and\ \bibinfo {author} {\bibfnamefont {J.~M.}\
  \bibnamefont {Martinis}},\ }\bibfield  {title} {\bibinfo {title} {{Quantum
  supremacy using a programmable superconducting processor}},\ }\href
  {https://doi.org/10.1038/s41586-019-1666-5} {\bibfield  {journal} {\bibinfo
  {journal} {Nature}\ }\textbf {\bibinfo {volume} {574}},\ \bibinfo {pages}
  {505} (\bibinfo {year} {2019})}\BibitemShut {NoStop}%
\bibitem [{\citenamefont {Ahsan}\ \emph {et~al.}(2016)\citenamefont {Ahsan},
  \citenamefont {Meter},\ and\ \citenamefont {Kim}}]{Ahsan2015}%
  \BibitemOpen
  \bibfield  {author} {\bibinfo {author} {\bibfnamefont {M.}~\bibnamefont
  {Ahsan}}, \bibinfo {author} {\bibfnamefont {R.~V.}\ \bibnamefont {Meter}},\
  and\ \bibinfo {author} {\bibfnamefont {J.}~\bibnamefont {Kim}},\ }\bibfield
  {title} {\bibinfo {title} {{Designing a Million-Qubit Quantum Computer Using
  Resource Performance Simulator}},\ }\href {https://doi.org/10.1145/2830570}
  {\bibfield  {journal} {\bibinfo  {journal} {ACM Journal on Emerging
  Technologies in Computing Systems}\ }\textbf {\bibinfo {volume} {12}},\
  \bibinfo {pages} {1} (\bibinfo {year} {2016})},\ \Eprint
  {https://arxiv.org/abs/1512.00796} {arXiv:1512.00796} \BibitemShut {NoStop}%
\bibitem [{\citenamefont {Gouzien}\ and\ \citenamefont
  {Sangouard}(2021)}]{Gouzien2021}%
  \BibitemOpen
  \bibfield  {author} {\bibinfo {author} {\bibfnamefont {{\'{E}}.}~\bibnamefont
  {Gouzien}}\ and\ \bibinfo {author} {\bibfnamefont {N.}~\bibnamefont
  {Sangouard}},\ }\bibfield  {title} {\bibinfo {title} {{Factoring 2048-bit RSA
  Integers in 177 Days with 13 436 Qubits and a Multimode Memory}},\ }\href
  {https://doi.org/10.1103/PhysRevLett.127.140503} {\bibfield  {journal}
  {\bibinfo  {journal} {Physical Review Letters}\ }\textbf {\bibinfo {volume}
  {127}},\ \bibinfo {pages} {140503} (\bibinfo {year} {2021})},\ \Eprint
  {https://arxiv.org/abs/2103.06159} {arXiv:2103.06159} \BibitemShut {NoStop}%
\bibitem [{\citenamefont {Yimsiriwattana}\ and\ \citenamefont {{Lomonaco
  Jr.}}(2004)}]{Yimsiriwattana2004}%
  \BibitemOpen
  \bibfield  {author} {\bibinfo {author} {\bibfnamefont {A.}~\bibnamefont
  {Yimsiriwattana}}\ and\ \bibinfo {author} {\bibfnamefont {S.~J.}\
  \bibnamefont {{Lomonaco Jr.}}},\ }\bibfield  {title} {\bibinfo {title}
  {{Distributed quantum computing: a distributed Shor algorithm}},\ }in\ \href
  {https://doi.org/10.1117/12.546504} {\emph {\bibinfo {booktitle} {Quantum
  Information and Computation II}}},\ Vol.\ \bibinfo {volume} {5436},\ \bibinfo
  {editor} {edited by\ \bibinfo {editor} {\bibfnamefont {E.}~\bibnamefont
  {Donkor}}, \bibinfo {editor} {\bibfnamefont {A.~R.}\ \bibnamefont {Pirich}},\
  and\ \bibinfo {editor} {\bibfnamefont {H.~E.}\ \bibnamefont {Brandt}}}\
  (\bibinfo {year} {2004})\ p.\ \bibinfo {pages} {360},\ \Eprint
  {https://arxiv.org/abs/0403146} {arXiv:0403146 [quant-ph]} \BibitemShut
  {NoStop}%
\bibitem [{\citenamefont {{Van Meter}}\ and\ \citenamefont
  {Devitt}(2016)}]{VanMeter2016}%
  \BibitemOpen
  \bibfield  {author} {\bibinfo {author} {\bibfnamefont {R.}~\bibnamefont {{Van
  Meter}}}\ and\ \bibinfo {author} {\bibfnamefont {S.~J.}\ \bibnamefont
  {Devitt}},\ }\bibfield  {title} {\bibinfo {title} {{The Path to Scalable
  Distributed Quantum Computing}},\ }\href
  {https://doi.org/10.1109/MC.2016.291} {\bibfield  {journal} {\bibinfo
  {journal} {Computer}\ }\textbf {\bibinfo {volume} {49}},\ \bibinfo {pages}
  {31} (\bibinfo {year} {2016})},\ \Eprint {https://arxiv.org/abs/1605.06951}
  {arXiv:1605.06951} \BibitemShut {NoStop}%
\bibitem [{\citenamefont {Cuomo}\ \emph {et~al.}(2020)\citenamefont {Cuomo},
  \citenamefont {Caleffi},\ and\ \citenamefont {Cacciapuoti}}]{Cuomo2020}%
  \BibitemOpen
  \bibfield  {author} {\bibinfo {author} {\bibfnamefont {D.}~\bibnamefont
  {Cuomo}}, \bibinfo {author} {\bibfnamefont {M.}~\bibnamefont {Caleffi}},\
  and\ \bibinfo {author} {\bibfnamefont {A.~S.}\ \bibnamefont {Cacciapuoti}},\
  }\bibfield  {title} {\bibinfo {title} {{Towards a distributed quantum
  computing ecosystem}},\ }\href {https://doi.org/10.1049/iet-qtc.2020.0002}
  {\bibfield  {journal} {\bibinfo  {journal} {IET Quantum Communication}\
  }\textbf {\bibinfo {volume} {1}},\ \bibinfo {pages} {3} (\bibinfo {year}
  {2020})},\ \Eprint {https://arxiv.org/abs/2002.11808} {arXiv:2002.11808}
  \BibitemShut {NoStop}%
\bibitem [{\citenamefont {Lambert}\ \emph {et~al.}(2020)\citenamefont
  {Lambert}, \citenamefont {Rueda}, \citenamefont {Sedlmeir},\ and\
  \citenamefont {Schwefel}}]{Lambert2020}%
  \BibitemOpen
  \bibfield  {author} {\bibinfo {author} {\bibfnamefont {N.~J.}\ \bibnamefont
  {Lambert}}, \bibinfo {author} {\bibfnamefont {A.}~\bibnamefont {Rueda}},
  \bibinfo {author} {\bibfnamefont {F.}~\bibnamefont {Sedlmeir}},\ and\
  \bibinfo {author} {\bibfnamefont {H.~G.~L.}\ \bibnamefont {Schwefel}},\
  }\bibfield  {title} {\bibinfo {title} {{Coherent Conversion Between Microwave
  and Optical Photons-An Overview of Physical Implementations}},\ }\href
  {https://doi.org/10.1002/qute.201900077} {\bibfield  {journal} {\bibinfo
  {journal} {Advanced Quantum Technologies}\ }\textbf {\bibinfo {volume} {3}},\
  \bibinfo {pages} {1900077} (\bibinfo {year} {2020})},\ \Eprint
  {https://arxiv.org/abs/1906.10255} {arXiv:1906.10255} \BibitemShut {NoStop}%
\bibitem [{\citenamefont {Lauk}\ \emph {et~al.}(2020)\citenamefont {Lauk},
  \citenamefont {Sinclair}, \citenamefont {Barzanjeh}, \citenamefont {Covey},
  \citenamefont {Saffman}, \citenamefont {Spiropulu},\ and\ \citenamefont
  {Simon}}]{Lauk2020}%
  \BibitemOpen
  \bibfield  {author} {\bibinfo {author} {\bibfnamefont {N.}~\bibnamefont
  {Lauk}}, \bibinfo {author} {\bibfnamefont {N.}~\bibnamefont {Sinclair}},
  \bibinfo {author} {\bibfnamefont {S.}~\bibnamefont {Barzanjeh}}, \bibinfo
  {author} {\bibfnamefont {J.~P.}\ \bibnamefont {Covey}}, \bibinfo {author}
  {\bibfnamefont {M.}~\bibnamefont {Saffman}}, \bibinfo {author} {\bibfnamefont
  {M.}~\bibnamefont {Spiropulu}},\ and\ \bibinfo {author} {\bibfnamefont
  {C.}~\bibnamefont {Simon}},\ }\bibfield  {title} {\bibinfo {title}
  {{Perspectives on quantum transduction}},\ }\href
  {https://doi.org/10.1088/2058-9565/ab788a} {\bibfield  {journal} {\bibinfo
  {journal} {Quantum Science and Technology}\ }\textbf {\bibinfo {volume}
  {5}},\ \bibinfo {pages} {020501} (\bibinfo {year} {2020})}\BibitemShut
  {NoStop}%
\bibitem [{\citenamefont {Zhong}\ \emph {et~al.}(2020)\citenamefont {Zhong},
  \citenamefont {Wang}, \citenamefont {Zou}, \citenamefont {Zhang},
  \citenamefont {Han}, \citenamefont {Fu}, \citenamefont {Xu}, \citenamefont
  {Shankar}, \citenamefont {Devoret}, \citenamefont {Tang},\ and\ \citenamefont
  {Jiang}}]{Zhong2020}%
  \BibitemOpen
  \bibfield  {author} {\bibinfo {author} {\bibfnamefont {C.}~\bibnamefont
  {Zhong}}, \bibinfo {author} {\bibfnamefont {Z.}~\bibnamefont {Wang}},
  \bibinfo {author} {\bibfnamefont {C.}~\bibnamefont {Zou}}, \bibinfo {author}
  {\bibfnamefont {M.}~\bibnamefont {Zhang}}, \bibinfo {author} {\bibfnamefont
  {X.}~\bibnamefont {Han}}, \bibinfo {author} {\bibfnamefont {W.}~\bibnamefont
  {Fu}}, \bibinfo {author} {\bibfnamefont {M.}~\bibnamefont {Xu}}, \bibinfo
  {author} {\bibfnamefont {S.}~\bibnamefont {Shankar}}, \bibinfo {author}
  {\bibfnamefont {M.~H.}\ \bibnamefont {Devoret}}, \bibinfo {author}
  {\bibfnamefont {H.~X.}\ \bibnamefont {Tang}},\ and\ \bibinfo {author}
  {\bibfnamefont {L.}~\bibnamefont {Jiang}},\ }\bibfield  {title} {\bibinfo
  {title} {{Proposal for Heralded Generation and Detection of Entangled
  Microwave^^e2^^80^^93Optical-Photon Pairs}},\ }\href
  {https://doi.org/10.1103/PhysRevLett.124.010511} {\bibfield  {journal}
  {\bibinfo  {journal} {Physical Review Letters}\ }\textbf {\bibinfo {volume}
  {124}},\ \bibinfo {pages} {010511} (\bibinfo {year} {2020})}\BibitemShut
  {NoStop}%
\bibitem [{\citenamefont {Krastanov}\ \emph {et~al.}(2021)\citenamefont
  {Krastanov}, \citenamefont {Raniwala}, \citenamefont {Holzgrafe},
  \citenamefont {Jacobs}, \citenamefont {Lon{\v{c}}ar}, \citenamefont
  {Reagor},\ and\ \citenamefont {Englund}}]{Krastanov2021}%
  \BibitemOpen
  \bibfield  {author} {\bibinfo {author} {\bibfnamefont {S.}~\bibnamefont
  {Krastanov}}, \bibinfo {author} {\bibfnamefont {H.}~\bibnamefont {Raniwala}},
  \bibinfo {author} {\bibfnamefont {J.}~\bibnamefont {Holzgrafe}}, \bibinfo
  {author} {\bibfnamefont {K.}~\bibnamefont {Jacobs}}, \bibinfo {author}
  {\bibfnamefont {M.}~\bibnamefont {Lon{\v{c}}ar}}, \bibinfo {author}
  {\bibfnamefont {M.~J.}\ \bibnamefont {Reagor}},\ and\ \bibinfo {author}
  {\bibfnamefont {D.~R.}\ \bibnamefont {Englund}},\ }\bibfield  {title}
  {\bibinfo {title} {{Optically Heralded Entanglement of Superconducting
  Systems in Quantum Networks}},\ }\href
  {https://doi.org/10.1103/PhysRevLett.127.040503} {\bibfield  {journal}
  {\bibinfo  {journal} {Physical Review Letters}\ }\textbf {\bibinfo {volume}
  {127}},\ \bibinfo {pages} {040503} (\bibinfo {year} {2021})},\ \Eprint
  {https://arxiv.org/abs/2012.13408} {arXiv:2012.13408} \BibitemShut {NoStop}%
\bibitem [{\citenamefont {Wu}\ \emph {et~al.}()\citenamefont {Wu},
  \citenamefont {Cui}, \citenamefont {Fan},\ and\ \citenamefont
  {Zhuang}}]{Wu2021}%
  \BibitemOpen
  \bibfield  {author} {\bibinfo {author} {\bibfnamefont {J.}~\bibnamefont
  {Wu}}, \bibinfo {author} {\bibfnamefont {C.}~\bibnamefont {Cui}}, \bibinfo
  {author} {\bibfnamefont {L.}~\bibnamefont {Fan}},\ and\ \bibinfo {author}
  {\bibfnamefont {Q.}~\bibnamefont {Zhuang}},\ }\bibfield  {title} {\bibinfo
  {title} {{Deterministic microwave-optical transduction based on quantum
  teleportation}},\ }\href {http://arxiv.org/abs/2106.14037} {\ }\Eprint
  {https://arxiv.org/abs/2106.14037} {arXiv:2106.14037} \BibitemShut {NoStop}%
\bibitem [{\citenamefont {Yamamoto}\ \emph {et~al.}(2003)\citenamefont
  {Yamamoto}, \citenamefont {Koashi}, \citenamefont {{\"{O}}zdemir},\ and\
  \citenamefont {Imoto}}]{Yamamoto2003}%
  \BibitemOpen
  \bibfield  {author} {\bibinfo {author} {\bibfnamefont {T.}~\bibnamefont
  {Yamamoto}}, \bibinfo {author} {\bibfnamefont {M.}~\bibnamefont {Koashi}},
  \bibinfo {author} {\bibfnamefont {\c{S}.~K.}\ \bibnamefont {{\"{O}}zdemir}},\
  and\ \bibinfo {author} {\bibfnamefont {N.}~\bibnamefont {Imoto}},\ }\bibfield
   {title} {\bibinfo {title} {{Experimental extraction of an entangled photon
  pair from two identically decohered pairs}},\ }\href
  {https://doi.org/10.1038/nature01358} {\bibfield  {journal} {\bibinfo
  {journal} {Nature}\ }\textbf {\bibinfo {volume} {421}},\ \bibinfo {pages}
  {343} (\bibinfo {year} {2003})}\BibitemShut {NoStop}%
\bibitem [{\citenamefont {Li}\ \emph {et~al.}(2004)\citenamefont {Li},
  \citenamefont {Chen}, \citenamefont {Voss}, \citenamefont {Sharping},\ and\
  \citenamefont {Kumar}}]{Li2004}%
  \BibitemOpen
  \bibfield  {author} {\bibinfo {author} {\bibfnamefont {X.}~\bibnamefont
  {Li}}, \bibinfo {author} {\bibfnamefont {J.}~\bibnamefont {Chen}}, \bibinfo
  {author} {\bibfnamefont {P.}~\bibnamefont {Voss}}, \bibinfo {author}
  {\bibfnamefont {J.}~\bibnamefont {Sharping}},\ and\ \bibinfo {author}
  {\bibfnamefont {P.}~\bibnamefont {Kumar}},\ }\bibfield  {title} {\bibinfo
  {title} {{All-fiber photon-pair source for quantum communications: Improved
  generation of correlated photons}},\ }\href
  {https://doi.org/10.1364/OPEX.12.003737} {\bibfield  {journal} {\bibinfo
  {journal} {Optics Express}\ }\textbf {\bibinfo {volume} {12}},\ \bibinfo
  {pages} {3737} (\bibinfo {year} {2004})}\BibitemShut {NoStop}%
\bibitem [{\citenamefont {Peng}\ \emph {et~al.}(2005)\citenamefont {Peng},
  \citenamefont {Yang}, \citenamefont {Bao}, \citenamefont {Zhang},
  \citenamefont {Jin}, \citenamefont {Feng}, \citenamefont {Yang},
  \citenamefont {Yang}, \citenamefont {Yin}, \citenamefont {Zhang},
  \citenamefont {Li}, \citenamefont {Tian},\ and\ \citenamefont
  {Pan}}]{Peng2005}%
  \BibitemOpen
  \bibfield  {author} {\bibinfo {author} {\bibfnamefont {C.-Z.}\ \bibnamefont
  {Peng}}, \bibinfo {author} {\bibfnamefont {T.}~\bibnamefont {Yang}}, \bibinfo
  {author} {\bibfnamefont {X.-H.}\ \bibnamefont {Bao}}, \bibinfo {author}
  {\bibfnamefont {J.}~\bibnamefont {Zhang}}, \bibinfo {author} {\bibfnamefont
  {X.-M.}\ \bibnamefont {Jin}}, \bibinfo {author} {\bibfnamefont {F.-Y.}\
  \bibnamefont {Feng}}, \bibinfo {author} {\bibfnamefont {B.}~\bibnamefont
  {Yang}}, \bibinfo {author} {\bibfnamefont {J.}~\bibnamefont {Yang}}, \bibinfo
  {author} {\bibfnamefont {J.}~\bibnamefont {Yin}}, \bibinfo {author}
  {\bibfnamefont {Q.}~\bibnamefont {Zhang}}, \bibinfo {author} {\bibfnamefont
  {N.}~\bibnamefont {Li}}, \bibinfo {author} {\bibfnamefont {B.-L.}\
  \bibnamefont {Tian}},\ and\ \bibinfo {author} {\bibfnamefont {J.-W.}\
  \bibnamefont {Pan}},\ }\bibfield  {title} {\bibinfo {title} {{Experimental
  Free-Space Distribution of Entangled Photon Pairs Over 13 km: Towards
  Satellite-Based Global Quantum Communication}},\ }\href
  {https://doi.org/10.1103/PhysRevLett.94.150501} {\bibfield  {journal}
  {\bibinfo  {journal} {Physical Review Letters}\ }\textbf {\bibinfo {volume}
  {94}},\ \bibinfo {pages} {150501} (\bibinfo {year} {2005})}\BibitemShut
  {NoStop}%
\bibitem [{\citenamefont {Duan}\ \emph {et~al.}(2001)\citenamefont {Duan},
  \citenamefont {Lukin}, \citenamefont {Cirac},\ and\ \citenamefont
  {Zoller}}]{Duan2001}%
  \BibitemOpen
  \bibfield  {author} {\bibinfo {author} {\bibfnamefont {L.-M.}\ \bibnamefont
  {Duan}}, \bibinfo {author} {\bibfnamefont {M.~D.}\ \bibnamefont {Lukin}},
  \bibinfo {author} {\bibfnamefont {J.~I.}\ \bibnamefont {Cirac}},\ and\
  \bibinfo {author} {\bibfnamefont {P.}~\bibnamefont {Zoller}},\ }\bibfield
  {title} {\bibinfo {title} {{Long-distance quantum communication with atomic
  ensembles and linear optics}},\ }\href {https://doi.org/10.1038/35106500}
  {\bibfield  {journal} {\bibinfo  {journal} {Nature}\ }\textbf {\bibinfo
  {volume} {414}},\ \bibinfo {pages} {413} (\bibinfo {year}
  {2001})}\BibitemShut {NoStop}%
\bibitem [{\citenamefont {Moehring}\ \emph {et~al.}(2007)\citenamefont
  {Moehring}, \citenamefont {Maunz}, \citenamefont {Olmschenk}, \citenamefont
  {Younge}, \citenamefont {Matsukevich}, \citenamefont {Duan},\ and\
  \citenamefont {Monroe}}]{Moehring2007}%
  \BibitemOpen
  \bibfield  {author} {\bibinfo {author} {\bibfnamefont {D.~L.}\ \bibnamefont
  {Moehring}}, \bibinfo {author} {\bibfnamefont {P.}~\bibnamefont {Maunz}},
  \bibinfo {author} {\bibfnamefont {S.}~\bibnamefont {Olmschenk}}, \bibinfo
  {author} {\bibfnamefont {K.~C.}\ \bibnamefont {Younge}}, \bibinfo {author}
  {\bibfnamefont {D.~N.}\ \bibnamefont {Matsukevich}}, \bibinfo {author}
  {\bibfnamefont {L.-M.}\ \bibnamefont {Duan}},\ and\ \bibinfo {author}
  {\bibfnamefont {C.}~\bibnamefont {Monroe}},\ }\bibfield  {title} {\bibinfo
  {title} {{Entanglement of single-atom quantum bits at a distance}},\ }\href
  {https://doi.org/10.1038/nature06118} {\bibfield  {journal} {\bibinfo
  {journal} {Nature}\ }\textbf {\bibinfo {volume} {449}},\ \bibinfo {pages}
  {68} (\bibinfo {year} {2007})}\BibitemShut {NoStop}%
\bibitem [{\citenamefont {Hofmann}\ \emph {et~al.}(2012)\citenamefont
  {Hofmann}, \citenamefont {Krug}, \citenamefont {Ortegel}, \citenamefont
  {G{\'{e}}rard}, \citenamefont {Weber}, \citenamefont {Rosenfeld},\ and\
  \citenamefont {Weinfurter}}]{Hofmann2012}%
  \BibitemOpen
  \bibfield  {author} {\bibinfo {author} {\bibfnamefont {J.}~\bibnamefont
  {Hofmann}}, \bibinfo {author} {\bibfnamefont {M.}~\bibnamefont {Krug}},
  \bibinfo {author} {\bibfnamefont {N.}~\bibnamefont {Ortegel}}, \bibinfo
  {author} {\bibfnamefont {L.}~\bibnamefont {G{\'{e}}rard}}, \bibinfo {author}
  {\bibfnamefont {M.}~\bibnamefont {Weber}}, \bibinfo {author} {\bibfnamefont
  {W.}~\bibnamefont {Rosenfeld}},\ and\ \bibinfo {author} {\bibfnamefont
  {H.}~\bibnamefont {Weinfurter}},\ }\bibfield  {title} {\bibinfo {title}
  {{Heralded Entanglement Between Widely Separated Atoms}},\ }\href
  {https://doi.org/10.1126/science.1221856} {\bibfield  {journal} {\bibinfo
  {journal} {Science}\ }\textbf {\bibinfo {volume} {337}},\ \bibinfo {pages}
  {72} (\bibinfo {year} {2012})}\BibitemShut {NoStop}%
\bibitem [{\citenamefont {Bernien}\ \emph {et~al.}(2013)\citenamefont
  {Bernien}, \citenamefont {Hensen}, \citenamefont {Pfaff}, \citenamefont
  {Koolstra}, \citenamefont {Blok}, \citenamefont {Robledo}, \citenamefont
  {Taminiau}, \citenamefont {Markham}, \citenamefont {Twitchen}, \citenamefont
  {Childress},\ and\ \citenamefont {Hanson}}]{Bernien2013}%
  \BibitemOpen
  \bibfield  {author} {\bibinfo {author} {\bibfnamefont {H.}~\bibnamefont
  {Bernien}}, \bibinfo {author} {\bibfnamefont {B.}~\bibnamefont {Hensen}},
  \bibinfo {author} {\bibfnamefont {W.}~\bibnamefont {Pfaff}}, \bibinfo
  {author} {\bibfnamefont {G.}~\bibnamefont {Koolstra}}, \bibinfo {author}
  {\bibfnamefont {M.~S.}\ \bibnamefont {Blok}}, \bibinfo {author}
  {\bibfnamefont {L.}~\bibnamefont {Robledo}}, \bibinfo {author} {\bibfnamefont
  {T.~H.}\ \bibnamefont {Taminiau}}, \bibinfo {author} {\bibfnamefont
  {M.}~\bibnamefont {Markham}}, \bibinfo {author} {\bibfnamefont {D.~J.}\
  \bibnamefont {Twitchen}}, \bibinfo {author} {\bibfnamefont {L.}~\bibnamefont
  {Childress}},\ and\ \bibinfo {author} {\bibfnamefont {R.}~\bibnamefont
  {Hanson}},\ }\bibfield  {title} {\bibinfo {title} {{Heralded entanglement
  between solid-state qubits separated by three metres}},\ }\href
  {https://doi.org/10.1038/nature12016} {\bibfield  {journal} {\bibinfo
  {journal} {Nature}\ }\textbf {\bibinfo {volume} {497}},\ \bibinfo {pages}
  {86} (\bibinfo {year} {2013})},\ \Eprint {https://arxiv.org/abs/1212.6136}
  {arXiv:1212.6136} \BibitemShut {NoStop}%
\bibitem [{\citenamefont {Delteil}\ \emph {et~al.}(2016)\citenamefont
  {Delteil}, \citenamefont {Sun}, \citenamefont {Gao}, \citenamefont {Togan},
  \citenamefont {Faelt},\ and\ \citenamefont {Imamo^^c4^^9flu}}]{Delteil2016}%
  \BibitemOpen
  \bibfield  {author} {\bibinfo {author} {\bibfnamefont {A.}~\bibnamefont
  {Delteil}}, \bibinfo {author} {\bibfnamefont {Z.}~\bibnamefont {Sun}},
  \bibinfo {author} {\bibfnamefont {W.-b.}\ \bibnamefont {Gao}}, \bibinfo
  {author} {\bibfnamefont {E.}~\bibnamefont {Togan}}, \bibinfo {author}
  {\bibfnamefont {S.}~\bibnamefont {Faelt}},\ and\ \bibinfo {author}
  {\bibfnamefont {A.}~\bibnamefont {Imamo^^c4^^9flu}},\ }\bibfield  {title}
  {\bibinfo {title} {{Generation of heralded entanglement between distant hole
  spins}},\ }\href {https://doi.org/10.1038/nphys3605} {\bibfield  {journal}
  {\bibinfo  {journal} {Nature Physics}\ }\textbf {\bibinfo {volume} {12}},\
  \bibinfo {pages} {218} (\bibinfo {year} {2016})},\ \Eprint
  {https://arxiv.org/abs/1507.00465} {arXiv:1507.00465} \BibitemShut {NoStop}%
\bibitem [{\citenamefont {Neuman}\ \emph {et~al.}(2021)\citenamefont {Neuman},
  \citenamefont {Eichenfield}, \citenamefont {Trusheim}, \citenamefont
  {Hackett}, \citenamefont {Narang},\ and\ \citenamefont
  {Englund}}]{Neuman2021}%
  \BibitemOpen
  \bibfield  {author} {\bibinfo {author} {\bibfnamefont {T.}~\bibnamefont
  {Neuman}}, \bibinfo {author} {\bibfnamefont {M.}~\bibnamefont {Eichenfield}},
  \bibinfo {author} {\bibfnamefont {M.~E.}\ \bibnamefont {Trusheim}}, \bibinfo
  {author} {\bibfnamefont {L.}~\bibnamefont {Hackett}}, \bibinfo {author}
  {\bibfnamefont {P.}~\bibnamefont {Narang}},\ and\ \bibinfo {author}
  {\bibfnamefont {D.}~\bibnamefont {Englund}},\ }\bibfield  {title} {\bibinfo
  {title} {{A phononic interface between a superconducting quantum processor
  and quantum networked spin memories}},\ }\href
  {https://doi.org/10.1038/s41534-021-00457-4} {\bibfield  {journal} {\bibinfo
  {journal} {npj Quantum Information}\ }\textbf {\bibinfo {volume} {7}},\
  \bibinfo {pages} {121} (\bibinfo {year} {2021})}\BibitemShut {NoStop}%
\bibitem [{\citenamefont {Lee}\ \emph {et~al.}(2016)\citenamefont {Lee},
  \citenamefont {Lee}, \citenamefont {Ovartchaiyapong}, \citenamefont
  {Minguzzi}, \citenamefont {Maze},\ and\ \citenamefont {{Bleszynski
  Jayich}}}]{Lee2016}%
  \BibitemOpen
  \bibfield  {author} {\bibinfo {author} {\bibfnamefont {K.~W.}\ \bibnamefont
  {Lee}}, \bibinfo {author} {\bibfnamefont {D.}~\bibnamefont {Lee}}, \bibinfo
  {author} {\bibfnamefont {P.}~\bibnamefont {Ovartchaiyapong}}, \bibinfo
  {author} {\bibfnamefont {J.}~\bibnamefont {Minguzzi}}, \bibinfo {author}
  {\bibfnamefont {J.~R.}\ \bibnamefont {Maze}},\ and\ \bibinfo {author}
  {\bibfnamefont {A.~C.}\ \bibnamefont {{Bleszynski Jayich}}},\ }\bibfield
  {title} {\bibinfo {title} {{Strain Coupling of a Mechanical Resonator to a
  Single Quantum Emitter in Diamond}},\ }\href
  {https://doi.org/10.1103/PhysRevApplied.6.034005} {\bibfield  {journal}
  {\bibinfo  {journal} {Physical Review Applied}\ }\textbf {\bibinfo {volume}
  {6}},\ \bibinfo {pages} {034005} (\bibinfo {year} {2016})},\ \Eprint
  {https://arxiv.org/abs/1603.07680} {arXiv:1603.07680} \BibitemShut {NoStop}%
\bibitem [{\citenamefont {Meesala}\ \emph {et~al.}(2018)\citenamefont
  {Meesala}, \citenamefont {Sohn}, \citenamefont {Pingault}, \citenamefont
  {Shao}, \citenamefont {Atikian}, \citenamefont {Holzgrafe}, \citenamefont
  {G{\"{u}}ndo^^c4^^9fan}, \citenamefont {Stavrakas}, \citenamefont
  {Sipahigil}, \citenamefont {Chia}, \citenamefont {Evans}, \citenamefont
  {Burek}, \citenamefont {Zhang}, \citenamefont {Wu}, \citenamefont {Pacheco},
  \citenamefont {Abraham}, \citenamefont {Bielejec}, \citenamefont {Lukin},
  \citenamefont {Atat{\"{u}}re},\ and\ \citenamefont
  {Lon{\v{c}}ar}}]{Meesala2018}%
  \BibitemOpen
  \bibfield  {author} {\bibinfo {author} {\bibfnamefont {S.}~\bibnamefont
  {Meesala}}, \bibinfo {author} {\bibfnamefont {Y.-I.}\ \bibnamefont {Sohn}},
  \bibinfo {author} {\bibfnamefont {B.}~\bibnamefont {Pingault}}, \bibinfo
  {author} {\bibfnamefont {L.}~\bibnamefont {Shao}}, \bibinfo {author}
  {\bibfnamefont {H.~A.}\ \bibnamefont {Atikian}}, \bibinfo {author}
  {\bibfnamefont {J.}~\bibnamefont {Holzgrafe}}, \bibinfo {author}
  {\bibfnamefont {M.}~\bibnamefont {G{\"{u}}ndo^^c4^^9fan}}, \bibinfo {author}
  {\bibfnamefont {C.}~\bibnamefont {Stavrakas}}, \bibinfo {author}
  {\bibfnamefont {A.}~\bibnamefont {Sipahigil}}, \bibinfo {author}
  {\bibfnamefont {C.}~\bibnamefont {Chia}}, \bibinfo {author} {\bibfnamefont
  {R.}~\bibnamefont {Evans}}, \bibinfo {author} {\bibfnamefont {M.~J.}\
  \bibnamefont {Burek}}, \bibinfo {author} {\bibfnamefont {M.}~\bibnamefont
  {Zhang}}, \bibinfo {author} {\bibfnamefont {L.}~\bibnamefont {Wu}}, \bibinfo
  {author} {\bibfnamefont {J.~L.}\ \bibnamefont {Pacheco}}, \bibinfo {author}
  {\bibfnamefont {J.}~\bibnamefont {Abraham}}, \bibinfo {author} {\bibfnamefont
  {E.}~\bibnamefont {Bielejec}}, \bibinfo {author} {\bibfnamefont {M.~D.}\
  \bibnamefont {Lukin}}, \bibinfo {author} {\bibfnamefont {M.}~\bibnamefont
  {Atat{\"{u}}re}},\ and\ \bibinfo {author} {\bibfnamefont {M.}~\bibnamefont
  {Lon{\v{c}}ar}},\ }\bibfield  {title} {\bibinfo {title} {{Strain engineering
  of the silicon-vacancy center in diamond}},\ }\href
  {https://doi.org/10.1103/PhysRevB.97.205444} {\bibfield  {journal} {\bibinfo
  {journal} {Physical Review B}\ }\textbf {\bibinfo {volume} {97}},\ \bibinfo
  {pages} {205444} (\bibinfo {year} {2018})},\ \Eprint
  {https://arxiv.org/abs/1801.09833} {arXiv:1801.09833} \BibitemShut {NoStop}%
\bibitem [{\citenamefont {Humphreys}\ \emph {et~al.}(2018)\citenamefont
  {Humphreys}, \citenamefont {Kalb}, \citenamefont {Morits}, \citenamefont
  {Schouten}, \citenamefont {Vermeulen}, \citenamefont {Twitchen},
  \citenamefont {Markham},\ and\ \citenamefont {Hanson}}]{Humphreys2018}%
  \BibitemOpen
  \bibfield  {author} {\bibinfo {author} {\bibfnamefont {P.~C.}\ \bibnamefont
  {Humphreys}}, \bibinfo {author} {\bibfnamefont {N.}~\bibnamefont {Kalb}},
  \bibinfo {author} {\bibfnamefont {J.~P.~J.}\ \bibnamefont {Morits}}, \bibinfo
  {author} {\bibfnamefont {R.~N.}\ \bibnamefont {Schouten}}, \bibinfo {author}
  {\bibfnamefont {R.~F.~L.}\ \bibnamefont {Vermeulen}}, \bibinfo {author}
  {\bibfnamefont {D.~J.}\ \bibnamefont {Twitchen}}, \bibinfo {author}
  {\bibfnamefont {M.}~\bibnamefont {Markham}},\ and\ \bibinfo {author}
  {\bibfnamefont {R.}~\bibnamefont {Hanson}},\ }\bibfield  {title} {\bibinfo
  {title} {{Deterministic delivery of remote entanglement on a quantum
  network}},\ }\href {https://doi.org/10.1038/s41586-018-0200-5} {\bibfield
  {journal} {\bibinfo  {journal} {Nature}\ }\textbf {\bibinfo {volume} {558}},\
  \bibinfo {pages} {268} (\bibinfo {year} {2018})},\ \Eprint
  {https://arxiv.org/abs/1712.07567} {arXiv:1712.07567} \BibitemShut {NoStop}%
\bibitem [{\citenamefont {Pompili}\ \emph {et~al.}(2021)\citenamefont
  {Pompili}, \citenamefont {Hermans}, \citenamefont {Baier}, \citenamefont
  {Beukers}, \citenamefont {Humphreys}, \citenamefont {Schouten}, \citenamefont
  {Vermeulen}, \citenamefont {Tiggelman}, \citenamefont {{dos Santos Martins}},
  \citenamefont {Dirkse}, \citenamefont {Wehner},\ and\ \citenamefont
  {Hanson}}]{Pompili2021}%
  \BibitemOpen
  \bibfield  {author} {\bibinfo {author} {\bibfnamefont {M.}~\bibnamefont
  {Pompili}}, \bibinfo {author} {\bibfnamefont {S.~L.~N.}\ \bibnamefont
  {Hermans}}, \bibinfo {author} {\bibfnamefont {S.}~\bibnamefont {Baier}},
  \bibinfo {author} {\bibfnamefont {H.~K.~C.}\ \bibnamefont {Beukers}},
  \bibinfo {author} {\bibfnamefont {P.~C.}\ \bibnamefont {Humphreys}}, \bibinfo
  {author} {\bibfnamefont {R.~N.}\ \bibnamefont {Schouten}}, \bibinfo {author}
  {\bibfnamefont {R.~F.~L.}\ \bibnamefont {Vermeulen}}, \bibinfo {author}
  {\bibfnamefont {M.~J.}\ \bibnamefont {Tiggelman}}, \bibinfo {author}
  {\bibfnamefont {L.}~\bibnamefont {{dos Santos Martins}}}, \bibinfo {author}
  {\bibfnamefont {B.}~\bibnamefont {Dirkse}}, \bibinfo {author} {\bibfnamefont
  {S.}~\bibnamefont {Wehner}},\ and\ \bibinfo {author} {\bibfnamefont
  {R.}~\bibnamefont {Hanson}},\ }\bibfield  {title} {\bibinfo {title}
  {{Realization of a multinode quantum network of remote solid-state qubits}},\
  }\href {https://doi.org/10.1126/science.abg1919} {\bibfield  {journal}
  {\bibinfo  {journal} {Science}\ }\textbf {\bibinfo {volume} {372}},\ \bibinfo
  {pages} {259} (\bibinfo {year} {2021})},\ \Eprint
  {https://arxiv.org/abs/2102.04471} {arXiv:2102.04471} \BibitemShut {NoStop}%
\bibitem [{\citenamefont {Levonian}\ \emph {et~al.}()\citenamefont {Levonian},
  \citenamefont {Riedinger}, \citenamefont {Machielse}, \citenamefont {Knall},
  \citenamefont {Bhaskar}, \citenamefont {Knaut}, \citenamefont {Bekenstein},
  \citenamefont {Park}, \citenamefont {Loncar},\ and\ \citenamefont
  {Lukin}}]{Levonian2021}%
  \BibitemOpen
  \bibfield  {author} {\bibinfo {author} {\bibfnamefont {D.~S.}\ \bibnamefont
  {Levonian}}, \bibinfo {author} {\bibfnamefont {R.}~\bibnamefont {Riedinger}},
  \bibinfo {author} {\bibfnamefont {B.}~\bibnamefont {Machielse}}, \bibinfo
  {author} {\bibfnamefont {E.~N.}\ \bibnamefont {Knall}}, \bibinfo {author}
  {\bibfnamefont {M.~K.}\ \bibnamefont {Bhaskar}}, \bibinfo {author}
  {\bibfnamefont {C.~M.}\ \bibnamefont {Knaut}}, \bibinfo {author}
  {\bibfnamefont {R.}~\bibnamefont {Bekenstein}}, \bibinfo {author}
  {\bibfnamefont {H.}~\bibnamefont {Park}}, \bibinfo {author} {\bibfnamefont
  {M.}~\bibnamefont {Loncar}},\ and\ \bibinfo {author} {\bibfnamefont {M.~D.}\
  \bibnamefont {Lukin}},\ }\bibfield  {title} {\bibinfo {title} {{Optical
  Entanglement of Distinguishable Quantum Emitters}},\ }\href
  {http://arxiv.org/abs/2108.10928} {\ }\Eprint
  {https://arxiv.org/abs/2108.10928v1} {arXiv:2108.10928v1} \BibitemShut
  {NoStop}%
\bibitem [{\citenamefont {Kurpiers}\ \emph {et~al.}(2019)\citenamefont
  {Kurpiers}, \citenamefont {Pechal}, \citenamefont {Royer}, \citenamefont
  {Magnard}, \citenamefont {Walter}, \citenamefont {Heinsoo}, \citenamefont
  {Salath{\'{e}}}, \citenamefont {Akin}, \citenamefont {Storz}, \citenamefont
  {Besse}, \citenamefont {Gasparinetti}, \citenamefont {Blais},\ and\
  \citenamefont {Wallraff}}]{Kurpiers2019}%
  \BibitemOpen
  \bibfield  {author} {\bibinfo {author} {\bibfnamefont {P.}~\bibnamefont
  {Kurpiers}}, \bibinfo {author} {\bibfnamefont {M.}~\bibnamefont {Pechal}},
  \bibinfo {author} {\bibfnamefont {B.}~\bibnamefont {Royer}}, \bibinfo
  {author} {\bibfnamefont {P.}~\bibnamefont {Magnard}}, \bibinfo {author}
  {\bibfnamefont {T.}~\bibnamefont {Walter}}, \bibinfo {author} {\bibfnamefont
  {J.}~\bibnamefont {Heinsoo}}, \bibinfo {author} {\bibfnamefont
  {Y.}~\bibnamefont {Salath{\'{e}}}}, \bibinfo {author} {\bibfnamefont
  {A.}~\bibnamefont {Akin}}, \bibinfo {author} {\bibfnamefont {S.}~\bibnamefont
  {Storz}}, \bibinfo {author} {\bibfnamefont {J.-C.}\ \bibnamefont {Besse}},
  \bibinfo {author} {\bibfnamefont {S.}~\bibnamefont {Gasparinetti}}, \bibinfo
  {author} {\bibfnamefont {A.}~\bibnamefont {Blais}},\ and\ \bibinfo {author}
  {\bibfnamefont {A.}~\bibnamefont {Wallraff}},\ }\bibfield  {title} {\bibinfo
  {title} {{Quantum Communication with Time-Bin Encoded Microwave Photons}},\
  }\href {https://doi.org/10.1103/PhysRevApplied.12.044067} {\bibfield
  {journal} {\bibinfo  {journal} {Physical Review Applied}\ }\textbf {\bibinfo
  {volume} {12}},\ \bibinfo {pages} {044067} (\bibinfo {year} {2019})},\
  \Eprint {https://arxiv.org/abs/1811.07604} {arXiv:1811.07604} \BibitemShut
  {NoStop}%
\bibitem [{\citenamefont {Kurpiers}\ \emph
  {et~al.}(2017{\natexlab{a}})\citenamefont {Kurpiers}, \citenamefont
  {Magnard}, \citenamefont {Walter}, \citenamefont {Royer}, \citenamefont
  {Pechal}, \citenamefont {Heinsoo}, \citenamefont {Salath{\'{e}}},
  \citenamefont {Akin}, \citenamefont {Storz}, \citenamefont {Besse},
  \citenamefont {Gasparinetti}, \citenamefont {Blais},\ and\ \citenamefont
  {Wallraff}}]{Kurpiers2018}%
  \BibitemOpen
  \bibfield  {author} {\bibinfo {author} {\bibfnamefont {P.}~\bibnamefont
  {Kurpiers}}, \bibinfo {author} {\bibfnamefont {P.}~\bibnamefont {Magnard}},
  \bibinfo {author} {\bibfnamefont {T.}~\bibnamefont {Walter}}, \bibinfo
  {author} {\bibfnamefont {B.}~\bibnamefont {Royer}}, \bibinfo {author}
  {\bibfnamefont {M.}~\bibnamefont {Pechal}}, \bibinfo {author} {\bibfnamefont
  {J.}~\bibnamefont {Heinsoo}}, \bibinfo {author} {\bibfnamefont
  {Y.}~\bibnamefont {Salath{\'{e}}}}, \bibinfo {author} {\bibfnamefont
  {A.}~\bibnamefont {Akin}}, \bibinfo {author} {\bibfnamefont {S.}~\bibnamefont
  {Storz}}, \bibinfo {author} {\bibfnamefont {J.-C.}\ \bibnamefont {Besse}},
  \bibinfo {author} {\bibfnamefont {S.}~\bibnamefont {Gasparinetti}}, \bibinfo
  {author} {\bibfnamefont {A.}~\bibnamefont {Blais}},\ and\ \bibinfo {author}
  {\bibfnamefont {A.}~\bibnamefont {Wallraff}},\ }\bibfield  {title} {\bibinfo
  {title} {{Deterministic Quantum State Transfer and Generation of Remote
  Entanglement using Microwave Photons}},\ }\href
  {https://doi.org/10.1038/s41586-018-0195-y} {\bibfield  {journal} {\bibinfo
  {journal} {Nature}\ }\textbf {\bibinfo {volume} {558}},\ \bibinfo {pages}
  {264} (\bibinfo {year} {2017}{\natexlab{a}})},\ \Eprint
  {https://arxiv.org/abs/1712.08593} {arXiv:1712.08593} \BibitemShut {NoStop}%
\bibitem [{\citenamefont {Douce}\ \emph {et~al.}(2015)\citenamefont {Douce},
  \citenamefont {Stern}, \citenamefont {Zagury}, \citenamefont {Bertet},\ and\
  \citenamefont {Milman}}]{Douce2015}%
  \BibitemOpen
  \bibfield  {author} {\bibinfo {author} {\bibfnamefont {T.}~\bibnamefont
  {Douce}}, \bibinfo {author} {\bibfnamefont {M.}~\bibnamefont {Stern}},
  \bibinfo {author} {\bibfnamefont {N.}~\bibnamefont {Zagury}}, \bibinfo
  {author} {\bibfnamefont {P.}~\bibnamefont {Bertet}},\ and\ \bibinfo {author}
  {\bibfnamefont {P.}~\bibnamefont {Milman}},\ }\bibfield  {title} {\bibinfo
  {title} {{Coupling a single nitrogen-vacancy center to a superconducting flux
  qubit in the far-off-resonance regime}},\ }\href
  {https://doi.org/10.1103/PhysRevA.92.052335} {\bibfield  {journal} {\bibinfo
  {journal} {Physical Review A}\ }\textbf {\bibinfo {volume} {92}},\ \bibinfo
  {pages} {052335} (\bibinfo {year} {2015})},\ \Eprint
  {https://arxiv.org/abs/1507.08099} {arXiv:1507.08099} \BibitemShut {NoStop}%
\bibitem [{\citenamefont {Li}\ and\ \citenamefont {Li}(2018)}]{Li2018}%
  \BibitemOpen
  \bibfield  {author} {\bibinfo {author} {\bibfnamefont {C.-H.}\ \bibnamefont
  {Li}}\ and\ \bibinfo {author} {\bibfnamefont {P.-B.}\ \bibnamefont {Li}},\
  }\bibfield  {title} {\bibinfo {title} {{Coupling a single nitrogen-vacancy
  center with a superconducting qubit via the electro-optic effect}},\ }\href
  {https://doi.org/10.1103/PhysRevA.97.052319} {\bibfield  {journal} {\bibinfo
  {journal} {Physical Review A}\ }\textbf {\bibinfo {volume} {97}},\ \bibinfo
  {pages} {052319} (\bibinfo {year} {2018})},\ \Eprint
  {https://arxiv.org/abs/1804.10722} {arXiv:1804.10722} \BibitemShut {NoStop}%
\bibitem [{\citenamefont {Park}\ \emph {et~al.}(2006)\citenamefont {Park},
  \citenamefont {Cook},\ and\ \citenamefont {Wang}}]{Park2006}%
  \BibitemOpen
  \bibfield  {author} {\bibinfo {author} {\bibfnamefont {Y.-S.}\ \bibnamefont
  {Park}}, \bibinfo {author} {\bibfnamefont {A.~K.}\ \bibnamefont {Cook}},\
  and\ \bibinfo {author} {\bibfnamefont {H.}~\bibnamefont {Wang}},\ }\bibfield
  {title} {\bibinfo {title} {{Cavity QED with Diamond Nanocrystals and Silica
  Microspheres}},\ }\href {https://doi.org/10.1021/nl061342r} {\bibfield
  {journal} {\bibinfo  {journal} {Nano Letters}\ }\textbf {\bibinfo {volume}
  {6}},\ \bibinfo {pages} {2075} (\bibinfo {year} {2006})}\BibitemShut
  {NoStop}%
\bibitem [{\citenamefont {Larsson}\ \emph {et~al.}(2009)\citenamefont
  {Larsson}, \citenamefont {Dinyari},\ and\ \citenamefont
  {Wang}}]{Larsson2009}%
  \BibitemOpen
  \bibfield  {author} {\bibinfo {author} {\bibfnamefont {M.}~\bibnamefont
  {Larsson}}, \bibinfo {author} {\bibfnamefont {K.~N.}\ \bibnamefont
  {Dinyari}},\ and\ \bibinfo {author} {\bibfnamefont {H.}~\bibnamefont
  {Wang}},\ }\bibfield  {title} {\bibinfo {title} {{Composite Optical
  Microcavity of Diamond Nanopillar and Silica Microsphere}},\ }\href
  {https://doi.org/10.1021/nl8032944} {\bibfield  {journal} {\bibinfo
  {journal} {Nano Letters}\ }\textbf {\bibinfo {volume} {9}},\ \bibinfo {pages}
  {1447} (\bibinfo {year} {2009})}\BibitemShut {NoStop}%
\bibitem [{\citenamefont {Schietinger}\ \emph {et~al.}(2008)\citenamefont
  {Schietinger}, \citenamefont {Schr{\"{o}}der},\ and\ \citenamefont
  {Benson}}]{Schietinger2008}%
  \BibitemOpen
  \bibfield  {author} {\bibinfo {author} {\bibfnamefont {S.}~\bibnamefont
  {Schietinger}}, \bibinfo {author} {\bibfnamefont {T.}~\bibnamefont
  {Schr{\"{o}}der}},\ and\ \bibinfo {author} {\bibfnamefont {O.}~\bibnamefont
  {Benson}},\ }\bibfield  {title} {\bibinfo {title} {{One-by-One Coupling of
  Single Defect Centers in Nanodiamonds to High-Q Modes of an Optical
  Microresonator}},\ }\href {https://doi.org/10.1021/nl8023627} {\bibfield
  {journal} {\bibinfo  {journal} {Nano Letters}\ }\textbf {\bibinfo {volume}
  {8}},\ \bibinfo {pages} {3911} (\bibinfo {year} {2008})}\BibitemShut
  {NoStop}%
\bibitem [{\citenamefont {Li}\ \emph {et~al.}(2011)\citenamefont {Li},
  \citenamefont {Gao},\ and\ \citenamefont {Li}}]{Li2011}%
  \BibitemOpen
  \bibfield  {author} {\bibinfo {author} {\bibfnamefont {P.-B.}\ \bibnamefont
  {Li}}, \bibinfo {author} {\bibfnamefont {S.-Y.}\ \bibnamefont {Gao}},\ and\
  \bibinfo {author} {\bibfnamefont {F.-L.}\ \bibnamefont {Li}},\ }\bibfield
  {title} {\bibinfo {title} {{Quantum-information transfer with
  nitrogen-vacancy centers coupled to a whispering-gallery microresonator}},\
  }\href {https://doi.org/10.1103/PhysRevA.83.054306} {\bibfield  {journal}
  {\bibinfo  {journal} {Physical Review A}\ }\textbf {\bibinfo {volume} {83}},\
  \bibinfo {pages} {054306} (\bibinfo {year} {2011})},\ \Eprint
  {https://arxiv.org/abs/1010.6138} {arXiv:1010.6138} \BibitemShut {NoStop}%
\bibitem [{\citenamefont {Ilves}\ \emph {et~al.}(2020)\citenamefont {Ilves},
  \citenamefont {Kono}, \citenamefont {Sunada}, \citenamefont {Yamazaki},
  \citenamefont {Kim}, \citenamefont {Koshino},\ and\ \citenamefont
  {Nakamura}}]{Ilves2020}%
  \BibitemOpen
  \bibfield  {author} {\bibinfo {author} {\bibfnamefont {J.}~\bibnamefont
  {Ilves}}, \bibinfo {author} {\bibfnamefont {S.}~\bibnamefont {Kono}},
  \bibinfo {author} {\bibfnamefont {Y.}~\bibnamefont {Sunada}}, \bibinfo
  {author} {\bibfnamefont {S.}~\bibnamefont {Yamazaki}}, \bibinfo {author}
  {\bibfnamefont {M.}~\bibnamefont {Kim}}, \bibinfo {author} {\bibfnamefont
  {K.}~\bibnamefont {Koshino}},\ and\ \bibinfo {author} {\bibfnamefont
  {Y.}~\bibnamefont {Nakamura}},\ }\bibfield  {title} {\bibinfo {title}
  {{On-demand generation and characterization of a microwave time-bin qubit}},\
  }\href {https://doi.org/10.1038/s41534-020-0266-4} {\bibfield  {journal}
  {\bibinfo  {journal} {npj Quantum Information}\ }\textbf {\bibinfo {volume}
  {6}},\ \bibinfo {pages} {34} (\bibinfo {year} {2020})},\ \Eprint
  {https://arxiv.org/abs/1912.03006} {arXiv:1912.03006} \BibitemShut {NoStop}%
\bibitem [{\citenamefont {Somoroff}\ \emph {et~al.}()\citenamefont {Somoroff},
  \citenamefont {Ficheux}, \citenamefont {Mencia}, \citenamefont {Xiong},
  \citenamefont {Kuzmin},\ and\ \citenamefont {Manucharyan}}]{Somoroff2021}%
  \BibitemOpen
  \bibfield  {author} {\bibinfo {author} {\bibfnamefont {A.}~\bibnamefont
  {Somoroff}}, \bibinfo {author} {\bibfnamefont {Q.}~\bibnamefont {Ficheux}},
  \bibinfo {author} {\bibfnamefont {R.~A.}\ \bibnamefont {Mencia}}, \bibinfo
  {author} {\bibfnamefont {H.}~\bibnamefont {Xiong}}, \bibinfo {author}
  {\bibfnamefont {R.}~\bibnamefont {Kuzmin}},\ and\ \bibinfo {author}
  {\bibfnamefont {V.~E.}\ \bibnamefont {Manucharyan}},\ }\bibfield  {title}
  {\bibinfo {title} {{Millisecond coherence in a superconducting qubit}},\
  }\href {http://arxiv.org/abs/2103.08578} {\ }\Eprint
  {https://arxiv.org/abs/2103.08578} {arXiv:2103.08578} \BibitemShut {NoStop}%
\bibitem [{\citenamefont {Shim}\ \emph {et~al.}(2013)\citenamefont {Shim},
  \citenamefont {Niemeyer}, \citenamefont {Zhang},\ and\ \citenamefont
  {Suter}}]{Shim2013}%
  \BibitemOpen
  \bibfield  {author} {\bibinfo {author} {\bibfnamefont {J.~H.}\ \bibnamefont
  {Shim}}, \bibinfo {author} {\bibfnamefont {I.}~\bibnamefont {Niemeyer}},
  \bibinfo {author} {\bibfnamefont {J.}~\bibnamefont {Zhang}},\ and\ \bibinfo
  {author} {\bibfnamefont {D.}~\bibnamefont {Suter}},\ }\bibfield  {title}
  {\bibinfo {title} {{Room-temperature high-speed nuclear-spin quantum memory
  in diamond}},\ }\href {https://doi.org/10.1103/PhysRevA.87.012301} {\bibfield
   {journal} {\bibinfo  {journal} {Physical Review A}\ }\textbf {\bibinfo
  {volume} {87}},\ \bibinfo {pages} {012301} (\bibinfo {year} {2013})},\
  \Eprint {https://arxiv.org/abs/1210.8278} {arXiv:1210.8278} \BibitemShut
  {NoStop}%
\bibitem [{\citenamefont {Hegde}\ \emph {et~al.}(2020)\citenamefont {Hegde},
  \citenamefont {Zhang},\ and\ \citenamefont {Suter}}]{Hegde2020}%
  \BibitemOpen
  \bibfield  {author} {\bibinfo {author} {\bibfnamefont {S.~S.}\ \bibnamefont
  {Hegde}}, \bibinfo {author} {\bibfnamefont {J.}~\bibnamefont {Zhang}},\ and\
  \bibinfo {author} {\bibfnamefont {D.}~\bibnamefont {Suter}},\ }\bibfield
  {title} {\bibinfo {title} {{Efficient Quantum Gates for Individual Nuclear
  Spin Qubits by Indirect Control}},\ }\href
  {https://doi.org/10.1103/PhysRevLett.124.220501} {\bibfield  {journal}
  {\bibinfo  {journal} {Physical Review Letters}\ }\textbf {\bibinfo {volume}
  {124}},\ \bibinfo {pages} {220501} (\bibinfo {year} {2020})},\ \Eprint
  {https://arxiv.org/abs/1905.01649} {arXiv:1905.01649} \BibitemShut {NoStop}%
\bibitem [{\citenamefont {Zu}\ \emph {et~al.}(2014)\citenamefont {Zu},
  \citenamefont {Wang}, \citenamefont {He}, \citenamefont {Zhang},
  \citenamefont {Dai}, \citenamefont {Wang},\ and\ \citenamefont
  {Duan}}]{Zu2014}%
  \BibitemOpen
  \bibfield  {author} {\bibinfo {author} {\bibfnamefont {C.}~\bibnamefont
  {Zu}}, \bibinfo {author} {\bibfnamefont {W.-B.}\ \bibnamefont {Wang}},
  \bibinfo {author} {\bibfnamefont {L.}~\bibnamefont {He}}, \bibinfo {author}
  {\bibfnamefont {W.-G.}\ \bibnamefont {Zhang}}, \bibinfo {author}
  {\bibfnamefont {C.-Y.}\ \bibnamefont {Dai}}, \bibinfo {author} {\bibfnamefont
  {F.}~\bibnamefont {Wang}},\ and\ \bibinfo {author} {\bibfnamefont {L.-M.}\
  \bibnamefont {Duan}},\ }\bibfield  {title} {\bibinfo {title} {{Experimental
  realization of universal geometric quantum gates with solid-state spins}},\
  }\href {https://doi.org/10.1038/nature13729} {\bibfield  {journal} {\bibinfo
  {journal} {Nature}\ }\textbf {\bibinfo {volume} {514}},\ \bibinfo {pages}
  {72} (\bibinfo {year} {2014})},\ \Eprint {https://arxiv.org/abs/1411.3157}
  {arXiv:1411.3157} \BibitemShut {NoStop}%
\bibitem [{\citenamefont {Nagata}\ \emph {et~al.}(2018)\citenamefont {Nagata},
  \citenamefont {Kuramitani}, \citenamefont {Sekiguchi},\ and\ \citenamefont
  {Kosaka}}]{Nagata2018}%
  \BibitemOpen
  \bibfield  {author} {\bibinfo {author} {\bibfnamefont {K.}~\bibnamefont
  {Nagata}}, \bibinfo {author} {\bibfnamefont {K.}~\bibnamefont {Kuramitani}},
  \bibinfo {author} {\bibfnamefont {Y.}~\bibnamefont {Sekiguchi}},\ and\
  \bibinfo {author} {\bibfnamefont {H.}~\bibnamefont {Kosaka}},\ }\bibfield
  {title} {\bibinfo {title} {{Universal holonomic quantum gates over geometric
  spin qubits with polarised microwaves}},\ }\href
  {https://doi.org/10.1038/s41467-018-05664-w} {\bibfield  {journal} {\bibinfo
  {journal} {Nature Communications}\ }\textbf {\bibinfo {volume} {9}},\
  \bibinfo {pages} {3227} (\bibinfo {year} {2018})}\BibitemShut {NoStop}%
\bibitem [{\citenamefont {Bhaskar}\ \emph {et~al.}(2020)\citenamefont
  {Bhaskar}, \citenamefont {Riedinger}, \citenamefont {Machielse},
  \citenamefont {Levonian}, \citenamefont {Nguyen}, \citenamefont {Knall},
  \citenamefont {Park}, \citenamefont {Englund}, \citenamefont {Lon{\v{c}}ar},
  \citenamefont {Sukachev},\ and\ \citenamefont {Lukin}}]{Bhaskar2020}%
  \BibitemOpen
  \bibfield  {author} {\bibinfo {author} {\bibfnamefont {M.~K.}\ \bibnamefont
  {Bhaskar}}, \bibinfo {author} {\bibfnamefont {R.}~\bibnamefont {Riedinger}},
  \bibinfo {author} {\bibfnamefont {B.}~\bibnamefont {Machielse}}, \bibinfo
  {author} {\bibfnamefont {D.~S.}\ \bibnamefont {Levonian}}, \bibinfo {author}
  {\bibfnamefont {C.~T.}\ \bibnamefont {Nguyen}}, \bibinfo {author}
  {\bibfnamefont {E.~N.}\ \bibnamefont {Knall}}, \bibinfo {author}
  {\bibfnamefont {H.}~\bibnamefont {Park}}, \bibinfo {author} {\bibfnamefont
  {D.}~\bibnamefont {Englund}}, \bibinfo {author} {\bibfnamefont
  {M.}~\bibnamefont {Lon{\v{c}}ar}}, \bibinfo {author} {\bibfnamefont {D.~D.}\
  \bibnamefont {Sukachev}},\ and\ \bibinfo {author} {\bibfnamefont {M.~D.}\
  \bibnamefont {Lukin}},\ }\bibfield  {title} {\bibinfo {title} {{Experimental
  demonstration of memory-enhanced quantum communication}},\ }\href
  {https://doi.org/10.1038/s41586-020-2103-5} {\bibfield  {journal} {\bibinfo
  {journal} {Nature}\ }\textbf {\bibinfo {volume} {580}},\ \bibinfo {pages}
  {60} (\bibinfo {year} {2020})},\ \Eprint {https://arxiv.org/abs/1909.01323}
  {arXiv:1909.01323} \BibitemShut {NoStop}%
\bibitem [{\citenamefont {Nguyen}\ \emph {et~al.}(2019)\citenamefont {Nguyen},
  \citenamefont {Sukachev}, \citenamefont {Bhaskar}, \citenamefont {Machielse},
  \citenamefont {Levonian}, \citenamefont {Knall}, \citenamefont {Stroganov},
  \citenamefont {Chia}, \citenamefont {Burek}, \citenamefont {Riedinger},
  \citenamefont {Park}, \citenamefont {Lon{\v{c}}ar},\ and\ \citenamefont
  {Lukin}}]{Nguyen2019a}%
  \BibitemOpen
  \bibfield  {author} {\bibinfo {author} {\bibfnamefont {C.~T.}\ \bibnamefont
  {Nguyen}}, \bibinfo {author} {\bibfnamefont {D.~D.}\ \bibnamefont
  {Sukachev}}, \bibinfo {author} {\bibfnamefont {M.~K.}\ \bibnamefont
  {Bhaskar}}, \bibinfo {author} {\bibfnamefont {B.}~\bibnamefont {Machielse}},
  \bibinfo {author} {\bibfnamefont {D.~S.}\ \bibnamefont {Levonian}}, \bibinfo
  {author} {\bibfnamefont {E.~N.}\ \bibnamefont {Knall}}, \bibinfo {author}
  {\bibfnamefont {P.}~\bibnamefont {Stroganov}}, \bibinfo {author}
  {\bibfnamefont {C.}~\bibnamefont {Chia}}, \bibinfo {author} {\bibfnamefont
  {M.~J.}\ \bibnamefont {Burek}}, \bibinfo {author} {\bibfnamefont
  {R.}~\bibnamefont {Riedinger}}, \bibinfo {author} {\bibfnamefont
  {H.}~\bibnamefont {Park}}, \bibinfo {author} {\bibfnamefont {M.}~\bibnamefont
  {Lon{\v{c}}ar}},\ and\ \bibinfo {author} {\bibfnamefont {M.~D.}\ \bibnamefont
  {Lukin}},\ }\bibfield  {title} {\bibinfo {title} {{An integrated nanophotonic
  quantum register based on silicon-vacancy spins in diamond}},\ }\href
  {https://doi.org/10.1103/PhysRevB.100.165428} {\bibfield  {journal} {\bibinfo
   {journal} {Physical Review B}\ }\textbf {\bibinfo {volume} {100}},\ \bibinfo
  {pages} {165428} (\bibinfo {year} {2019})},\ \Eprint
  {https://arxiv.org/abs/1907.13200} {arXiv:1907.13200} \BibitemShut {NoStop}%
\bibitem [{\citenamefont {Wang}\ \emph {et~al.}(2021)\citenamefont {Wang},
  \citenamefont {Wu}, \citenamefont {Bao}, \citenamefont {Li}, \citenamefont
  {Ma}, \citenamefont {Wang}, \citenamefont {Song}, \citenamefont {Zhang},\
  and\ \citenamefont {Duan}}]{Wang2021}%
  \BibitemOpen
  \bibfield  {author} {\bibinfo {author} {\bibfnamefont {Z.}~\bibnamefont
  {Wang}}, \bibinfo {author} {\bibfnamefont {Y.}~\bibnamefont {Wu}}, \bibinfo
  {author} {\bibfnamefont {Z.}~\bibnamefont {Bao}}, \bibinfo {author}
  {\bibfnamefont {Y.}~\bibnamefont {Li}}, \bibinfo {author} {\bibfnamefont
  {C.}~\bibnamefont {Ma}}, \bibinfo {author} {\bibfnamefont {H.}~\bibnamefont
  {Wang}}, \bibinfo {author} {\bibfnamefont {Y.}~\bibnamefont {Song}}, \bibinfo
  {author} {\bibfnamefont {H.}~\bibnamefont {Zhang}},\ and\ \bibinfo {author}
  {\bibfnamefont {L.}~\bibnamefont {Duan}},\ }\bibfield  {title} {\bibinfo
  {title} {{Experimental Realization of a Deterministic Quantum Router with
  Superconducting Quantum Circuits}},\ }\href
  {https://doi.org/10.1103/PhysRevApplied.15.014049} {\bibfield  {journal}
  {\bibinfo  {journal} {Physical Review Applied}\ }\textbf {\bibinfo {volume}
  {15}},\ \bibinfo {pages} {014049} (\bibinfo {year} {2021})}\BibitemShut
  {NoStop}%
\bibitem [{\citenamefont {Kurpiers}\ \emph
  {et~al.}(2017{\natexlab{b}})\citenamefont {Kurpiers}, \citenamefont {Walter},
  \citenamefont {Magnard}, \citenamefont {Salathe},\ and\ \citenamefont
  {Wallraff}}]{Kurpiers2017}%
  \BibitemOpen
  \bibfield  {author} {\bibinfo {author} {\bibfnamefont {P.}~\bibnamefont
  {Kurpiers}}, \bibinfo {author} {\bibfnamefont {T.}~\bibnamefont {Walter}},
  \bibinfo {author} {\bibfnamefont {P.}~\bibnamefont {Magnard}}, \bibinfo
  {author} {\bibfnamefont {Y.}~\bibnamefont {Salathe}},\ and\ \bibinfo {author}
  {\bibfnamefont {A.}~\bibnamefont {Wallraff}},\ }\bibfield  {title} {\bibinfo
  {title} {{Characterizing the attenuation of coaxial and rectangular
  microwave-frequency waveguides at cryogenic temperatures}},\ }\href
  {https://doi.org/10.1140/epjqt/s40507-017-0059-7} {\bibfield  {journal}
  {\bibinfo  {journal} {EPJ Quantum Technology}\ }\textbf {\bibinfo {volume}
  {4}},\ \bibinfo {pages} {8} (\bibinfo {year} {2017}{\natexlab{b}})},\ \Eprint
  {https://arxiv.org/abs/1612.07977} {arXiv:1612.07977} \BibitemShut {NoStop}%
\bibitem [{\citenamefont {Ruf}\ \emph {et~al.}(2021)\citenamefont {Ruf},
  \citenamefont {Wan}, \citenamefont {Choi}, \citenamefont {Englund},\ and\
  \citenamefont {Hanson}}]{Ruf2021}%
  \BibitemOpen
  \bibfield  {author} {\bibinfo {author} {\bibfnamefont {M.}~\bibnamefont
  {Ruf}}, \bibinfo {author} {\bibfnamefont {N.~H.}\ \bibnamefont {Wan}},
  \bibinfo {author} {\bibfnamefont {H.}~\bibnamefont {Choi}}, \bibinfo {author}
  {\bibfnamefont {D.}~\bibnamefont {Englund}},\ and\ \bibinfo {author}
  {\bibfnamefont {R.}~\bibnamefont {Hanson}},\ }\bibfield  {title} {\bibinfo
  {title} {{Quantum networks based on color centers in diamond}},\ }\href
  {https://doi.org/10.1063/5.0056534} {\bibfield  {journal} {\bibinfo
  {journal} {Journal of Applied Physics}\ }\textbf {\bibinfo {volume} {130}},\
  \bibinfo {pages} {070901} (\bibinfo {year} {2021})},\ \Eprint
  {https://arxiv.org/abs/2105.04341} {arXiv:2105.04341} \BibitemShut {NoStop}%
\bibitem [{\citenamefont {Sekiguchi}\ \emph {et~al.}(2016)\citenamefont
  {Sekiguchi}, \citenamefont {Komura}, \citenamefont {Mishima}, \citenamefont
  {Tanaka}, \citenamefont {Niikura},\ and\ \citenamefont
  {Kosaka}}]{Sekiguchi2016}%
  \BibitemOpen
  \bibfield  {author} {\bibinfo {author} {\bibfnamefont {Y.}~\bibnamefont
  {Sekiguchi}}, \bibinfo {author} {\bibfnamefont {Y.}~\bibnamefont {Komura}},
  \bibinfo {author} {\bibfnamefont {S.}~\bibnamefont {Mishima}}, \bibinfo
  {author} {\bibfnamefont {T.}~\bibnamefont {Tanaka}}, \bibinfo {author}
  {\bibfnamefont {N.}~\bibnamefont {Niikura}},\ and\ \bibinfo {author}
  {\bibfnamefont {H.}~\bibnamefont {Kosaka}},\ }\bibfield  {title} {\bibinfo
  {title} {{Geometric spin echo under zero field}},\ }\href
  {https://doi.org/10.1038/ncomms11668} {\bibfield  {journal} {\bibinfo
  {journal} {Nature Communications}\ }\textbf {\bibinfo {volume} {7}},\
  \bibinfo {pages} {11668} (\bibinfo {year} {2016})}\BibitemShut {NoStop}%
\bibitem [{\citenamefont {Sekiguchi}\ \emph {et~al.}(2017)\citenamefont
  {Sekiguchi}, \citenamefont {Niikura}, \citenamefont {Kuroiwa}, \citenamefont
  {Kano},\ and\ \citenamefont {Kosaka}}]{Sekiguchi2017}%
  \BibitemOpen
  \bibfield  {author} {\bibinfo {author} {\bibfnamefont {Y.}~\bibnamefont
  {Sekiguchi}}, \bibinfo {author} {\bibfnamefont {N.}~\bibnamefont {Niikura}},
  \bibinfo {author} {\bibfnamefont {R.}~\bibnamefont {Kuroiwa}}, \bibinfo
  {author} {\bibfnamefont {H.}~\bibnamefont {Kano}},\ and\ \bibinfo {author}
  {\bibfnamefont {H.}~\bibnamefont {Kosaka}},\ }\bibfield  {title} {\bibinfo
  {title} {{Optical holonomic single quantum gates with a geometric spin under
  a zero field}},\ }\href {https://doi.org/10.1038/nphoton.2017.40} {\bibfield
  {journal} {\bibinfo  {journal} {Nature Photonics}\ }\textbf {\bibinfo
  {volume} {11}},\ \bibinfo {pages} {309} (\bibinfo {year} {2017})}\BibitemShut
  {NoStop}%
\bibitem [{\citenamefont {Samkharadze}\ \emph {et~al.}(2016)\citenamefont
  {Samkharadze}, \citenamefont {Bruno}, \citenamefont {Scarlino}, \citenamefont
  {Zheng}, \citenamefont {DiVincenzo}, \citenamefont {DiCarlo},\ and\
  \citenamefont {Vandersypen}}]{Samkharadze2016}%
  \BibitemOpen
  \bibfield  {author} {\bibinfo {author} {\bibfnamefont {N.}~\bibnamefont
  {Samkharadze}}, \bibinfo {author} {\bibfnamefont {A.}~\bibnamefont {Bruno}},
  \bibinfo {author} {\bibfnamefont {P.}~\bibnamefont {Scarlino}}, \bibinfo
  {author} {\bibfnamefont {G.}~\bibnamefont {Zheng}}, \bibinfo {author}
  {\bibfnamefont {D.~P.}\ \bibnamefont {DiVincenzo}}, \bibinfo {author}
  {\bibfnamefont {L.}~\bibnamefont {DiCarlo}},\ and\ \bibinfo {author}
  {\bibfnamefont {L.~M.~K.}\ \bibnamefont {Vandersypen}},\ }\bibfield  {title}
  {\bibinfo {title} {{High-Kinetic-Inductance Superconducting Nanowire
  Resonators for Circuit QED in a Magnetic Field}},\ }\href
  {https://doi.org/10.1103/PhysRevApplied.5.044004} {\bibfield  {journal}
  {\bibinfo  {journal} {Physical Review Applied}\ }\textbf {\bibinfo {volume}
  {5}},\ \bibinfo {pages} {044004} (\bibinfo {year} {2016})},\ \Eprint
  {https://arxiv.org/abs/1511.01760} {arXiv:1511.01760} \BibitemShut {NoStop}%
\bibitem [{\citenamefont {Krause}\ \emph {et~al.}()\citenamefont {Krause},
  \citenamefont {Dickel}, \citenamefont {Vaal}, \citenamefont {Vielmetter},
  \citenamefont {Feng}, \citenamefont {Bounds}, \citenamefont {Catelani},
  \citenamefont {Fink},\ and\ \citenamefont {Ando}}]{Krause2021}%
  \BibitemOpen
  \bibfield  {author} {\bibinfo {author} {\bibfnamefont {J.}~\bibnamefont
  {Krause}}, \bibinfo {author} {\bibfnamefont {C.}~\bibnamefont {Dickel}},
  \bibinfo {author} {\bibfnamefont {E.}~\bibnamefont {Vaal}}, \bibinfo {author}
  {\bibfnamefont {M.}~\bibnamefont {Vielmetter}}, \bibinfo {author}
  {\bibfnamefont {J.}~\bibnamefont {Feng}}, \bibinfo {author} {\bibfnamefont
  {R.}~\bibnamefont {Bounds}}, \bibinfo {author} {\bibfnamefont
  {G.}~\bibnamefont {Catelani}}, \bibinfo {author} {\bibfnamefont {J.~M.}\
  \bibnamefont {Fink}},\ and\ \bibinfo {author} {\bibfnamefont
  {Y.}~\bibnamefont {Ando}},\ }\bibfield  {title} {\bibinfo {title}
  {{Magnetic-field resilience of 3D transmons with thin-film Al/AlOx/Al
  Josephson junctions approaching 1 T}},\ }\href
  {http://arxiv.org/abs/2111.01115} {\ }\Eprint
  {https://arxiv.org/abs/2111.01115} {arXiv:2111.01115} \BibitemShut {NoStop}%
\bibitem [{\citenamefont {Borregaard}\ \emph {et~al.}(2019)\citenamefont
  {Borregaard}, \citenamefont {S{\o}rensen},\ and\ \citenamefont
  {Lodahl}}]{Borregaard2019}%
  \BibitemOpen
  \bibfield  {author} {\bibinfo {author} {\bibfnamefont {J.}~\bibnamefont
  {Borregaard}}, \bibinfo {author} {\bibfnamefont {A.~S.}\ \bibnamefont
  {S{\o}rensen}},\ and\ \bibinfo {author} {\bibfnamefont {P.}~\bibnamefont
  {Lodahl}},\ }\bibfield  {title} {\bibinfo {title} {{Quantum Networks with
  Deterministic Spin-Photon Interfaces}},\ }\href
  {https://doi.org/10.1002/qute.201800091} {\bibfield  {journal} {\bibinfo
  {journal} {Advanced Quantum Technologies}\ }\textbf {\bibinfo {volume} {2}},\
  \bibinfo {pages} {1800091} (\bibinfo {year} {2019})},\ \Eprint
  {https://arxiv.org/abs/1811.08242} {arXiv:1811.08242} \BibitemShut {NoStop}%
\bibitem [{\citenamefont {Marsili}\ \emph {et~al.}(2013)\citenamefont
  {Marsili}, \citenamefont {Verma}, \citenamefont {Stern}, \citenamefont
  {Harrington}, \citenamefont {Lita}, \citenamefont {Gerrits}, \citenamefont
  {Vayshenker}, \citenamefont {Baek}, \citenamefont {Shaw}, \citenamefont
  {Mirin},\ and\ \citenamefont {Nam}}]{Marsili2013}%
  \BibitemOpen
  \bibfield  {author} {\bibinfo {author} {\bibfnamefont {F.}~\bibnamefont
  {Marsili}}, \bibinfo {author} {\bibfnamefont {V.~B.}\ \bibnamefont {Verma}},
  \bibinfo {author} {\bibfnamefont {J.~A.}\ \bibnamefont {Stern}}, \bibinfo
  {author} {\bibfnamefont {S.}~\bibnamefont {Harrington}}, \bibinfo {author}
  {\bibfnamefont {A.~E.}\ \bibnamefont {Lita}}, \bibinfo {author}
  {\bibfnamefont {T.}~\bibnamefont {Gerrits}}, \bibinfo {author} {\bibfnamefont
  {I.}~\bibnamefont {Vayshenker}}, \bibinfo {author} {\bibfnamefont
  {B.}~\bibnamefont {Baek}}, \bibinfo {author} {\bibfnamefont {M.~D.}\
  \bibnamefont {Shaw}}, \bibinfo {author} {\bibfnamefont {R.~P.}\ \bibnamefont
  {Mirin}},\ and\ \bibinfo {author} {\bibfnamefont {S.~W.}\ \bibnamefont
  {Nam}},\ }\bibfield  {title} {\bibinfo {title} {{Detecting single infrared
  photons with 93{\%} system efficiency}},\ }\href
  {https://doi.org/10.1038/nphoton.2013.13} {\bibfield  {journal} {\bibinfo
  {journal} {Nature Photonics}\ }\textbf {\bibinfo {volume} {7}},\ \bibinfo
  {pages} {210} (\bibinfo {year} {2013})},\ \Eprint
  {https://arxiv.org/abs/1209.5774} {arXiv:1209.5774} \BibitemShut {NoStop}%
\bibitem [{\citenamefont {Arrangoiz-Arriola}\ \emph {et~al.}(2018)\citenamefont
  {Arrangoiz-Arriola}, \citenamefont {Wollack}, \citenamefont {Pechal},
  \citenamefont {Witmer}, \citenamefont {Hill},\ and\ \citenamefont
  {Safavi-Naeini}}]{Arrangoiz-Arriola2018}%
  \BibitemOpen
  \bibfield  {author} {\bibinfo {author} {\bibfnamefont {P.}~\bibnamefont
  {Arrangoiz-Arriola}}, \bibinfo {author} {\bibfnamefont {E.~A.}\ \bibnamefont
  {Wollack}}, \bibinfo {author} {\bibfnamefont {M.}~\bibnamefont {Pechal}},
  \bibinfo {author} {\bibfnamefont {J.~D.}\ \bibnamefont {Witmer}}, \bibinfo
  {author} {\bibfnamefont {J.~T.}\ \bibnamefont {Hill}},\ and\ \bibinfo
  {author} {\bibfnamefont {A.~H.}\ \bibnamefont {Safavi-Naeini}},\ }\bibfield
  {title} {\bibinfo {title} {{Coupling a Superconducting Quantum Circuit to a
  Phononic Crystal Defect Cavity}},\ }\href
  {https://doi.org/10.1103/PhysRevX.8.031007} {\bibfield  {journal} {\bibinfo
  {journal} {Physical Review X}\ }\textbf {\bibinfo {volume} {8}},\ \bibinfo
  {pages} {031007} (\bibinfo {year} {2018})},\ \Eprint
  {https://arxiv.org/abs/1804.03625} {arXiv:1804.03625} \BibitemShut {NoStop}%
\bibitem [{\citenamefont {Mirhosseini}\ \emph {et~al.}(2020)\citenamefont
  {Mirhosseini}, \citenamefont {Sipahigil}, \citenamefont {Kalaee},\ and\
  \citenamefont {Painter}}]{Mirhosseini2020}%
  \BibitemOpen
  \bibfield  {author} {\bibinfo {author} {\bibfnamefont {M.}~\bibnamefont
  {Mirhosseini}}, \bibinfo {author} {\bibfnamefont {A.}~\bibnamefont
  {Sipahigil}}, \bibinfo {author} {\bibfnamefont {M.}~\bibnamefont {Kalaee}},\
  and\ \bibinfo {author} {\bibfnamefont {O.}~\bibnamefont {Painter}},\
  }\bibfield  {title} {\bibinfo {title} {{Superconducting qubit to optical
  photon transduction}},\ }\href {https://doi.org/10.1038/s41586-020-3038-6}
  {\bibfield  {journal} {\bibinfo  {journal} {Nature}\ }\textbf {\bibinfo
  {volume} {588}},\ \bibinfo {pages} {599} (\bibinfo {year}
  {2020})}\BibitemShut {NoStop}%
\bibitem [{\citenamefont {Golter}\ \emph {et~al.}(2016)\citenamefont {Golter},
  \citenamefont {Oo}, \citenamefont {Amezcua}, \citenamefont {Stewart},\ and\
  \citenamefont {Wang}}]{Golter2016}%
  \BibitemOpen
  \bibfield  {author} {\bibinfo {author} {\bibfnamefont {D.~A.}\ \bibnamefont
  {Golter}}, \bibinfo {author} {\bibfnamefont {T.}~\bibnamefont {Oo}}, \bibinfo
  {author} {\bibfnamefont {M.}~\bibnamefont {Amezcua}}, \bibinfo {author}
  {\bibfnamefont {K.~A.}\ \bibnamefont {Stewart}},\ and\ \bibinfo {author}
  {\bibfnamefont {H.}~\bibnamefont {Wang}},\ }\bibfield  {title} {\bibinfo
  {title} {{Optomechanical Quantum Control of a Nitrogen-Vacancy Center in
  Diamond}},\ }\href {https://doi.org/10.1103/PhysRevLett.116.143602}
  {\bibfield  {journal} {\bibinfo  {journal} {Physical Review Letters}\
  }\textbf {\bibinfo {volume} {116}},\ \bibinfo {pages} {143602} (\bibinfo
  {year} {2016})},\ \Eprint {https://arxiv.org/abs/1603.03804}
  {arXiv:1603.03804} \BibitemShut {NoStop}%
\bibitem [{\citenamefont {Leibfried}\ \emph {et~al.}(2003)\citenamefont
  {Leibfried}, \citenamefont {Blatt}, \citenamefont {Monroe},\ and\
  \citenamefont {Wineland}}]{Leibfried2003}%
  \BibitemOpen
  \bibfield  {author} {\bibinfo {author} {\bibfnamefont {D.}~\bibnamefont
  {Leibfried}}, \bibinfo {author} {\bibfnamefont {R.}~\bibnamefont {Blatt}},
  \bibinfo {author} {\bibfnamefont {C.}~\bibnamefont {Monroe}},\ and\ \bibinfo
  {author} {\bibfnamefont {D.}~\bibnamefont {Wineland}},\ }\bibfield  {title}
  {\bibinfo {title} {{Quantum dynamics of single trapped ions}},\ }\href
  {https://doi.org/10.1103/RevModPhys.75.281} {\bibfield  {journal} {\bibinfo
  {journal} {Reviews of Modern Physics}\ }\textbf {\bibinfo {volume} {75}},\
  \bibinfo {pages} {281} (\bibinfo {year} {2003})}\BibitemShut {NoStop}%
\bibitem [{\citenamefont {Zhou}\ \emph {et~al.}(2017)\citenamefont {Zhou},
  \citenamefont {Rasmita}, \citenamefont {Li}, \citenamefont {Xiong},
  \citenamefont {Aharonovich},\ and\ \citenamefont {Gao}}]{Zhou2017}%
  \BibitemOpen
  \bibfield  {author} {\bibinfo {author} {\bibfnamefont {Y.}~\bibnamefont
  {Zhou}}, \bibinfo {author} {\bibfnamefont {A.}~\bibnamefont {Rasmita}},
  \bibinfo {author} {\bibfnamefont {K.}~\bibnamefont {Li}}, \bibinfo {author}
  {\bibfnamefont {Q.}~\bibnamefont {Xiong}}, \bibinfo {author} {\bibfnamefont
  {I.}~\bibnamefont {Aharonovich}},\ and\ \bibinfo {author} {\bibfnamefont
  {W.-b.}\ \bibnamefont {Gao}},\ }\bibfield  {title} {\bibinfo {title}
  {{Coherent control of a strongly driven silicon vacancy optical transition in
  diamond}},\ }\href {https://doi.org/10.1038/ncomms14451} {\bibfield
  {journal} {\bibinfo  {journal} {Nature Communications}\ }\textbf {\bibinfo
  {volume} {8}},\ \bibinfo {pages} {14451} (\bibinfo {year}
  {2017})}\BibitemShut {NoStop}%
\bibitem [{\citenamefont {Baier}\ \emph {et~al.}(2020)\citenamefont {Baier},
  \citenamefont {Bradley}, \citenamefont {Middelburg}, \citenamefont
  {Dobrovitski}, \citenamefont {Taminiau},\ and\ \citenamefont
  {Hanson}}]{Baier2020}%
  \BibitemOpen
  \bibfield  {author} {\bibinfo {author} {\bibfnamefont {S.}~\bibnamefont
  {Baier}}, \bibinfo {author} {\bibfnamefont {C.~E.}\ \bibnamefont {Bradley}},
  \bibinfo {author} {\bibfnamefont {T.}~\bibnamefont {Middelburg}}, \bibinfo
  {author} {\bibfnamefont {V.~V.}\ \bibnamefont {Dobrovitski}}, \bibinfo
  {author} {\bibfnamefont {T.~H.}\ \bibnamefont {Taminiau}},\ and\ \bibinfo
  {author} {\bibfnamefont {R.}~\bibnamefont {Hanson}},\ }\bibfield  {title}
  {\bibinfo {title} {{Orbital and Spin Dynamics of Single Neutrally-Charged
  Nitrogen-Vacancy Centers in Diamond}},\ }\href
  {https://doi.org/10.1103/PhysRevLett.125.193601} {\bibfield  {journal}
  {\bibinfo  {journal} {Physical Review Letters}\ }\textbf {\bibinfo {volume}
  {125}},\ \bibinfo {pages} {193601} (\bibinfo {year} {2020})},\ \Eprint
  {https://arxiv.org/abs/2007.14673} {arXiv:2007.14673} \BibitemShut {NoStop}%
\bibitem [{\citenamefont {Barson}\ \emph {et~al.}(2019)\citenamefont {Barson},
  \citenamefont {Krausz}, \citenamefont {Manson},\ and\ \citenamefont
  {Doherty}}]{Barson2019}%
  \BibitemOpen
  \bibfield  {author} {\bibinfo {author} {\bibfnamefont {M.~S.}\ \bibnamefont
  {Barson}}, \bibinfo {author} {\bibfnamefont {E.}~\bibnamefont {Krausz}},
  \bibinfo {author} {\bibfnamefont {N.~B.}\ \bibnamefont {Manson}},\ and\
  \bibinfo {author} {\bibfnamefont {M.~W.}\ \bibnamefont {Doherty}},\
  }\bibfield  {title} {\bibinfo {title} {{The fine structure of the neutral
  nitrogen-vacancy center in diamond}},\ }\href
  {https://doi.org/10.1515/nanoph-2019-0142} {\bibfield  {journal} {\bibinfo
  {journal} {Nanophotonics}\ }\textbf {\bibinfo {volume} {8}},\ \bibinfo
  {pages} {1985} (\bibinfo {year} {2019})},\ \Eprint
  {https://arxiv.org/abs/1905.09037} {arXiv:1905.09037} \BibitemShut {NoStop}%
\bibitem [{\citenamefont {Dolde}\ \emph {et~al.}(2011)\citenamefont {Dolde},
  \citenamefont {Fedder}, \citenamefont {Doherty}, \citenamefont
  {N{\"{o}}bauer}, \citenamefont {Rempp}, \citenamefont {Balasubramanian},
  \citenamefont {Wolf}, \citenamefont {Reinhard}, \citenamefont {Hollenberg},
  \citenamefont {Jelezko},\ and\ \citenamefont {Wrachtrup}}]{Dolde2011}%
  \BibitemOpen
  \bibfield  {author} {\bibinfo {author} {\bibfnamefont {F.}~\bibnamefont
  {Dolde}}, \bibinfo {author} {\bibfnamefont {H.}~\bibnamefont {Fedder}},
  \bibinfo {author} {\bibfnamefont {M.~W.}\ \bibnamefont {Doherty}}, \bibinfo
  {author} {\bibfnamefont {T.}~\bibnamefont {N{\"{o}}bauer}}, \bibinfo {author}
  {\bibfnamefont {F.}~\bibnamefont {Rempp}}, \bibinfo {author} {\bibfnamefont
  {G.}~\bibnamefont {Balasubramanian}}, \bibinfo {author} {\bibfnamefont
  {T.}~\bibnamefont {Wolf}}, \bibinfo {author} {\bibfnamefont {F.}~\bibnamefont
  {Reinhard}}, \bibinfo {author} {\bibfnamefont {L.~C.~L.}\ \bibnamefont
  {Hollenberg}}, \bibinfo {author} {\bibfnamefont {F.}~\bibnamefont
  {Jelezko}},\ and\ \bibinfo {author} {\bibfnamefont {J.}~\bibnamefont
  {Wrachtrup}},\ }\bibfield  {title} {\bibinfo {title} {{Electric-field sensing
  using single diamond spins}},\ }\href {https://doi.org/10.1038/nphys1969}
  {\bibfield  {journal} {\bibinfo  {journal} {Nature Physics}\ }\textbf
  {\bibinfo {volume} {7}},\ \bibinfo {pages} {459} (\bibinfo {year}
  {2011})}\BibitemShut {NoStop}%
\bibitem [{\citenamefont {Knauer}\ \emph {et~al.}(2020)\citenamefont {Knauer},
  \citenamefont {Hadden},\ and\ \citenamefont {Rarity}}]{Knauer2020}%
  \BibitemOpen
  \bibfield  {author} {\bibinfo {author} {\bibfnamefont {S.}~\bibnamefont
  {Knauer}}, \bibinfo {author} {\bibfnamefont {J.~P.}\ \bibnamefont {Hadden}},\
  and\ \bibinfo {author} {\bibfnamefont {J.~G.}\ \bibnamefont {Rarity}},\
  }\bibfield  {title} {\bibinfo {title} {{In-situ measurements of fabrication
  induced strain in diamond photonic-structures using intrinsic colour
  centres}},\ }\href {https://doi.org/10.1038/s41534-020-0277-1} {\bibfield
  {journal} {\bibinfo  {journal} {npj Quantum Information}\ }\textbf {\bibinfo
  {volume} {6}},\ \bibinfo {pages} {50} (\bibinfo {year} {2020})}\BibitemShut
  {NoStop}%
\bibitem [{\citenamefont {Linskens}\ \emph {et~al.}(1996)\citenamefont
  {Linskens}, \citenamefont {Holleman}, \citenamefont {Dam},\ and\
  \citenamefont {Reuss}}]{Linskens1996}%
  \BibitemOpen
  \bibfield  {author} {\bibinfo {author} {\bibfnamefont {A.~F.}\ \bibnamefont
  {Linskens}}, \bibinfo {author} {\bibfnamefont {I.}~\bibnamefont {Holleman}},
  \bibinfo {author} {\bibfnamefont {N.}~\bibnamefont {Dam}},\ and\ \bibinfo
  {author} {\bibfnamefont {J.}~\bibnamefont {Reuss}},\ }\bibfield  {title}
  {\bibinfo {title} {{Two-photon Rabi oscillations}},\ }\href
  {https://doi.org/10.1103/PhysRevA.54.4854} {\bibfield  {journal} {\bibinfo
  {journal} {Physical Review A}\ }\textbf {\bibinfo {volume} {54}},\ \bibinfo
  {pages} {4854} (\bibinfo {year} {1996})}\BibitemShut {NoStop}%
\end{thebibliography}
%

	\appendix
	\renewcommand{\thefigure}{\Alph{section}.\arabic{table}}
	\section{Entanglement generation between a spin and photon}\label{app-sec:single-or-two-photon-protocol}
	\subsection{The single-photon protocol}
	Here we describe a protocol to generate entanglement between remote spins using a single photon. First, the color center spin in each node is prepared in the superposition state, $(1/\sqrt{2})(\ket{0}_\textrm{e}+\ket{1}_\textrm{e})$. We assume a situation where $\ket{0}_\textrm{e}$ can be optically excited and emit an optical photon while $\ket{1}_\textrm{e}$ cannot (for experiments about the generation of a coherent photon using a color center, see Refs.\cite{Bernien2013,Humphreys2018,Nguyen2019a}). Then, just after the emission of a coherent photon, the spin-photon entangled state is $(1/\sqrt{2})(\ket{0}_\textrm{e}\ket{1}_\textrm{opt}+\ket{1}_\textrm{e}\ket{0}_\textrm{opt})$. The state of the composite system (nodes A and B) is 
	\begin{equation}
		\begin{split}
			\frac{1}{2}\left(\ket{0}_\textrm{e}^\textrm{A}\ket{0}_\textrm{e}^\textrm{B}\ket{1}_\textrm{opt}^\textrm{A}\ket{1}_\textrm{opt}^\textrm{B}
			+\ket{0}_\textrm{e}^\textrm{A}\ket{1}_\textrm{e}^\textrm{B}\ket{1}_\textrm{opt}^\textrm{A}\ket{0}_\textrm{opt}^\textrm{B}\right. \\
			\left. +\ket{1}_\textrm{e}^\textrm{A}\ket{0}_\textrm{e}^\textrm{B}\ket{0}_\textrm{opt}^\textrm{A}\ket{1}_\textrm{opt}^\textrm{B}
			+\ket{1}_\textrm{e}^\textrm{A}\ket{1}_\textrm{e}^\textrm{B}\ket{0}_\textrm{opt}^\textrm{A}\ket{0}_\textrm{opt}^\textrm{B}
			\right), 
		\end{split}
	\end{equation}
	where $\ket{0}_\textrm{opt}$ ($\ket{1}_\textrm{opt}$) denotes the absence (presence) of an optical photon in the path from the spin to the detector. Then, the path information of the optical photon is erased using a beamsplitter before the detection of the photon at the photodetector. In a situation where only a single photon is detected, the state after the beamsplitter becomes 
	\begin{equation}
		\begin{split}
			\frac{1}{2}\left(\ket{0}_\textrm{e}^\textrm{A}\ket{1}_\textrm{e}^\textrm{B}\frac{1}{\sqrt{2}}(\ket{1}_\textrm{opt}^\textrm{A}\ket{0}_\textrm{opt}^\textrm{B}+\ket{0}_\textrm{opt}^\textrm{A}\ket{1}_\textrm{opt}^\textrm{B}) \right.\\	\left.+\ket{1}_\textrm{e}^\textrm{A}\ket{0}_\textrm{e}^\textrm{B}\frac{1}{\sqrt{2}}(\ket{1}_\textrm{opt}^\textrm{A}\ket{0}_\textrm{opt}^\textrm{B}-\ket{0}_\textrm{opt}^\textrm{A}\ket{1}_\textrm{opt}^\textrm{B})
			\right).
		\end{split}
	\end{equation}
	Here, the events detecting zero or two photons are ignored since they do not contribute to the entanglement generation, which is the reason for factor 1/2 in eq. (\ref{eq:entangle-rate}). 
	Thus, detection of a single photon by the detectors heralds the generation of the entangled state, $(1/\sqrt{2})(\ket{0}_\textrm{e}^\textrm{A}\ket{1}_\textrm{e}^\textrm{B}\pm\ket{1}_\textrm{e}^\textrm{A}\ket{0}_\textrm{e}^\textrm{B})$, depending on the side on which the detectors is clicked. It is noteworthy that this formula applies only to a situation in which the difference in the optical path length between each node and the beamsplitter is zero. Practically, we should consider an additional phase depending on the difference in the optical path length. Just before the incidence to the beamsplitter, $\ket{0}_\textrm{e}^\textrm{A}\ket{1}_\textrm{e}^\textrm{B}\ket{1}_\textrm{opt}^\textrm{A}\ket{0}_\textrm{opt}^\textrm{B}$ obtain a phase factor $e^{i\theta_\textrm{A}}$ while $\ket{1}_\textrm{e}^\textrm{A}\ket{0}_\textrm{e}^\textrm{B}\ket{0}_\textrm{opt}^\textrm{A}\ket{1}_\textrm{opt}^\textrm{B}$ obtain 
	$e^{i\theta_\textrm{B}}$ during the propagation of optical fiber. By factoring out $e^{i\theta_\textrm{A}}$, $e^{i\theta}\equiv e^{i(\theta_\textrm{B}-\theta_\textrm{A})}$ remains as a coefficient in the entangled states, $(1/\sqrt{2})(\ket{0}_\textrm{e}^\textrm{A}\ket{1}_\textrm{e}^\textrm{B}\pm e^{i\theta}\ket{1}_\textrm{e}^\textrm{A}\ket{0}_\textrm{e}^\textrm{B})$. To compensate this phase difference, the optical path length must be stabilized. If it is not, the phase of the resultant state becomes random, decreasing the fidelity of the entangled state.
	
	\subsection{The two-photon protocol}
	First, the two-photon protocol uses the same procedure as the single-photon protocol to generate the entangled state, with an additional phase as $(1/\sqrt{2})(\ket{0}_\textrm{e}^\textrm{A}\ket{1}_\textrm{e}^\textrm{B}\pm e^{i\theta}\ket{1}_\textrm{e}^\textrm{A}\ket{0}_\textrm{e}^\textrm{B})$. Then, $\pi$ pulse exchanging $\ket{0}_\textrm{e}$ and $\ket{1}_\textrm{e}$ is applied to the spin in each node, resulting in a state,
	\begin{equation}
		(1/\sqrt{2})(\ket{1}_\textrm{e}^\textrm{A}\ket{0}_\textrm{e}^\textrm{B}\pm e^{i\theta}\ket{0}_\textrm{e}^\textrm{A}\ket{1}_\textrm{e}^\textrm{B}).
	\end{equation}
	Then, the same gate operation is applied to emit a coherent photon . Thus, the state before a beamsplitter is 
	\begin{equation}
		\begin{split}
		(1/\sqrt{2})\left(e^{i\theta_\textrm{B}}\ket{1}_\textrm{e}^\textrm{A}\ket{0}_\textrm{e}^\textrm{B}\ket{0}_\textrm{opt}^\textrm{A}\ket{1}_\textrm{opt}^\textrm{B}\right. \\
		\left. \pm e^{i\theta_\textrm{A}}e^{i\theta}\ket{0}_\textrm{e}^\textrm{A}\ket{1}_\textrm{e}^\textrm{B}\ket{1}_\textrm{opt}^\textrm{A}\ket{0}_\textrm{opt}^\textrm{B}\right).
	\end{split}
	\end{equation}
	Since $e^{i\theta_\textrm{A}}e^{i\theta}=e^{i\theta_\textrm{B}}$, the additional phase in each term becomes the same,  $e^{i\theta_\textrm{B}}$. Thus, using the two-photon protocol, at the cost of the time it takes to interfere with the optical photon twice, we can treat $e^{i\theta}$ appearing in the single-photon protocol as a global phase, which we can ignore.

	\section{Transmission efficiency of optical photon emitted from a color center} \label{app-sec:trans-eff}
	The photon transmission efficiency from a color center to a detector, $\eta_\textrm{e}^\textrm{opt}$, can be defined as 
	\begin{equation}
		\eta_\textrm{e}^\textrm{opt}\equiv\eta_\textrm{int}\eta_\textrm{coupling}\eta_\textrm{loss}\eta_\textrm{det}
	\end{equation}where $\eta_\textrm{int}=C^\textrm{coh}_\textrm{opt}/(1+C_\textrm{coh}^\textrm{opt})$, $C_\textrm{opt}^\textrm{coh}=4g_\textrm{opt}^2/[\kappa(\gamma_\textrm{rad}+\gamma_\textrm{nonrad}+\gamma_\textrm{dp})]$ is the optical coherent cooperativity \cite{Borregaard2019} that characterizes the coherent photon emission from spin, $g_\textrm{opt}$ is the coupling rate between the emitter and the optical cavity, $\kappa$ is the total decay rate of the cavity field, $\gamma_\textrm{rad}$ is the free space radiative decay rate of the emitter, $\gamma_\textrm{nonrad}$ is the nonradiative decay rate of the emitter, $\gamma_\textrm{dp}$ is the pure dephasing rate of the emitter, $\eta_\textrm{coupling}$ is the coupling efficiency to the fiber, $\eta_\textrm{loss}$ is the transmission loss in the fiber, and $\eta_\textrm{det}$ is the photon detection efficiency at a detector. For the spin-cavity system with $C^\textrm{coh}_\textrm{opt}\gg1$, the coherent photon is emitted almost deterministically. 
	Experimentally, $C^\textrm{coh}_\textrm{opt}>100$ has already been demonstrated in a SiV$^-$ center in a 1 D photonic crystal cavity, indicating the generation of a coherent photon with a probability exceeding $100/(100+1)=0.99$ \cite{Bhaskar2020}. Besides, $\eta_\textrm{coupling}\sim 0.9$ has been achieved using double- and single-sided fiber tapers \cite{Ruf2021}. If we take the transmission loss at the fiber to be 10 dB/km for the visible light (600-800 nm light considering ZPL of various color centers) and the distance between dilution refrigerators to be 5 m, $\eta_\textrm{loss}\sim0.99$, which is negligible. Through the use of a superconducting single-photon detector, $\eta_\textrm{det}>0.99$ has been reported for a 737 nm photon \cite{Marsili2013}. Hence, considering the present state-of-the-art technologies, $\eta_\textrm{e}^\textrm{opt}\sim0.9$ is already achievable.

	\section{Numerical simulation of the spin interacting with a microwave photon}\label{app-sec:numerical_simulation}
	
	Here, we numerically calculate the absorption (emission) efficiency of a microwave time-bin photon by the electron spin following the formulation in Ref.\cite{Neuman2021}. As mentioned in the main text, the spin is coupled to a vibration mode of the optomechanical cavity via mechanical-spin coupling, $g_\textrm{m-e}$. The mechanical mode is coupled to the electromagnetic mode of the microwave resonator by the  electro-mechanical coupling, $g_\textrm{mw-m}$, resulting from the  piezoelectric effect. Thus, the Hamiltonian in this system can be modeled simply as 
	\begin{equation}\label{eq:hamiltonian}
		\begin{split}
			H/\hbar
			&= \omega_\textrm{mw}a_\textrm{mw}^\dagger a_\textrm{mw} + \omega_\textrm{m}b_\textrm{m}^\dagger b_\textrm{m}+ \omega_\textrm{e}\sigma_\textrm{e}^\dagger\sigma_\textrm{e}\\
			&+g_\textrm{mw-m}(a_\textrm{mw}b_\textrm{m}^\dagger+a_\textrm{mw}^\dagger b_\textrm{m})\\
			&+g_\textrm{m-e}(b_\textrm{m} \sigma_\textrm{e}^\dagger+b_\textrm{m}^\dagger \sigma_\textrm{e}),
		\end{split}
	\end{equation}
	where $a_\textrm{mw}$ ($a_\textrm{mw}^\dagger$) is the annihilation (creation) operator of the microwave photon at the microwave resonator, $b_\textrm{m}$ ($b_\textrm{m}^\dagger$) is the annihilation (creation) operator of the microwave phonon at the optomechanical cavity, and $\sigma_\textrm{e}$ ($\sigma_\textrm{e}^\dagger$)  is the electron spin lowering (raising) operator. The resonance frequencies of the microwave resonator, the phononic resonator, and the electron spin are $\omega_\textrm{mw}$, $\omega_\textrm{m}$, and $\omega_\textrm{e}$, respectively. In the following, we consider the resonant condition, $\omega_\textrm{mw}=\omega_\textrm{m}=\omega_\textrm{e}$.

	We calculated the time evolution of the density matrix, $\rho$, of this system based on the master equation,
	\begin{equation}\label{eq:master}
		\frac{\textrm{d}\rho}{\textrm{d}t}=\frac{1}{i\hbar}[H,\rho]+\sum_i	\left[ \frac{\gamma_{c_i}}{2}(2c_i\rho c_i^\dagger-\{c_i^\dagger c_i,\rho\})\right] ,
	\end{equation}
	where $c_i\in{\left\{a_\textrm{mw},b_\textrm{m},\sigma_\textrm{e}^\dagger\sigma_\textrm{e}\right\}}$ and $\gamma_{c_i}\in{\left\{\gamma_\textrm{mw},\gamma_\textrm{m},\gamma_\textrm{e}\right\}}$ corresponding to the decay (decoherence) rates of the respective excited states. For the microwave cavity and the optomechanical cavity, $T_1$ limits the experimentally feasible $T_2$. Thus, $T_1$ processes are considered for the microwave cavity and the optomechanical cavity. On the other hand, $T_2$ relaxation is considered for the electron spin, since $T_1\gg T_2$. 
	
	To estimate the absorption efficiency of microwave photons by the spin, we performed a calculation based on eq. (\ref{eq:master}) with an initial state in which a microwave photon is in the microwave cavity. Actually, the microwave photon comes from a waveguide coupled to the microwave cavity. For simplicity, we assumed that the coupling between the microwave cavity and the waveguide can be arbitrarily tuned. Once the microwave photon enters the microwave cavity, the coupling to the waveguide turns off and the waveguide can be ignored. Thus, we focus on only a system consisting of three qubits.

\begin{figure}
		\begin{center}
			\includegraphics[width=80mm]{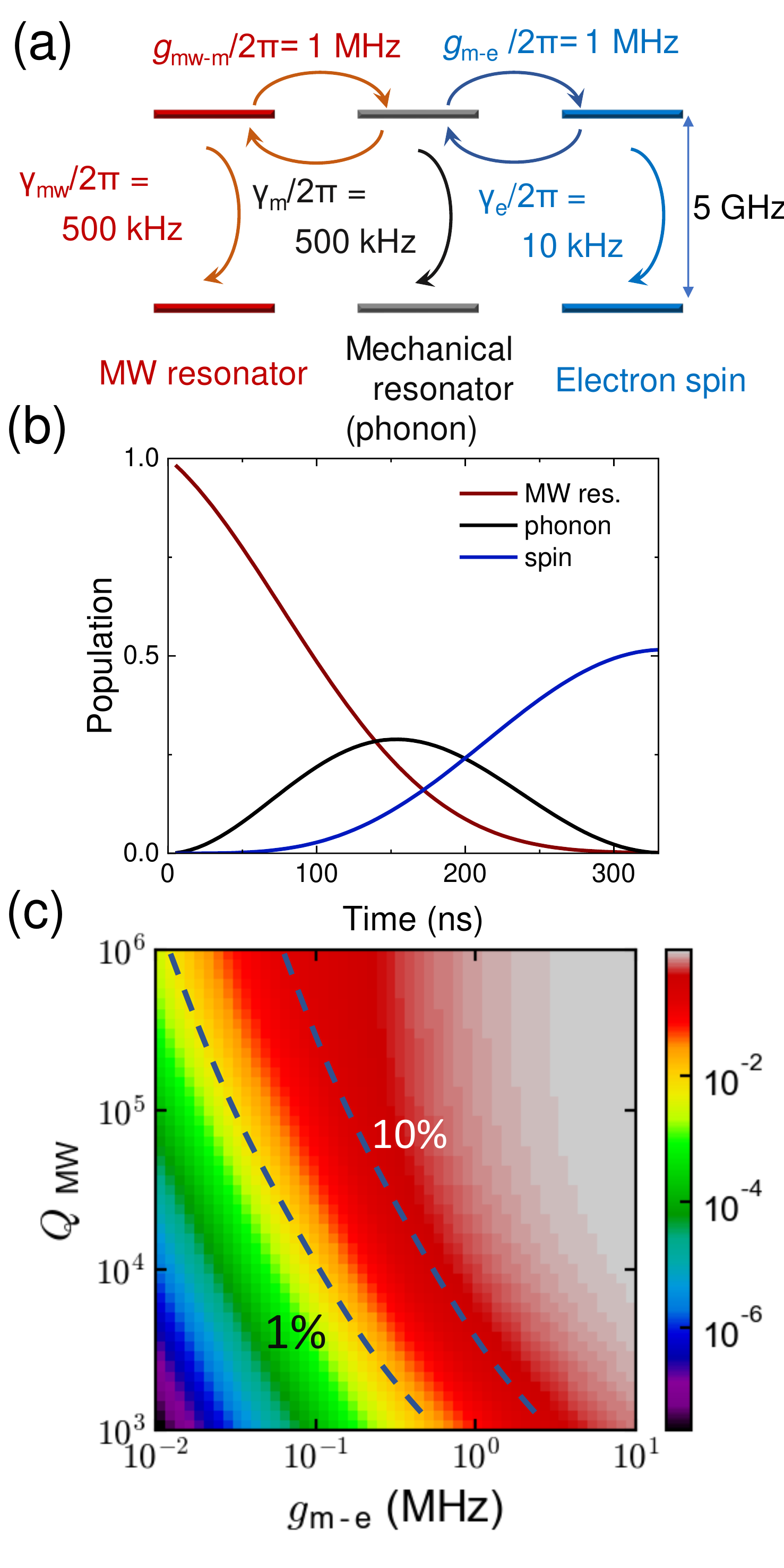}
		\end{center}
		\caption{(a) Schematic of the energy levels of three qubits with the related parameters. (b) Time evolution of the system described by eq. (\ref{eq:hamiltonian}) and (\ref{eq:master}) based on the parameters shown in (a). (c) Color map of the maximum population of the electron spin during the time evolution as a function of the mechanical-spin coupling and the quality factor of the microwave resonator, $Q_\textrm{mw}=\omega_\textrm{mw}/\gamma_\textrm{mw}$. 
		}
		\label{fig:app-master-eq}
	\end{figure}

	We chose the following parameters for the calculation:  $\omega_\textrm{mw}/2\pi=\omega_\textrm{m}/2\pi=\omega_\textrm{e}/2\pi=$ 5 GHz, $g_\textrm{mw-m}/2\pi=$ 1 MHz \cite{Arrangoiz-Arriola2018,Mirhosseini2020}, $g_\textrm{m-e}/2\pi$=1 MHz \cite{Neuman2021}, $\left\{\gamma_\textrm{mw}/2\pi,\gamma_\textrm{m}/2\pi,\gamma_\textrm{e}/2\pi \right\}=$ $\{$500 kHz, 500 kHz, 10 kHz$\}$ corresponding to the quality factor of $1\times10^4$ in the microwave cavity and the optomechanical cavity. These parameters are shown in Fig. \ref{fig:app-master-eq} (a). It should be noted that, $g_\textrm{m-e}/2\pi\sim1$ MHz is achievable only for color centers with high strain sensitivity, such as the excited state of NV$^{-}$ and the ground and excited states of SiV$^{-}$.
	
	
	Fig. \ref{fig:app-master-eq} (b) shows the time evolution of the population of the qubits described above assuming that one photon exists in the microwave resonator at the initial state. Based on the parameters shown in Fig. \ref{fig:app-master-eq} (a), the population transfer efficiency to the electron spin is $\sim$ 50$\%$ and the time for the transfer is $\sim$300 ns. The transfer efficiency corresponds to $\eta_\textrm{e}^\textrm{mw}$ in the main text. Thus, we consider that the absorption (emission) of a microwave photon with $\eta_\textrm{e}^\textrm{mw}>0.1$ is possible considering the current state-of-the-art technologies described in the main text. Besides, since the absorption (emission) of a microwave time-bin photon needs to repeat the time evolution twice, we roughly estimated that the time to absorb (emit) a microwave time-bin photon by the electron spin is $\sim$1 $\mu$s. Fig. \ref{fig:app-master-eq} (c) shows a color map of the maximum population of the electron spin during the time evolution as a function of the mechanical-spin coupling and the quality factor of the microwave resonator, $Q_\textrm{mw}=\omega_\textrm{mw}/\gamma_\textrm{mw}$. 
	If $g_\textrm{e-m}/2\pi\sim$ 1 MHz, it is possible to realize $\eta_\textrm{e}^\textrm{mw}>0.1$ with the quality factor of the microwave resonator of $10^3-10^4$, which is easily achieved using a superconducting resonator.

	\section{Entanglement generation between a superconducting qubit and spin} \label{app-sec:absorb_emit_microwave_using_spin}
	
	In this section, we describe in detail the process of absorbing (emitting) a microwave time-bin qubit by the spin.
	\subsection{Absorption of a microwave photon at the electron spin}
\subsubsection{A color center whose excited state has high strain sensitivity}\label{app-subsubsec:excited-state-absorption}
	\begin{figure*}
		\begin{center}
			\includegraphics[width=150mm]{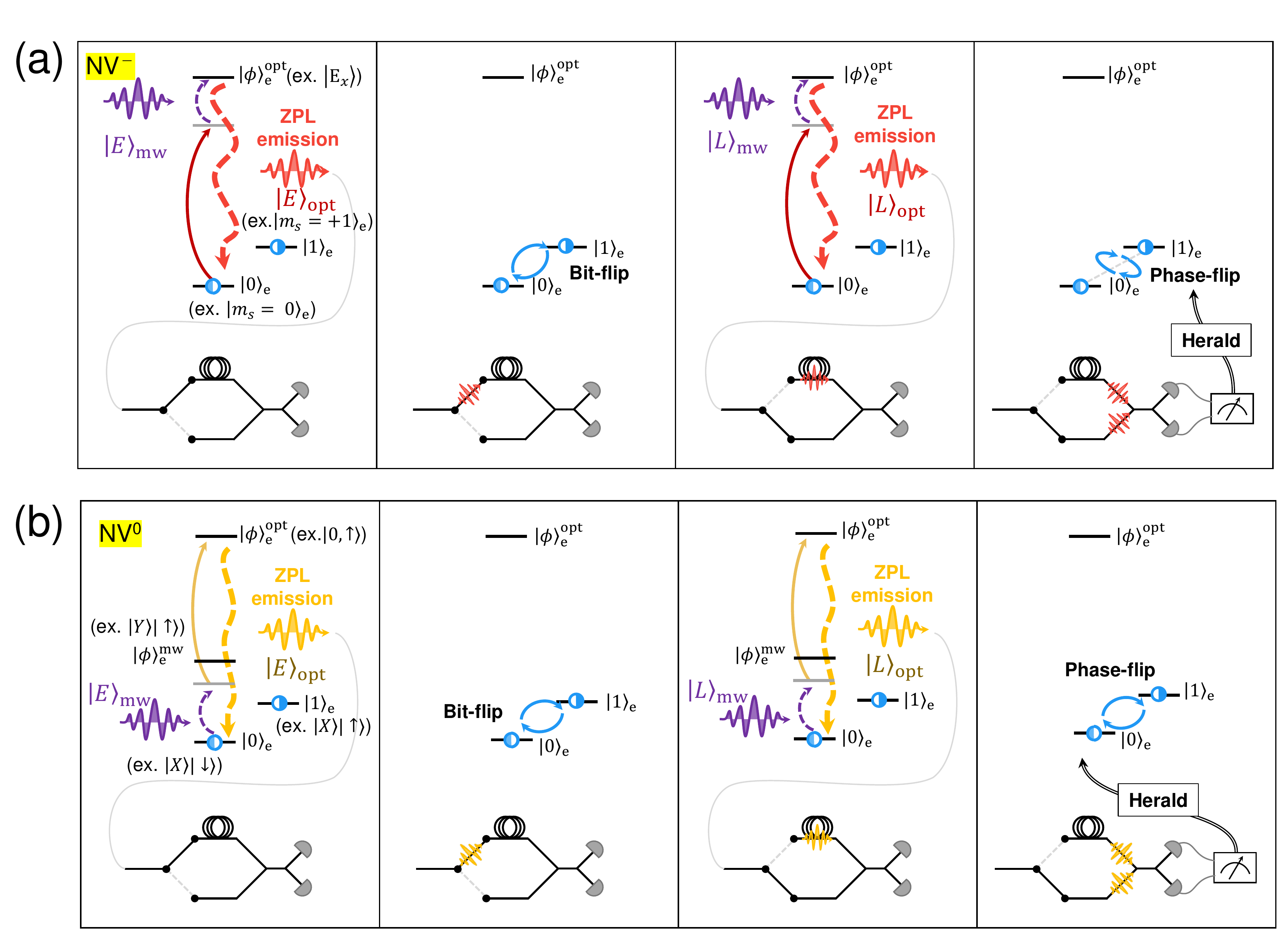}
		\end{center}
		\caption{(a) Schematic of the absorption of a microwave time-bin qubit by a spin whose optically excited state has high strain sensitivity. The energy level of NV$^-$ is considered as an example.
			(b) Schematic of the absorption of a microwave time-bin qubit by a spin whose ground state has high strain sensitivity. The energy level of NV$^0$ is considered as an example.
		}
		\label{fig:app-microwave-absorption-bySpin_1}
	\end{figure*}

	As mentioned in the main text, a superconducting qubit coupled to a cavity can emit a coherent microwave time-bin qubit, 
	\begin{equation}
		\frac{1}{\sqrt{2}}(\ket{g}_\textrm{sc}\ket{L}_\textrm{mw}+\ket{e}_\textrm{sc}\ket{E}_\textrm{mw}).
	\end{equation}
	We consider a process whereby the electron spin in diamond absorbs this microwave time-bin qubit and generates entanglement with the superconducting qubit. 
	
	For a case of a color center whose optically excited state has high strain sensitivity (e.g., NV$^-$), the procedure is shown in Fig.  \ref{fig:app-microwave-absorption-bySpin_1} (a) taking NV$^-$ as an example.	
	 First, the electron spin is prepared in the state,
	\begin{equation}
		\frac{1}{\sqrt{2}}(\ket{0}_\textrm{e}+\ket{1}_\textrm{e}),
	\end{equation}
	where $\ket{0}_\textrm{e}\equiv\ket{m_s=0}$ and $\ket{1}_\textrm{e}\equiv\ket{m_s=+1}$ for NV$^-$ and $m_s$ is the angular momentum of the spin. Then, $\ket{0}_\textrm{e}$ is optically virtually excited to $\ket{\phi}_\textrm{e}^\textrm{opt}$ with detuning equal to the frequency of the microwave phonon, where $\ket{\phi}_\textrm{e}^\textrm{opt}$ is the optically excited state of $\ket{0}_\textrm{e}$ (e.g., $\ket{E_\textrm{x}}$ for NV$^-$). This kind of excitation using a red sideband was previously demonstrated in NV$^-$ using a classical strain field \cite{Golter2016}. 	
	The spin that was initially at $\ket{0}_\textrm{e}$ interacts with $\ket{E}_\textrm{mw}$ in its virtually excited state and emits $\ket{E}_\textrm{opt}$ via spontaneous emission if $\ket{E}_\textrm{mw}$ is absorbed. After the application of the $\pi$ pulse exchanging $\ket{0}_\textrm{e}$ and $\ket{1}_\textrm{e}$, the state at $\ket{0}_\textrm{e}$ is optically excited again and emits $\ket{L}_\textrm{opt}$ conditioned on the successful absorption of $\ket{L}_\textrm{mw}$. Then, after the application of the second bit flip exchanging $\ket{0}_\textrm{e}$ and $\ket{1}_\textrm{e}$, $\ket{E}_\textrm{opt}$ and $\ket{L}_\textrm{opt}$ entangle with the spin as
	\begin{equation}
		\frac{1}{\sqrt{2}}(\ket{L}_\textrm{opt}\ket{1}_\textrm{e}+\ket{E}_\textrm{opt}\ket{0}_\textrm{e}).
	\end{equation}
	Following the measurement of the photonic time-bin qubit in the basis, $(1/\sqrt{2})(\ket{0}_\textrm{opt}^\textrm{A}\ket{1}_\textrm{opt}^\textrm{B}\pm\ket{1}_\textrm{opt}^\textrm{A}\ket{0}_\textrm{opt}^\textrm{B})$, using a delay-line and a beamsplitter, the application of the phase-flip gate to the spin depending on the measurement outcome teleports the quantum state of the phonon to the spin. Consequently, entanglement between the superconducting qubit and the microwave time-bin qubit is transferred to the spin as
	\begin{equation}
		\frac{1}{\sqrt{2}}(\ket{g}_\textrm{sc}\ket{1}_\textrm{e}+\ket{e}_\textrm{sc}\ket{0}_\textrm{e}).
	\end{equation}

    Next, we consider the effective mechanical-spin coupling obtained in this scheme. The effective interaction Hamiltonian, $H_\textrm{int}$, for the first red sideband transition can be expressed as
    \begin{equation}
    	H_\textrm{int}=i\frac{\hbar g_\textrm{m-e} \Omega_0}{2\omega_\textrm{m}}(b_\textrm{m}\sigma_\textrm{opt}^\dagger+b_\textrm{m}^\dagger\sigma_\textrm{opt}),
    \end{equation}
    where $\Omega_0$ is the Rabi frequency for the optical field and $\sigma_\textrm{opt}$ is the raising operator for the optical transition \cite{Leibfried2003,Golter2016}. Then, the effective Rabi frequency for the sideband transition, $\Omega_\textrm{sb}^\textrm{eff}$, is given by
    \begin{equation}
    	\Omega_\textrm{sb}^\textrm{eff}=g_\textrm{m-e}\sqrt{n}\Omega_0/\omega_\textrm{m},
    \end{equation}
    where $n$ is the average number of phonons. Since we consider excitation of the spin by the single phonon, $n=1$. Assuming that  $g_\textrm{m-e}/2\pi$ can reach 1-10 MHz using a color center with high strain susceptibility ($\sim$1 PHz/strain)  with a phononic crystal cavity \cite{Neuman2021}, $\Omega_0/2\pi\sim$ 1 GHz \cite{Zhou2017}, and $\omega_\textrm{m}/2\pi=5-10$ GHz, $\Omega_\textrm{sb}^\textrm{eff}/2\pi$ becomes 0.1-2 MHz. Here, $g_\textrm{m-e}$ shown in Fig. \ref{fig:app-master-eq} (a) and eq.(\ref{eq:hamiltonian}) is replaced by $\Omega_\textrm{sb}^\textrm{eff}/2$. As mentioned in Appendix \ref{app-sec:numerical_simulation}, mechanical-spin coupling on the order of 1 MHz is enough to transfer a quantum state. Thus, this scheme can transfer the state of the microwave time-bin qubit to the electron spin above the minimum required efficiency, $\eta_\textrm{e}^\textrm{mw}>0.02$.

\subsubsection{A color center whose ground state has high strain sensitivity}\label{app-subsubsec:1st-protocol_high_strain}

	For a color center whose ground state has high strain sensitivity (e.g., NV$^0$, SiV$^-$),  a procedure to absorb a microwave time-bin qubit is shown in Fig. \ref{fig:app-microwave-absorption-bySpin_1} (b). Here, we take NV$^0$ as an example since the ground state splitting due to the spin-orbit coupling is $\sim$ 10 GHz, which makes it easier to access the excited states than with the group-IV color centers, whose ground state splitting range is from 50 to thousands of gigahertz \cite{Ruf2021}. Besides, since its ground state has the orbital degree of freedom, it is considered to be sensitive to the strain and electric field, similar to the excited state of NV$^-$. High strain and electric field sensitivity are advantageous in the proposed protocol.

	Before explaining the detail of Fig. \ref{fig:app-microwave-absorption-bySpin_1} (b), we first consider the wavefunction of the ground state of NV$^0$ and show that the  energy level can be used for the protocol we describe. The ground state of  NV$^0$ is a spin doublet \cite{Baier2020}. The model Hamiltonian of NV$^0$ \cite{Barson2019,Baier2020} is
	\begin{equation}
		\begin{split}
			H_\textrm{NV}=&g\mu_\textrm{B}S_zB_z+l\mu_\textrm{B}L_zB_z+2\lambda L_zS_z\\
			&+\epsilon_{\perp}(L_-+L_+)+d_{\perp}(L_-+L_+),
		\end{split}	
	\end{equation}
	where $g$ is the spin g-factor, $\mu_\textrm{B}$ is the Bohr magneton, $l$ is the orbital g-factor, $\lambda$ is the spin-orbit interaction parameter, $\epsilon_{\perp}$ is the perpendicular strain parameter, and $d_{\perp}$ is the perpendicular electric field parameter. $L_z=\sigma_z$ and $S_z=(1/2)\sigma_z$ are the orbital and spin operators, while $L_\pm=\ket{\pm}\bra{\mp}$ with  $\ket{\pm}_\textrm{e}=\mp(1/\sqrt{2})(\ket{X}_\textrm{e}\pm i\ket{Y}_\textrm{e})$ are the orbital operators, and {$\ket{X}_\textrm{e}$, $\ket{Y}_\textrm{e}$} are the strain eigenstates. The z-axis is defined parallel to the NV axis. The last term is introduced phenomenologically considering that an electric field can be treated as a local strain \cite{Dolde2011,Knauer2020}. 

	The ground state with zero strain and zero electric field can be described by the eigenstates, $\ket{+}_\textrm{e}\ket{\downarrow}_\textrm{e}$, $\ket{-}_\textrm{e}\ket{\uparrow}_\textrm{e}$, $\ket{-}_\textrm{e}\ket{\downarrow}_\textrm{e}$, $\ket{+}_\textrm{e}\ket{\uparrow}_\textrm{e}$. Since the lower energy states, ($\ket{+}_\textrm{e}\ket{\downarrow}_\textrm{e}$, $\ket{-}_\textrm{e}\ket{\uparrow}_\textrm{e}$), have different spin and orbital characteristics, controlling these states using only a magnetic field or strain is inhibited in the first order. On the other hand, in a high-strain regime, the mixing of orbital states results in the eigenstates, $\ket{X}_\textrm{e}\ket{\downarrow}_\textrm{e}$, $\ket{X}_\textrm{e}\ket{\uparrow}_\textrm{e}$, $\ket{Y}_\textrm{e}\ket{\downarrow}_\textrm{e}$, $\ket{Y}_\textrm{e}\ket{\uparrow}$. Thus, flipping a spin using a magnetic field is possible in the lower energy states,  ($\ket{X}_\textrm{e}\ket{\downarrow}_\textrm{e}$, $\ket{X}_\textrm{e}\ket{\uparrow}_\textrm{e}$), since they have the same orbital characteristics. Besides, $\ket{X}_\textrm{e}\ket{\downarrow}_\textrm{e}$ is excited to $\ket{Y}_\textrm{e}\ket{\downarrow}_\textrm{e}$ using only the strain or an electric field. Then, we express these states using the notation in the main text,  $\ket{X}_\textrm{e}\ket{\downarrow}_\textrm{e}\equiv\ket{0}_\textrm{e}$, $\ket{X}_\textrm{e}\ket{\uparrow}_\textrm{e}\equiv\ket{1}_\textrm{e}$, and $\ket{Y}_\textrm{e}\ket{\downarrow}_\textrm{e}\equiv\ket{\phi}_\textrm{e}^\textrm{mw}$. We also apply a static magnetic field to split the degenerate energy levels, $\ket{\downarrow}_\textrm{e}$ and $\ket{\uparrow}_\textrm{e}$.

	To generate entanglement between a superconducting qubit and the spin, $\ket{E}_\textrm{mw}$ virtually excites $\ket{0}_\textrm{e}$ to $\ket{\phi}_\textrm{e}^\textrm{mw}$ with detuning, which mitigates orbital relaxation as shown in Fig. \ref{fig:app-microwave-absorption-bySpin_1} (b). Then, $\ket{\phi}_\textrm{e}^\textrm{mw}$ is optically excited with a laser pulse and emits $\ket{E}_\textrm{opt}$. This procedure is repeated for $\ket{L}_\textrm{mw}$, the same as in the case of NV$^-$, generating the entanglement between the optical photon and the spin, $(1/\sqrt{2})(\ket{E}_\textrm{opt}\ket{0}_\textrm{e}+\ket{L}_\textrm{opt}\ket{1}_\textrm{e})$. The rest the operation to generate entanglement between the superconducting qubit and the spin is the same as the protocol described in Appendix  \ref{app-subsubsec:excited-state-absorption}.

	The process is the same for the typical two-photon (phonon) absorption in a three-level atom. Thus, the effective Rabi frequency between $\ket{0}_\textrm{e}$ and $\ket{\phi}_\textrm{e}^\textrm{opt}$ can be expressed as $\Omega_\textrm{three}^\textrm{eff}=g_\textrm{m-e}\Omega_0/\Delta$ \cite{Linskens1996}, where $\Omega_0$ is the optical Rabi frequency exciting $\ket{\phi}_\textrm{e}^\textrm{mw}$ to $\ket{\phi}_\textrm{e}^\textrm{opt}$, and $\Delta$ is the detuning of $\omega_\textrm{m}$ from $\ket{\phi}_\textrm{e}^\textrm{mw}$.  If we take $g_\textrm{m-e}/2\pi=$ 1-10 MHz, $\Omega_0/2\pi=100$ MHz, $\Delta/2\pi=1$ GHz satisfying the adiabatic condition, $\Delta\gg g_\textrm{m-e}$, $\Omega_0$,  $\Omega_\textrm{three}^\textrm{eff}/2\pi$ becomes 0.1-1 MHz. In this case, $\Omega_\textrm{three}^\textrm{eff}/2$ replaces $g_\textrm{m-e}$ in Fig. \ref{fig:app-master-eq} (a) and eq.(\ref{eq:hamiltonian}). From Fig. \ref{fig:app-master-eq} (c), mechanical and strain coupling on the order of $\sim1$ MHz is enough to transfer the quantum state to the spin with the required efficiency, $\eta_\textrm{e}^\textrm{mw}>0.02$, suggesting that this protocol can be applied to the entanglement generation protocol. 	It should be noted that in a high-strain regime, the phonon absorption rate decreases to 0.1-0.5 of the rate with zero strain \cite{Meesala2018}.  Thus, it is desirable to achieve high mechanical-spin coupling of more than 10 MHz using an optimized phononic crystal cavity.

	\subsection{Emission of a microwave photon from the electron spin}
	Here, we describe a procedure to emit a microwave time-bin qubit from the spin. We take NV$^0$ as an example, as with the case shown in Appendix \ref{app-subsubsec:1st-protocol_high_strain}.  Different from the scheme discussed above, here we consider a moderate- to high-strain regime. Since an excitation to $\ket{\phi}_\textrm{e}^\textrm{mw}$ is forbidden using only a magnetic field, we consider operating NV$^0$ in the regime where the strain is not too strong so that $\ket{X}$ and $\ket{Y}$ are weakly mixed, $\ket{X}\rightarrow\ket{X'}$ and $\ket{Y}\rightarrow\ket{Y'}$. Then, it is possible to excite $\ket{X'}_\textrm{e}\ket{\downarrow}_\textrm{e}$ to $\ket{Y'}_\textrm{e}\ket{\uparrow}_\textrm{e}$ perturbatively using a magnetic field.
	We then put  $\ket{X'}_\textrm{e}\ket{\downarrow}_\textrm{e}\equiv\ket{0}_\textrm{e}$, $\ket{X'}_\textrm{e}\ket{\uparrow}_\textrm{e}\equiv\ket{1}_\textrm{e}$, and $\ket{Y'
	}_\textrm{e}\ket{\uparrow}_\textrm{e}\equiv\ket{\phi}_\textrm{e}^\textrm{mw}$.

	First, the electron spin is prepared in the state,
	\begin{equation}
		\frac{1}{\sqrt{2}}(\ket{0}_\textrm{e}+\ket{1}_\textrm{e}).
	\end{equation}
	A microwave magnetic field generated at an external signal source is then applied to the electron spin. Here, we assume the use of a microwave magnetic field to separate the external driving field and the generated microwave phononic time-bin qubit. 
	Since $\ket{\phi}_\textrm{e}^\textrm{mw}$ couples to the vibration mode (phonon) of the optomechanical crystal, $\ket{\phi}_\textrm{e}^\textrm{mw}$ emits the phonon to the optomechanical crystal, which is the reverse of Fig. \ref{fig:app-master-eq} (b). The microwave phonon is then converted to a microwave photon via the piezoelectric effect. The microwave cavity coupled to the optomechanical crystal receives the converted microwave photon and emits it to a waveguide as $\ket{E}_\textrm{mw}$. After the application of the $\pi$ pulse exchanging $\ket{0}_\textrm{e}$ and $\ket{1}_\textrm{e}$, the state at $\ket{0}_\textrm{e}$ is excited to $\ket{\phi}_\textrm{e}^\textrm{mw}$ again, emitting $\ket{L}_\textrm{mw}$ to the optomechanical crystal. Thus, the resultant microwave photon-spin entangled state is
	\begin{equation}
		\frac{1}{\sqrt{2}}(\ket{E}_\textrm{mw}\ket{0}_\textrm{e}+\ket{L}_\textrm{mw}\ket{1}_\textrm{e}).
	\end{equation}
	After a superconducting qubit, which is prepared in $\ket{g}_\textrm{sc}$, absorbs time-bin microwave qubit with appropriate gate operations \cite{Kurpiers2019}, the state becomes
	\begin{equation}
		\frac{1}{\sqrt{2}}(\ket{e}_\textrm{sc}\ket{0}_\textrm{e}+\ket{g}_\textrm{sc}\ket{1}_\textrm{e}).
	\end{equation}

\end{document}